\documentclass[final,3p]{CSP}
\usepackage{amsmath}
\usepackage{amssymb}
\usepackage{changepage}
\usepackage{graphicx}
\usepackage{subfig}
\usepackage{float}
\usepackage[bottom]{footmisc}
\usepackage{url}
\usepackage[T1]{fontenc}

\begin{document}

\begin{frontmatter}
\begin{large}{Team 44: The Oracles}\end{large}

\title{PECAIQR: A Model for Infectious Disease Applied to the Covid-19 Epidemic}
\author{Richard Bao, August Chen, Jethin Gowda, Shiva Mudide}

\begin{abstract}\rm
\begin{adjustwidth}{2cm}{2cm}{\itshape\textbf{Abstract:}} 
The Covid-19 pandemic has made clear the need to improve modern multivariate time-series forecasting models. Current state of the art predictions of future daily deaths and, especially, hospital resource usage have confidence intervals that are unacceptably wide. Policy makers and hospitals require accurate forecasts to make informed decisions on passing legislation and allocating resources. We used US county-level data on daily deaths and population statistics to forecast future deaths. We extended the SIR epidemiological model to a novel model we call the  PECAIQR model. It adds several new variables and parameters to the naive SIR model by taking into account the ramifications of the partial quarantining implemented in the US. We fitted data to the model parameters with numerical integration. Because of the fit degeneracy in parameter space and non-constant nature of the parameters, we developed several methods to optimize our fit, such as training on the data tail and training on specific policy regimes. We use cross-validation to tune our hyper parameters at the county level and generate a CDF for future daily deaths. For predictions made from training data up to May 25th, we consistently obtained an averaged pinball loss score of 0.096 on a 14 day forecast. We finally present examples of possible avenues for utility from our model. We generate longer-time horizon predictions over various 1-month windows in the past, forecast how many medical resources such as ventilators and ICU beds will be needed in counties, and evaluate the efficacy of our model in other countries.\footnote{This work was done as part of the CS156 course at Caltech in the spring of 2020.} \\

\end{adjustwidth}
\end{abstract}
\end{frontmatter}

\section{Data Usage and Preprocessing}
\noindent We used the county-level cumulative and daily death reporting from the New York times \cite{nyc}, as well as the county-level active cases reporting from Johns Hopkins University \cite{jhu}. We also used county-level population statistics from the 2017 American Community Survey and county-level policy dating information from Johns Hopkins University. Specifically, we loaded the policy dating information regarding the implementation of the stay at home orders. We loaded these data into a data frame by matching the county FIPS codes from each source, and computed a moving average death statistic as an additional feature, using a window of size three days. We dealt with missing values in the FIPS codes and death reporting by removing the corresponding observations. The only exception to this was the county 36061, which represents New York City. This county had a missing FIPS code, but had well reported data. Since New York City is the most active Covid-19 hotspot in the United States, its manual inclusion was necessary. In addition to the aforementioned data sources, we experimented with mobility data, but this did not make it into our final epidemiological model because there was no simple mapping to any of the model variables, and we believed the policy dating information was sufficient to establish distinct regimes in the data.
\section{PECAIQR Epidemiological Model}
\noindent
The traditional SIR epidemiological model breaks up a region’s population into three separate groups: Susceptible, Infected, and Removed \cite{sir}. The primary flaw of applying this model to the current Covid-19 pandemic is the poor approximation of assuming that all people in each group have uniform experiences. Clumping the population into only three groups is an example of omitted-variable bias, excluding major realities caused by the scale of the pandemic. The PECAIQR model that is described below takes each group in the SIR model and breaks them down into another level of classification. The variables in our model are developed following simple logical arguments based on current global realities and widely accepted scientific and epidemiological results.\\

\noindent
We first break down the Susceptible class of the SIR model. Due to policies implemented in the US, many individuals have been self-quarantining \cite{stayhome}. It is sensible then that each day only a fraction of the Susceptible population are exposed to potentially getting the virus from the outside world. Following this logic, our model breaks up the Susceptible population into three classes: Protected, Exposed, and Carriers (explained in detail below). Next, we consider the Infected group of SIR. There is evidence from the large data sets gathered in South Korea and other well-respected global scientific efforts that a significant fraction of people infected with Covid-19 do not display visible symptoms \cite{wang2020unique}. Using this line of logic our model breaks the Infected group into two classes: Asymptomatic and Infectious. An important note is that we assume these Asymptomatic and Infectious people still actively participate in the community, enabling them to come in contact with Exposed people. Finally, we are left with the Removed group. Our model incorporates two lines of logic in the breakdown of this group. First, it is sensible that a portion of the members of the Asymptomatic and Infectious classes end up self-quarantining (at least in effect) as a result of: getting tested, showing initial symptoms, or having an “intuition” they contracted the disease. Second, much of the evidence about Covid-19 so far shows that an individual who contracts the disease cannot get the disease again (at least on the order of a few months) \cite{recovered}. Following these arguments, the model breaks the Removed group into: Quarantined and Removed classes. Removed further has subclasses of Dead people and Recovered people, who are assumed to be immune from the virus. Each of the variables we describe change continuously with time. We describe the variables for a given time $t$ (associated with some given day $T$).\\

\noindent
\textit{Protected}: People who did not go outside to expose themselves to any infected individuals (Asymptomatic or Infectious) on t, though they have a chance of coming in contact with a Carrier. Any person in Protected also has a chance of joining the Exposed class at a later time $t + \delta$ (interpreted as the next day: $T + 1$), as they may wish to travel outside that day for any reason.
Exposed: People who went outside on t and therefore had a chance of coming into contact with an infected individual (Asymptomatic or Infectious) to become a Carrier. Any member also has a chance of joining the Protected class on the following day $t + \delta$, as they may wish to self-isolate the next day for any reason.\\

\noindent
\textit{Exposed}: People who went outside on t and therefore had a chance of coming into contact with an infected individual (Asymptomatic or Infectious) to become a Carrier. Any member also has a chance of joining the Protected class on the following day $t + \delta$, as they may wish to self-isolate the next day for any reason.\\

\noindent
\textit{Carrier}: People in the model in a purposely temporary position. They are people in the Exposed class who came into contact with an Asymptomatic or Infectious person and got the disease “on their hands” to some degree. Carriers returning home on t have a chance of spreading the disease to some Protected people living at their home, making those people either Asymptomatic or Infectious on $t + \delta$. Carriers at time t can also have a chance to either “touch their face” and contract the disease themselves (becoming Asymptomatic or Infectious on $t + \delta$), or “wash their hands” and return to being a member of the Exposed class on $t + \delta$.\\

\noindent
\textit{Asymptomatic}: People who were infected by Covid-19 and are contagious on $t$, but show no symptoms. These people are assumed to be active in the public, in that they have a chance to spread the virus to Exposed people in public areas on $t$. An Asymptomatic person on $t + \delta$ has a chance of becoming a member of Infectious on after showing initial symptoms, a member of Quarantined after somehow finding that they contracted the virus, or a member of Removed after having the disease pass through their immune system. Here, all the people going from Asymptomatic to Removed would go to the Recovered subclass.\\

\noindent
\textit{Infectious}: People who were infected by Covid-19, are contagious on $t$, and show symptoms. These people are assumed to be active in the public, in that they have a chance to spread the virus to Exposed people in public areas on $t$. An Infectious person on $t + \delta$ has a chance of becoming a member of Quarantined after figuring that they contracted the virus and a chance of becoming a member of Removed after having the disease pass through their immune system. When these Infectious people become a member of Removed they have a chance of dying, to become a member of the Dead, or living and becoming a member of Recovered.\\

\noindent
\textit{Quarantined}: People who were originally Asymptomatic or Infectious and subsequently removed themselves from the public and are self-quarantining on t . These people are assumed to not be able to spread the disease to any people, so this group includes people who have Covid-19 but are in a non-contagious stage. A Quarantined person on $t + \delta$ has a chance of becoming a member of Removed. When these Quarantined people become a member of Removed they have a chance of dying, to become a member of the Dead, or living and becoming a member of Recovered.\\

\noindent
\textit{Removed}: People who were originally Asymptomatic, Infectious, or Quarantined, and had the disease fully pass through their immune system on or before $t$. These people belong to one of two subclasses: Dead and Recovered. Recovered people are assumed to have survived the disease, and will not get infected again.\\

\begin{flalign*}
\frac{dP}{dt}+\frac{dE}{dt}+\frac{dC}{dt}+\frac{dA}{dt}+\frac{dI}{dt}+\frac{dQ}{dt}+\frac{dR}{dt}=0
\end{flalign*}
\begin{equation}
\frac{dP}{dt}=-\left(\alpha_{1}+\alpha_{2}\right) \frac{CP}{N}+\left(-\alpha_{3} P+\beta_{4} E\right) \frac{N}{P+E} \\
\end{equation}
\begin{equation}
\frac{dE}{dt}=-\left(\beta_{1} A+\beta_{2} I\right) \frac{E}{N}+\beta_{3} C+\left(\alpha_{3} P-\beta_{4} E\right) \frac{N}{P+E} \\
\end{equation}
\begin{equation}
\frac{dC}{dt}=-\left(\gamma_{A}+\gamma_{I}\right) C+\left(\beta_{1} A+\beta_{2} I\right) \frac{E}{N}-\beta_{3} C \\
\end{equation}
\begin{equation}
\frac{dA}{dt}=\alpha_{1} \frac{CP}{N}+\gamma_{A} C-\left(r_{A}+\delta_{A}+\theta\right) A \\
\end{equation}
\begin{equation}
\frac{dI}{dt}=\alpha_{2} \frac{CP}{N}+\gamma_{I} C-\left(\left(r_{I}+d_{I}\right)+\delta_{I}\right) I+\theta A \\
\end{equation}
\begin{equation}
\frac{dQ}{dt}=\delta_{A} A+\delta_{I} I-\left(r_{Q}+d_{Q}\right) Q \\
\end{equation}
\begin{equation}
\frac{dR}{dt}=r_{A} A+\left(r_{I}+d_{I}\right) I+\left(r_{Q}+d_{Q}\right) Q \\
\end{equation}
\begin{equation}
\frac{dD}{dt}=d_{I} I+d_{Q} Q\\
\end{equation}\\

\noindent \underline{Parameters}:\\
$\alpha_{1}:$ Rate of Protected becoming Asymptomatic from Carrier\\
$\alpha_{2}:$ Rate of Protected becoming Infectious from Carrier\\
$\alpha_{3}:$ Rate of Protected becoming Exposed\\
$\beta_{1}:$ Rate of Exposed becoming Carrier from Asymptomatic\\
$\beta_{2}:$ Rate of Exposed becoming Carrier from Infectious\\
$\beta_{3}:$ Rate of Carrier returning back to Exposed\\
$\beta_{4}:$ Rate of Exposed becoming Protected\\
$\gamma_{1}:$ Rate of Carrier becoming Asymptomatic\\
$\gamma_{2}:$ Rate of Carrier becoming Infectious\\
$\theta:$ Rate of Asymptomatic becoming Infectious\\
$r_{A}:$ Rate of Asymptomatic becoming Removed \& Recovering\\
$r_{I}:$ Rate of Infectious becoming Removed \& Recovering\\
$r_{Q}:$ Rate of Quarantined becoming Removed \& Recovering\\
$d_{I}:$ Rate of Infectious becoming Removed \& Dying\\
$d_{Q}:$ Rate of Quarantined becoming Removed \& Dying\\

\section{Training Strategy}
\noindent To fit the deaths data to the system of differential equations in the PECAIQR model, we performed numerical integration using the scipy odeint package \cite{odeint}, and traversed the parameter space to find a set of parameters that minimized the least squares error of each fit variable in relation to its observed variables. Due to the size of the parameter space, this requires an initial guess for the parameters and the initial conditions of each of the 7 PECAIQR variables, as well as defined ranges to restrict the size of the parameter space. Of course, we can simply examine the entire logical parameter space (with allowed values ranging from 0 to 1) for the each of the parameters, and for each of the PECAIQR variables as well, and then use a random guess within that space to initialize the least squares minimization. However, this is unnecessarily inefficient. The actual parameters may differ widely from county to county, but they all share a similar order of magnitude. A better approach is to use this exhaustive, unconstrained search only once, on a relatively mature curve like Lombardi in Italy or New York in the United States, to extract a reasonable guess for these orders of magnitude, and then simply feed in this guess for the other counties as well, to provide a more logical starting point in the least squares minimization. Due to the complexity of our model, however, there are a few more nuanced details that we will discuss later in section. First, we will elaborate more on the fitting procedure.\\
\subsection{Fitting}
\noindent
Initially, we only fit the death variable (D) to the observed death reporting. Early on, we decided not to fit our infection curves to daily case statistics, as they have inconsistent and unreliable reporting, so this would only diminish the accuracy of the fit to observed deaths, which is a much more reliable statistic. However, later on, we realized that we could also effectively fit the quarantined variable (Q) in the PECAIQR model to the active cases reporting. The intuition behind this is the assumption that those who test positive would either self quarantine at home or be forcefully quarantined in a hospital if their condition is severe enough. Of course, this does not capture all the self-quarantined individuals, so we permitted a large degree of fuzziness in the fit of the quarantined variable (Q) to active cases. We achieved this with a bias scaled factor that gave much more weight to observed deaths in the fit of the death variable (D). This causes the model to prioritize the deaths fit over the active cases fit, so the active cases fit becomes more of a suggestion rather than a constraint. The hope is that the least squares error on the active cases is not large enough in magnitude to compromise the deaths fit and force the parameter space into a different minimum, but will rather provide a subtle correction around the local minimum discovered by the minimization of the least squares error on the deaths. Unfortunately, a fit to active cases is not helpful for most counties, as most counties do not maintain their active cases reporting well, and moreover we discovered that the criteria for an active case may vary widely across different states.\\

\noindent We also realized that the PECAIQR epidemiological model parameters are not static, due to the dynamic and rapidly evolving state of the pandemic, caused mainly by external forces such as social distancing protocols and lock down policies. A naive fit on the data would only yield some sort of average of the parameters. But since we only care about the most recent characteristics of the death and infection curves when making predictions into the future, we can do better than this by affording the more recent data points more weight in the least squares error calculation. This forces the minimization to favor a solution that fits more heavily on a more recent window of time, while still retaining the effects of the past data to some degree. We can achieve this with a geometric progression of weights, as well as a bias term that sets a maximum weight for all data points before a certain cutoff. We called this method \textit{training on the tail}.\\

\noindent Another method to separate more recent parameters from the data is to use the assumption that there are distinct parameter regimes correlating to the start of policies - the stay at home orders in particular. Note that these policies directly affect the infection curves because they limit the spread of disease. For the death curve, there is typically an offset between the date of policy implementation and the date at which the effects become apparent in the death data, this offset approximately equal to the average time till death for Covid-19. With this assumption, we separate the deaths data into two training regimes -- one for dates before the policy implementation plus the offset, and one for dates after the policy implementation plus the offset. Then, we train one fit on the the first regime, and feed the fitted parameters and predicted variables on the date of the policy implementation as the guesses and initial conditions, respectively of a second fit on the second regime. We called this method \textit{training on the policy regime}.\\

\subsection{Predictions and Errors}
\noindent
We perform procedures described above on a county-level, and then use the fitted parameters with the numerical integration to extrapolate into the future to get county-specific predictions. Note that the predictions and their errors are in cumulative deaths, but for the sake of visualization, we converted to daily deaths later.\\

\noindent To get the errors, we initially used a method that calculated error bounds by finding the parameter variance from the residual variance using the covariance matrix of residuals and the Jacobian around the fitted parameters. Having obtained the mean and standard deviation for each of the parameters, we assumed that each parameter had normally distributed values, and so we sampled 100 parameter sets. We could not simply apply calculate the parameters for each confidence interval from their mean and standard deviation because it is not obvious which parameters are positively correlated or negatively correlated with the deaths predictions. This method turned out not to be ideal in all cases, as there are certain regions in the parameter space that are invalid and cause the error bars to spike. \\

\noindent We then developed a bounding method that allowed a reliable way to accurately tighten our confidence intervals. This method infers the error bars from the deviation relative to the predicted fit, of a smoothed version of the residuals calculated from the moving average of daily deaths, which is equivalent to the moving average of the slope of the cumulative deaths. Generate a PDF of deaths around the best fit prediction. For each time t in the training range, compute the difference in “slope” between the fit and the actual data. The slope in the fit is simply the current predicted death minus the previous death on the fit curve. For the slope in the actual data, to mitigate the dominating effects of outliers, we find the slope as the difference between consecutive points of the moving average (window of 3 days) instead. We then define a normal distribution of slope ratios. For each time $t$ in the training data, we find the ratio of the actual (moving average) slope and the fit slope, and fill a list of these ratios. Next we find the mean and standard deviation of the distribution as the mean and (sample) standard deviation of the ratios list. Outliers, which we remove, are defined as values that are above three standard deviations from the mean. This is necessary because the ratios are unrealistically high in the small number limit. For each time t in the extrapolated prediction fit we generate a confidence range of 100 points. We find the slope at the extrapolated time as the predicted death minus the predicted death at the previous time. This slope is multiplied by a random scaler sampled from the normal distribution of ratios and denoted as s. We multiply the predicted death at the previous time $t-1$ in the fit by $1+s$ and add this as a point in the PDF at time t. This multiplication is repeated 100 times so that 100 points are generated. We get the discrete 10, 20, 30, 40, 50, 60, 70, 80, 90 CDF percentiles by sampling the PDF. \\

\subsection{Validation}
\noindent The methods described in the past two subsections are implemented as options that can be activated with hyper parameters, and collectively they provide several different ways to fit the PECAIQR model and generate the confidence intervals. Due to the time constraint of the competition, we were not able to develop a sophisticated blending method to optimize on the hyper-parameter space. However, we were able to sample a few different combinations of hyper-parameters and determine on a county-level which one works the best for each county, by modifying the evaluation script provided by the TAs. This script essentially computes pinball \cite{pinball} loss for the submission file of predictions, scored against the most recent data. To determine a good set of hyper-parameters for each county, we simply trained using a two week cutoff in the data, and scored the predictions for the subsequent two weeks using the evaluation script.    
\noindent

% Talk about initial conditions, guesses, international data

% Later on, talk about how different guesses get different results, proving that there are several similarly valid solutions in the parameter space
% ODE stiffness, degeneracy, parameters to explain
% Visualize parameters
\section{Model Robustness}
\subsection{Model Performance}
\noindent For predictions made from training data up to May 25th, we were able to obtain an average pinball loss of 0.096 for all counties with scoring on a 14 day forecast.\\

\noindent Our greatest weakness was the lack of a second working model with which we could cross validate, as well as the lack of a sophisticated blending method to optimize on the hyper-parameter space for the single model. We were not able to develop an alternative working model due to the time constraints of our group members, but a detailed description of our attempts is available in section 5 \textbf{Failed Models}. Having two different models would have allowed us to mitigate their individual weaknesses and account for their individual edge cases with the other's strengths.\\

\noindent The epidemiological model has several major weaknesses. Although it works well for counties with well recorded data, it fails for the vast majority of counties, which have low or noisy death statistics. We called these counties the "non-convergent counties" because the epidemiological model was not able to converge to a parameter set that yielded predictions which could consistently score better than the naive all zeros prediction.\\
\subsection{Results for Selected Counties}
\begin{figure}[H]%
    \centering
    \subfloat[]{{\includegraphics[width=7.5cm, height=4cm]{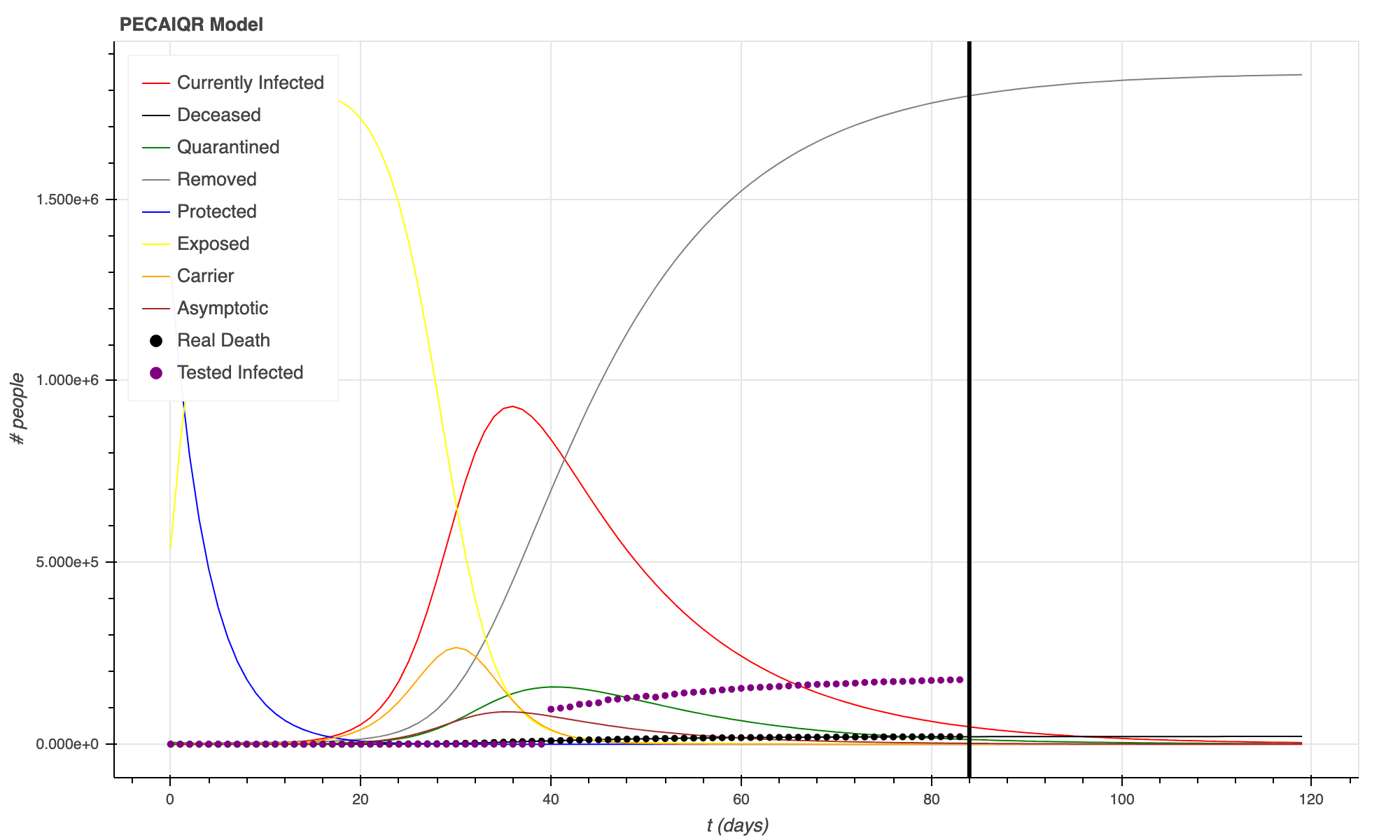}}}
    \qquad
    \subfloat[]{{\includegraphics[width=7.5cm, height=4cm]{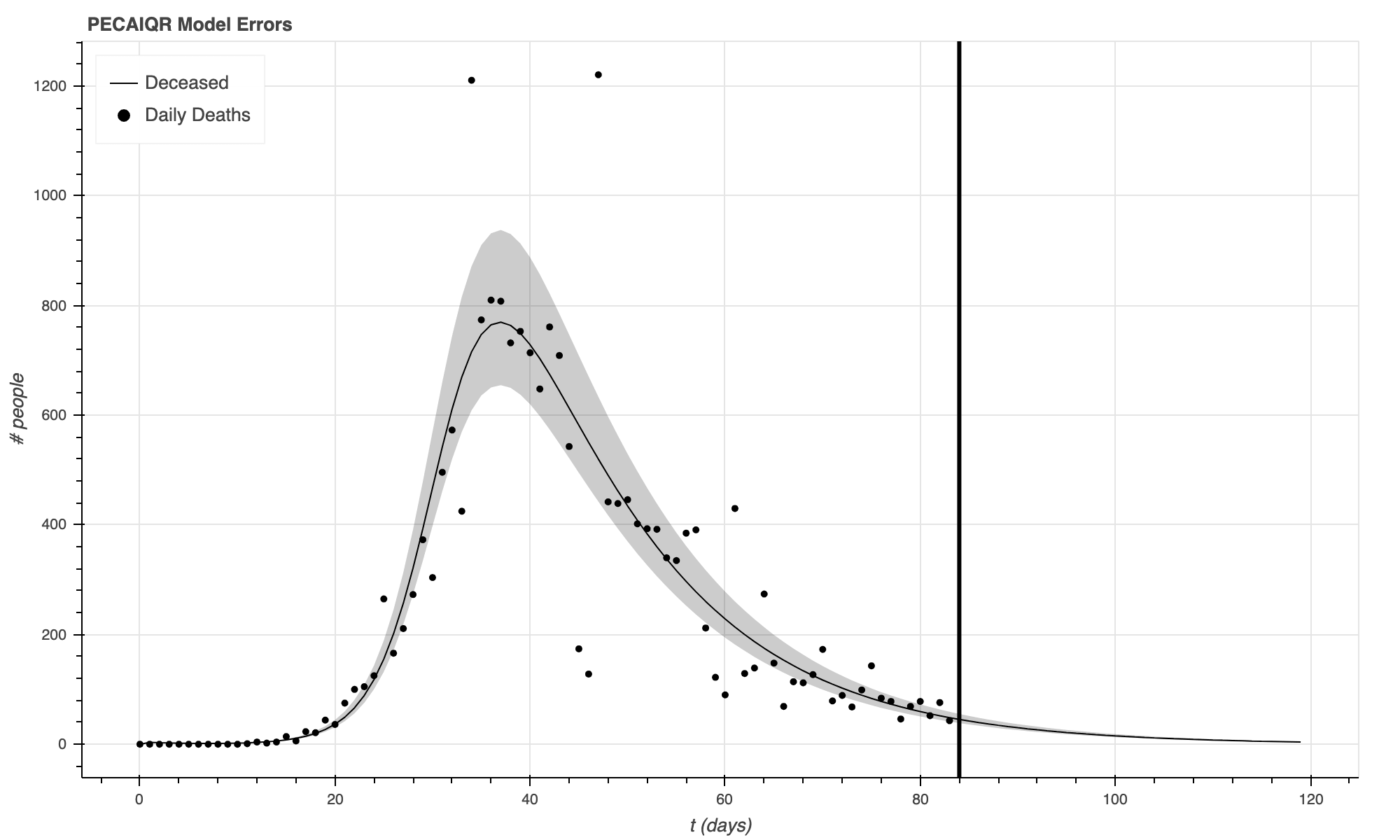}}}
    \qquad
    \subfloat[]{{\includegraphics[width=7.5cm, height=4cm]{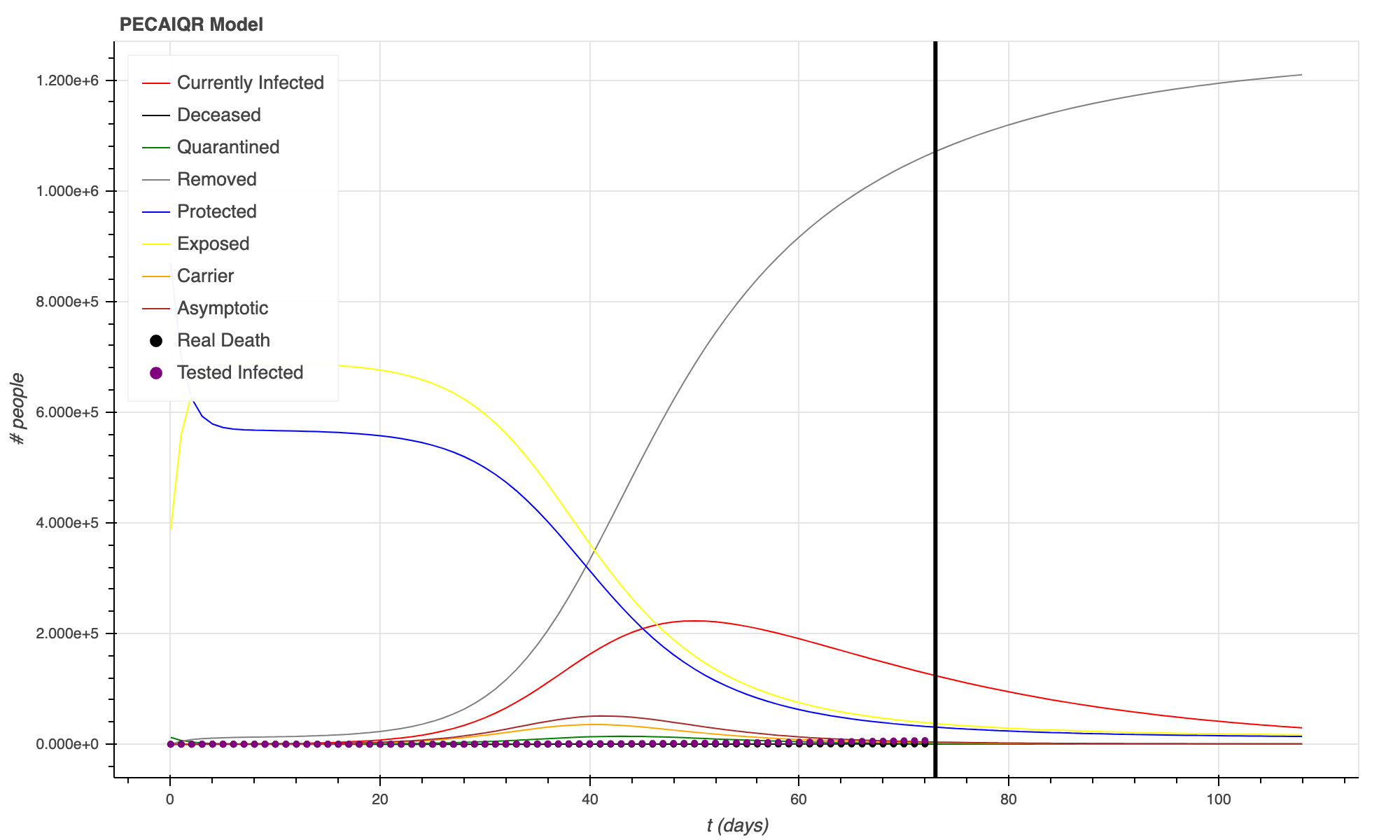}}}
    \qquad
    \subfloat[]{{\includegraphics[width=7.5cm, height=4cm]{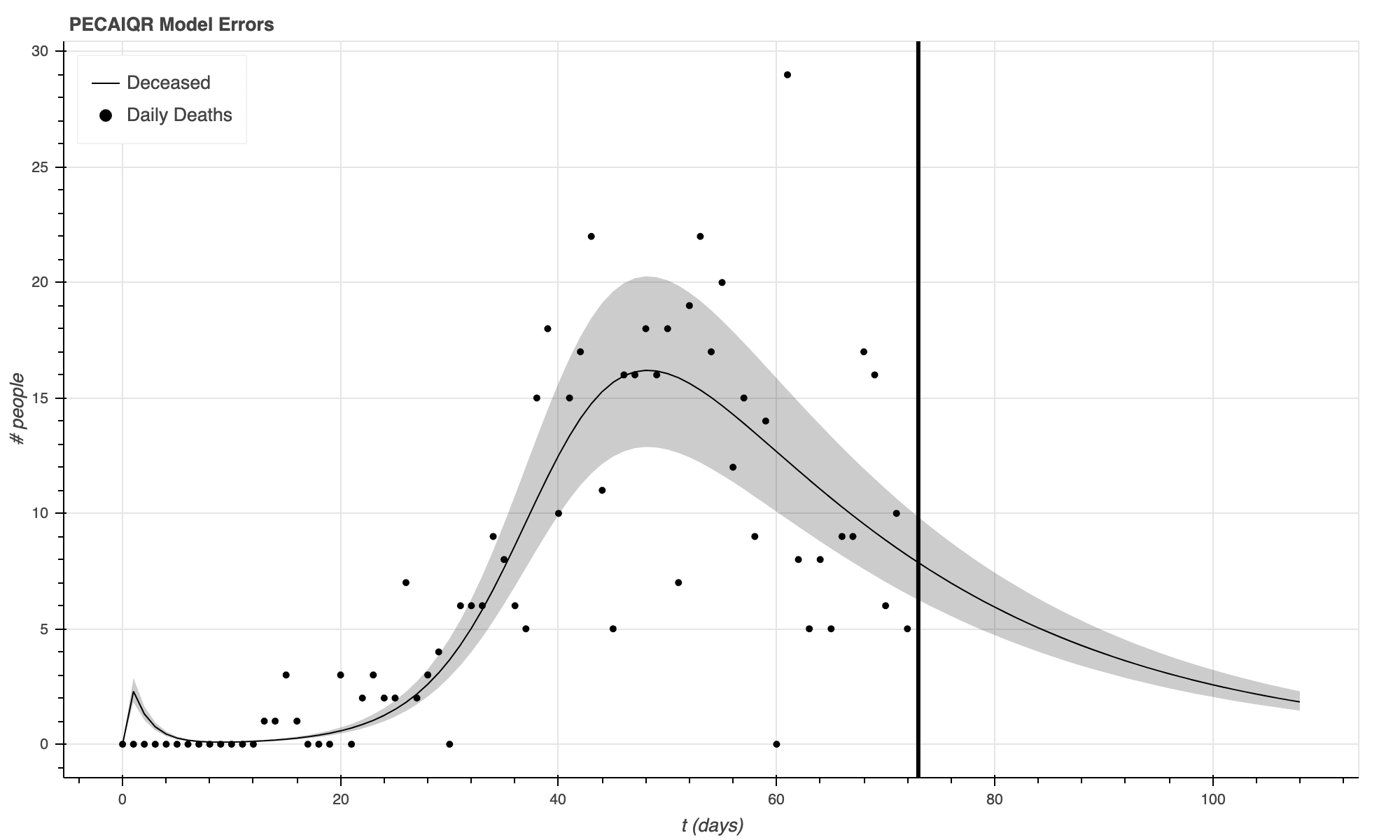}}}
    \qquad
    \subfloat[]{{\includegraphics[width=7.5cm, height=4cm]{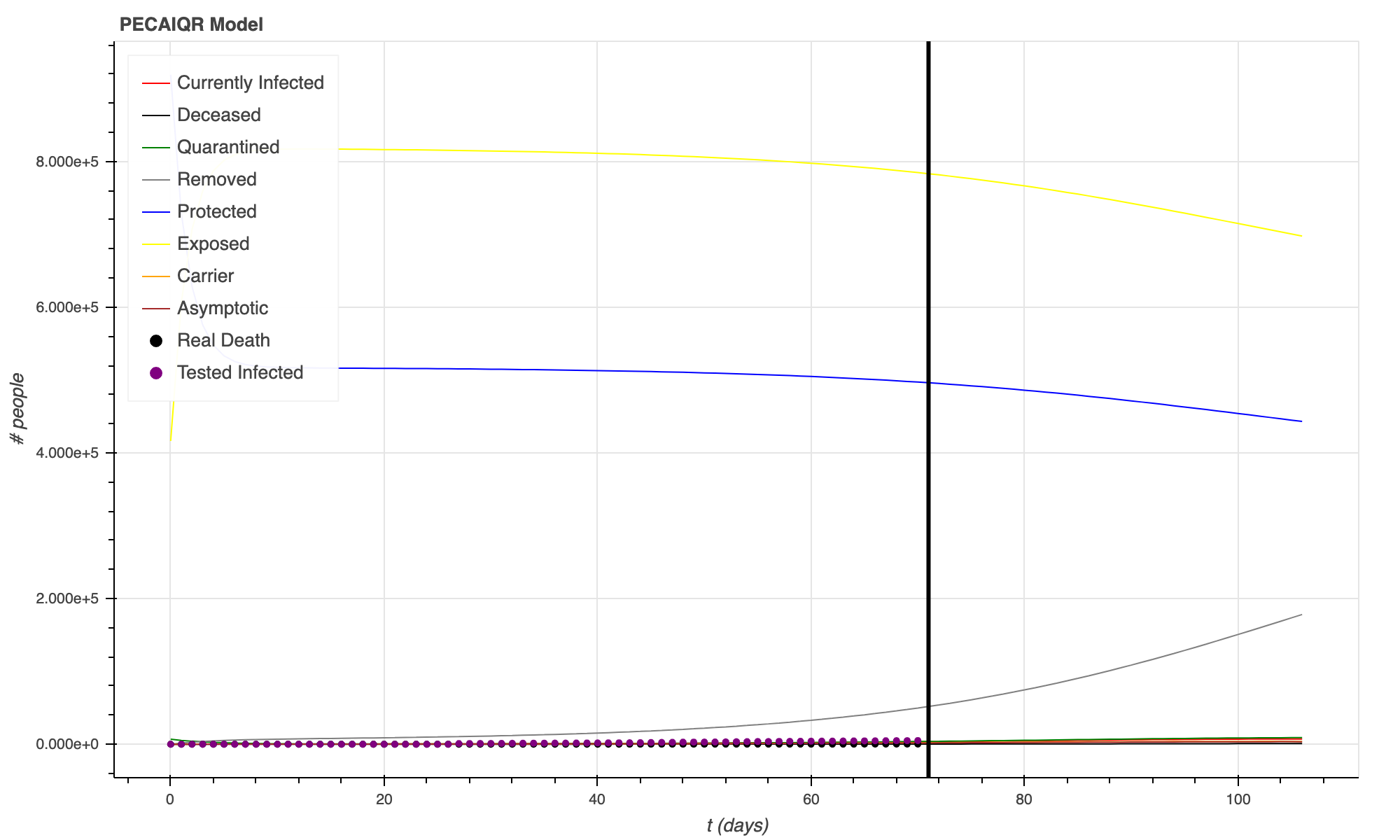}}}
    \qquad
    \subfloat[]{{\includegraphics[width=7.5cm, height=4cm]{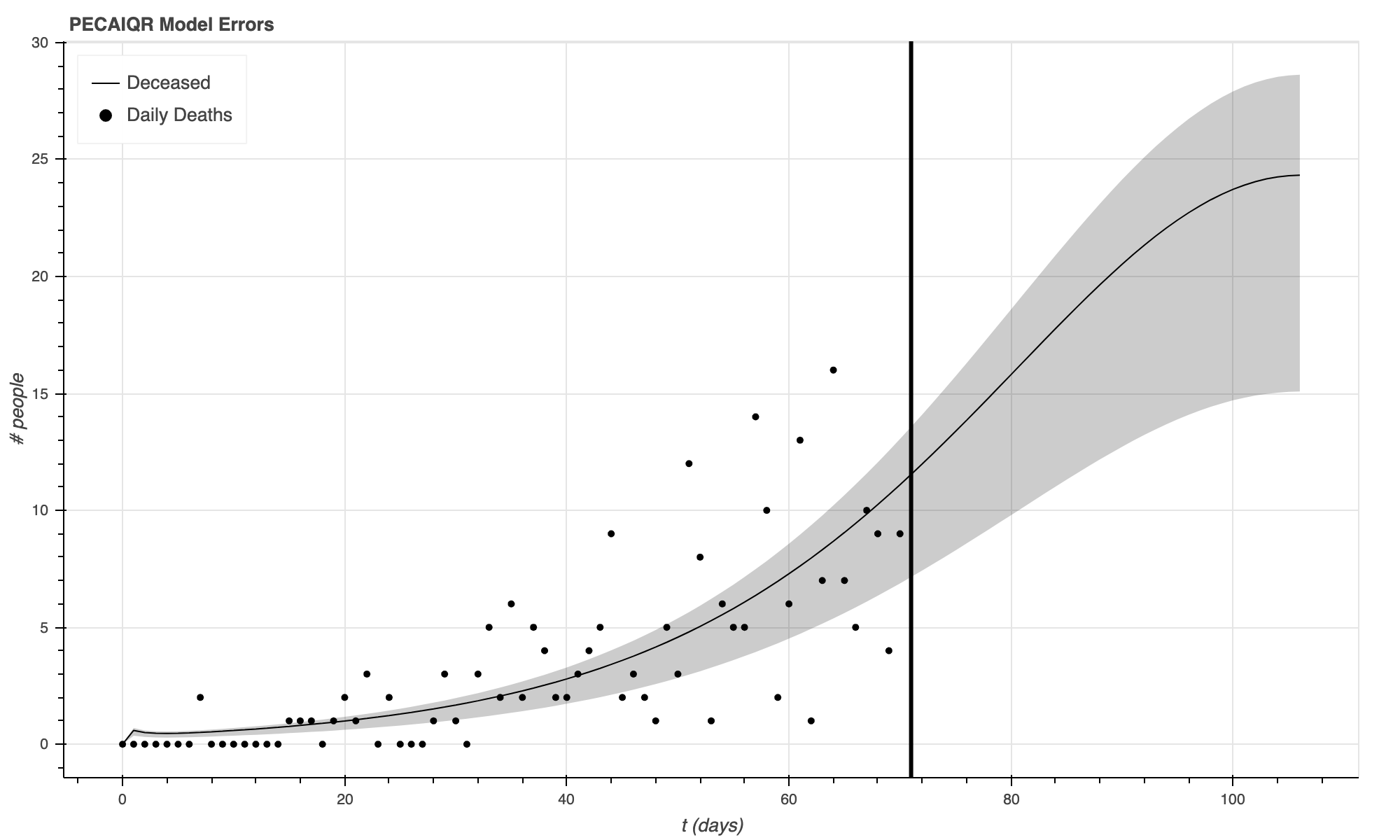}}}
    \qquad
    \subfloat[]{{\includegraphics[width=7.5cm, height=4cm]{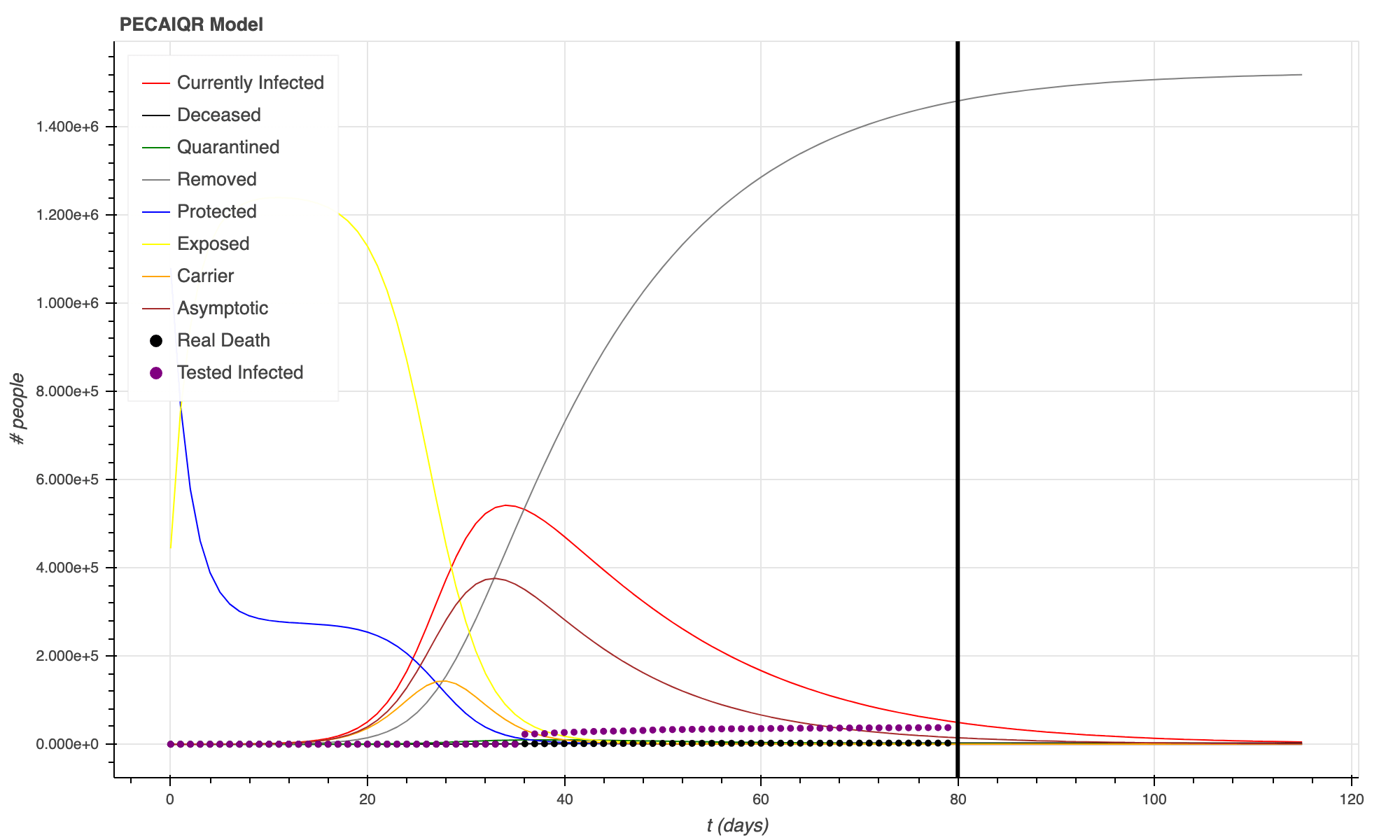}}}
    \qquad
    \subfloat[]{{\includegraphics[width=7.5cm, height=4cm]{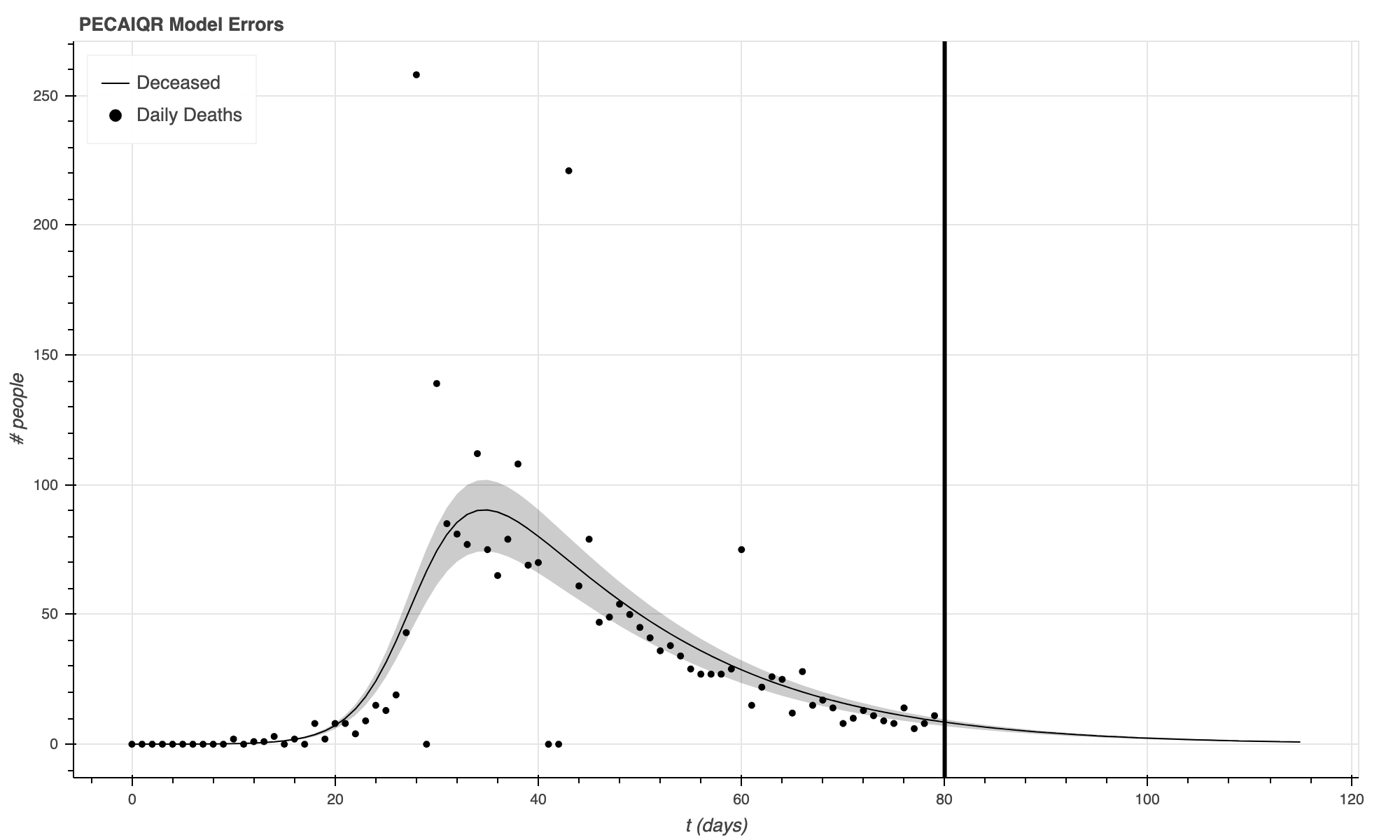}}}
    \qquad
    \caption{The first row of figures is county 36061. The second row is county 27053. The third row is county 39049. The fourth row is county 36059.}
\end{figure}

\begin{figure}[H]%
    \centering
    \subfloat[]{{\includegraphics[width=7.5cm, height=4cm]{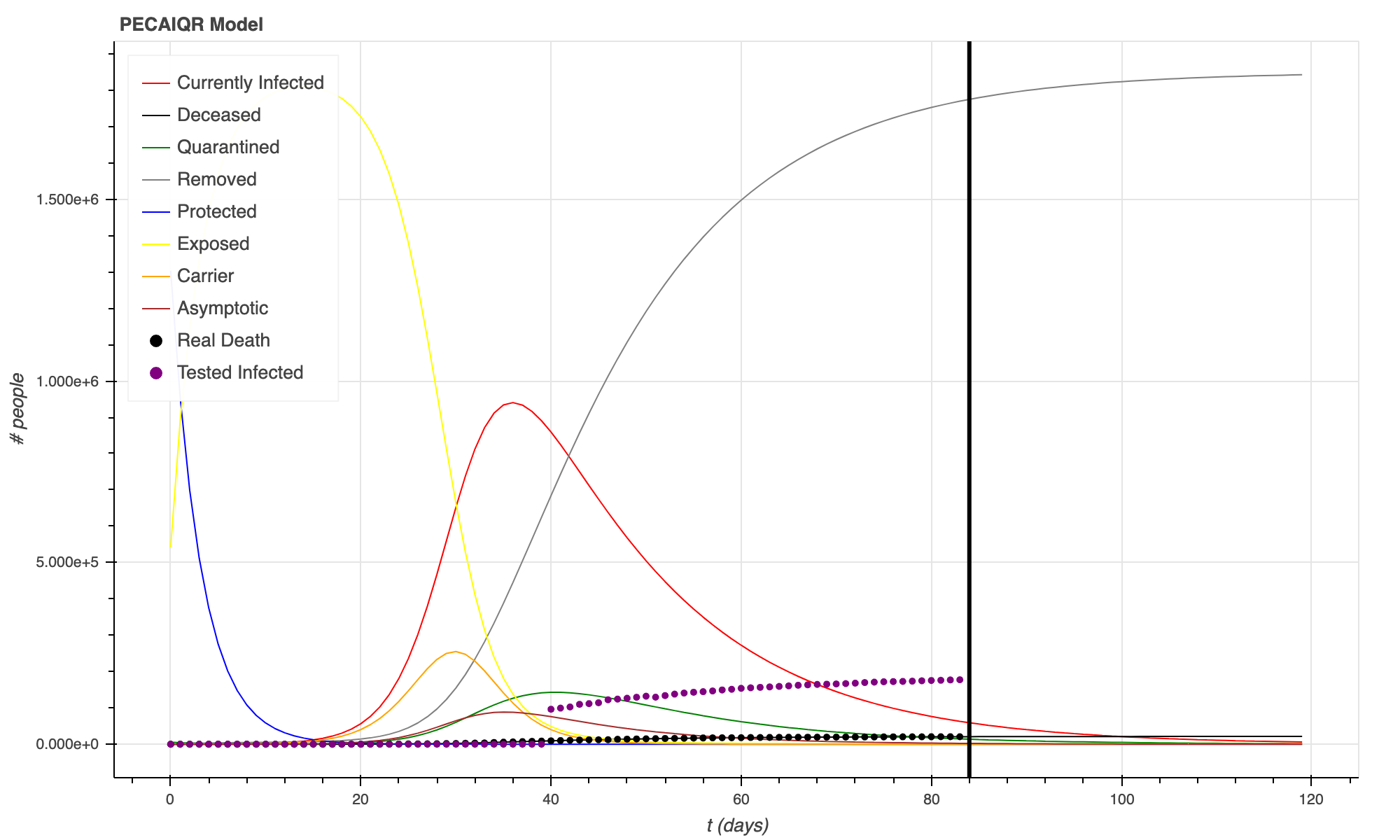}}}%
    \qquad
    \subfloat[]{{\includegraphics[width=7.5cm, height=4cm]{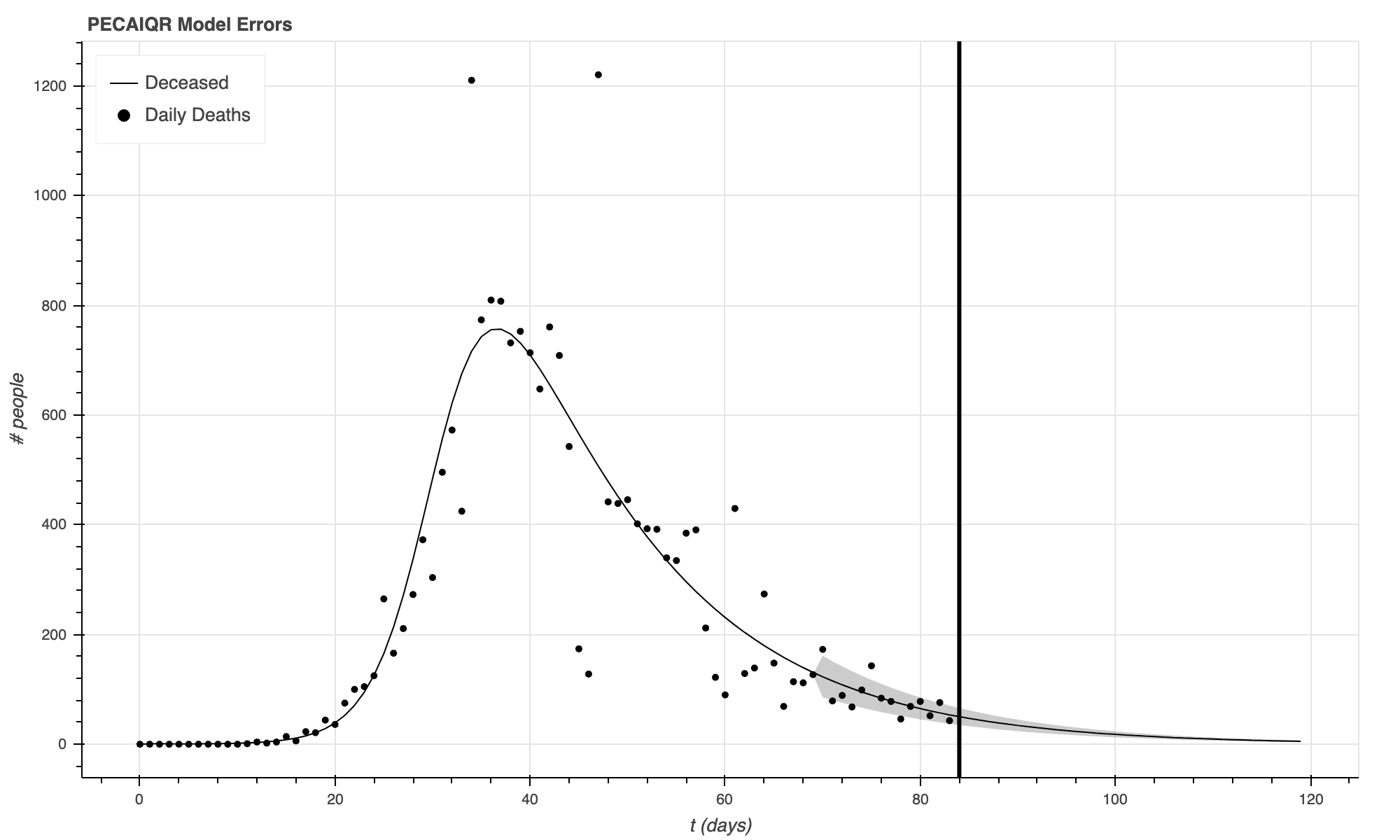}}}%
    \qquad
    \subfloat[]{{\includegraphics[width=7.5cm, height=4cm]{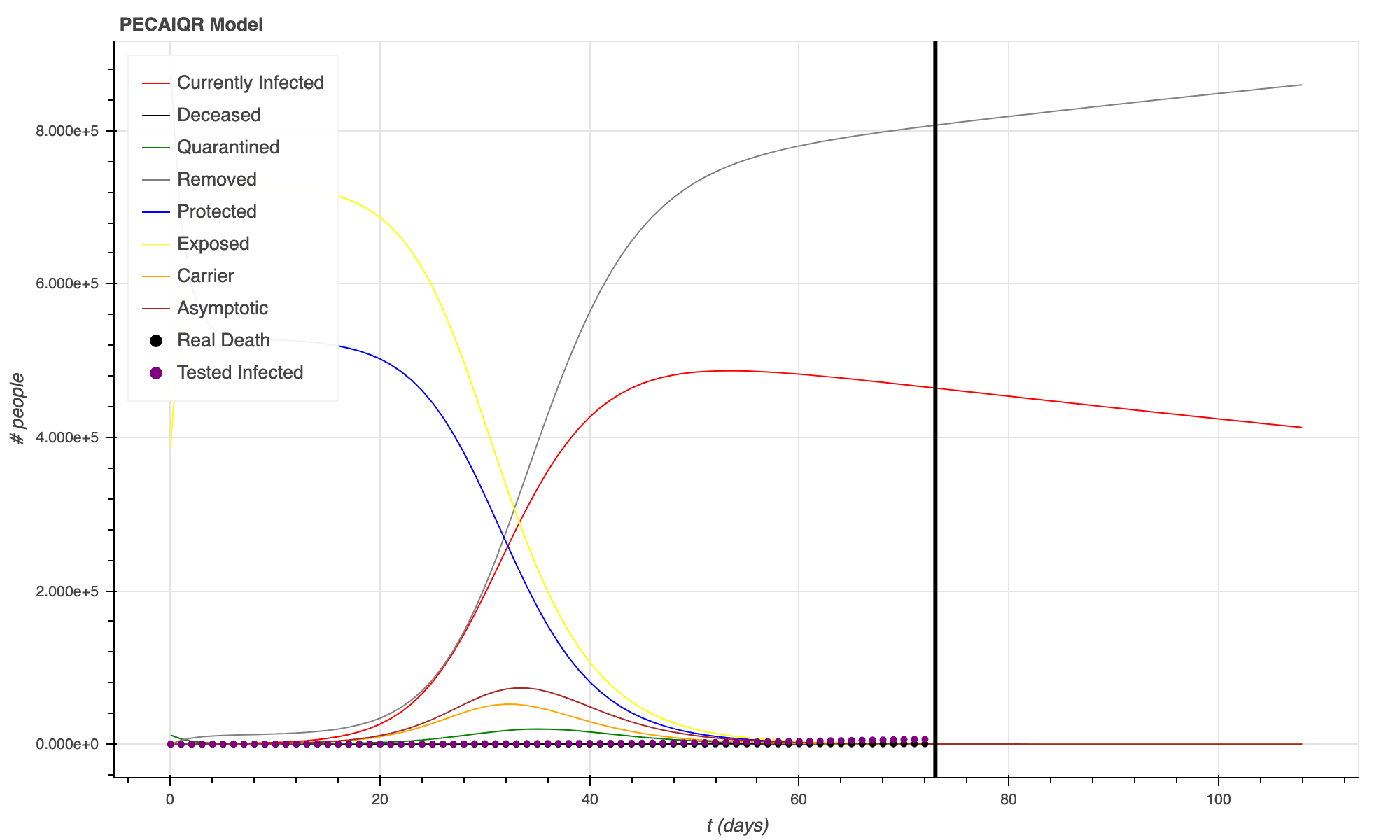} }}%
    \qquad
    \subfloat[]{{\includegraphics[width=7.5cm, height=4cm]{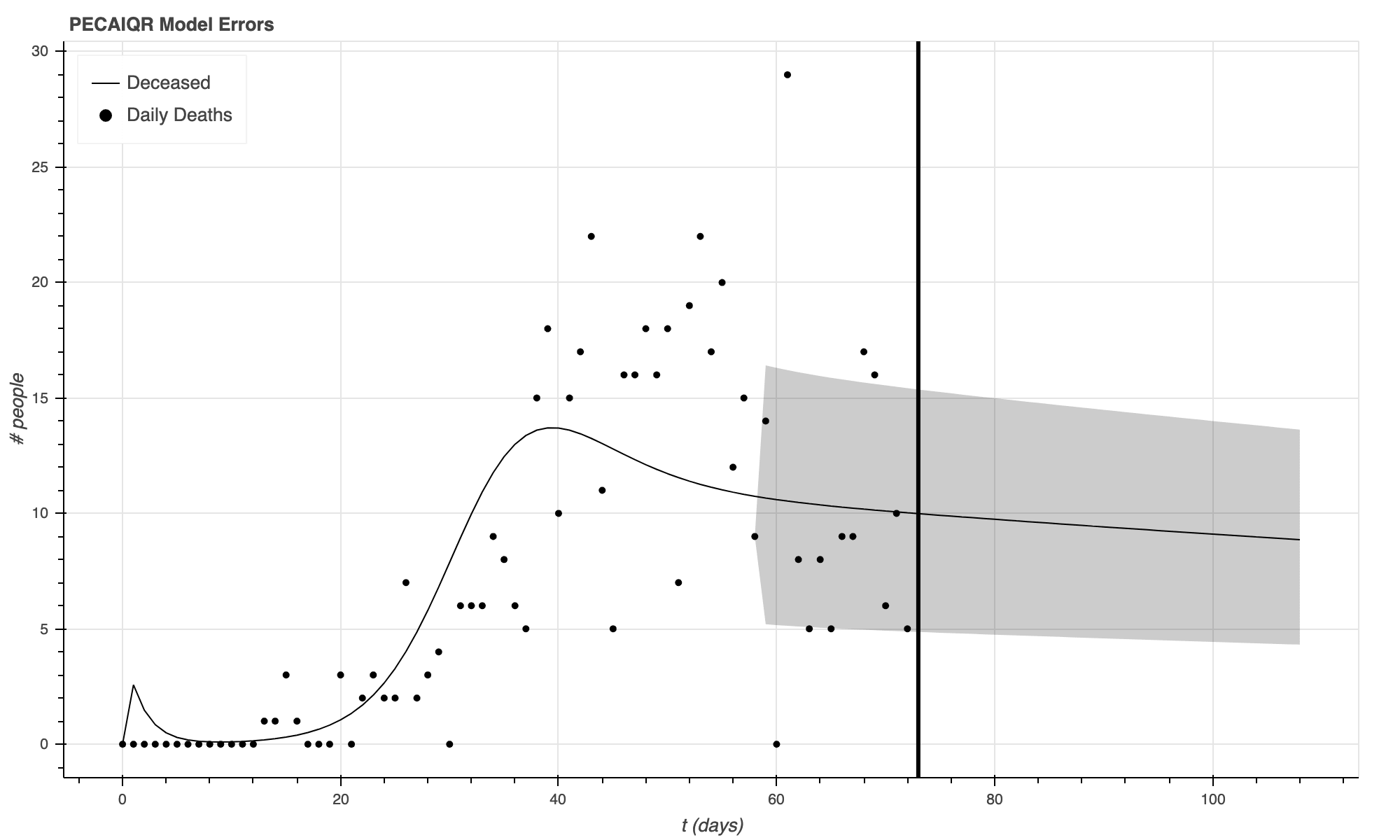}}}%
    \qquad
    \subfloat[]{{\includegraphics[width=7.5cm, height=4cm]{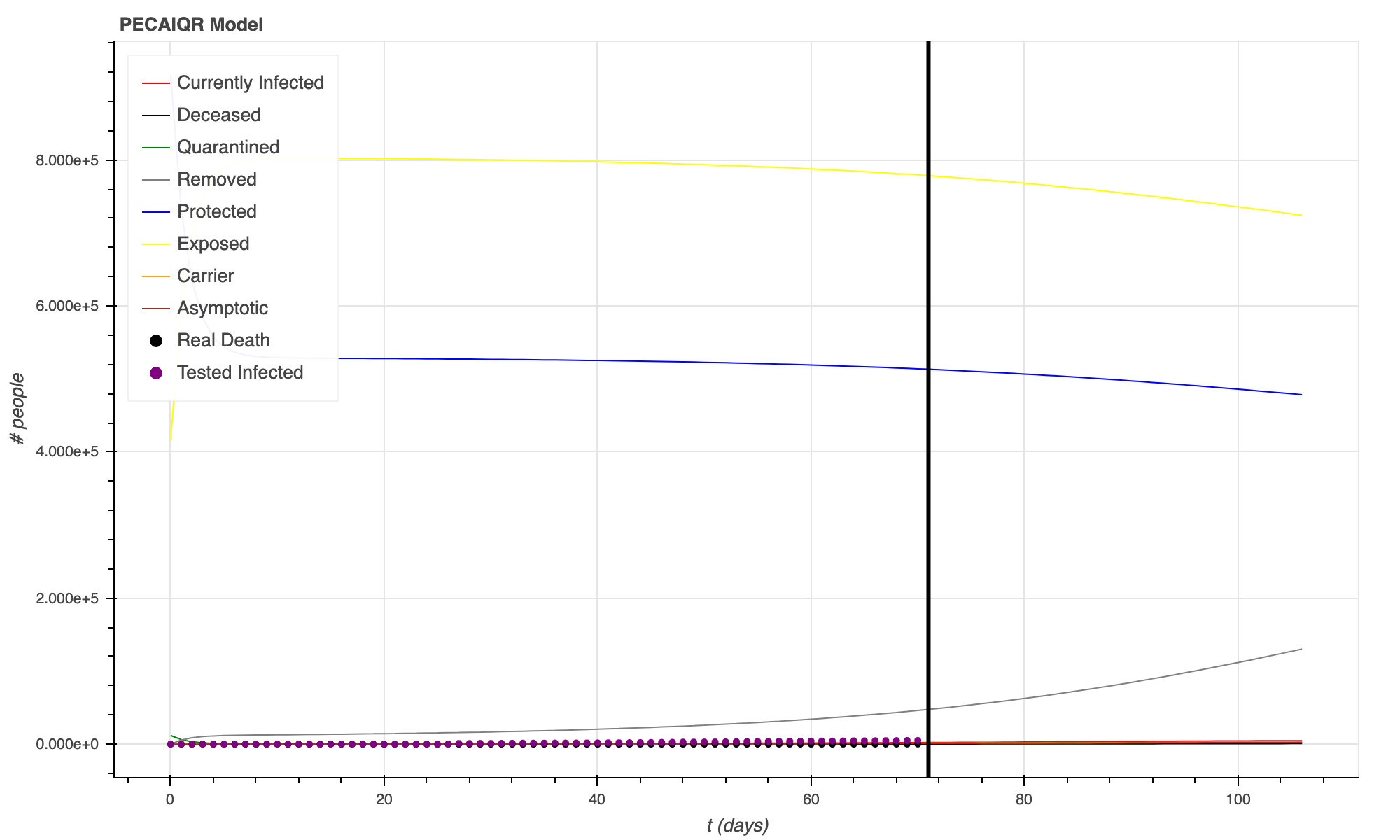}}}%
    \qquad
    \subfloat[]{{\includegraphics[width=7.5cm, height=4cm]{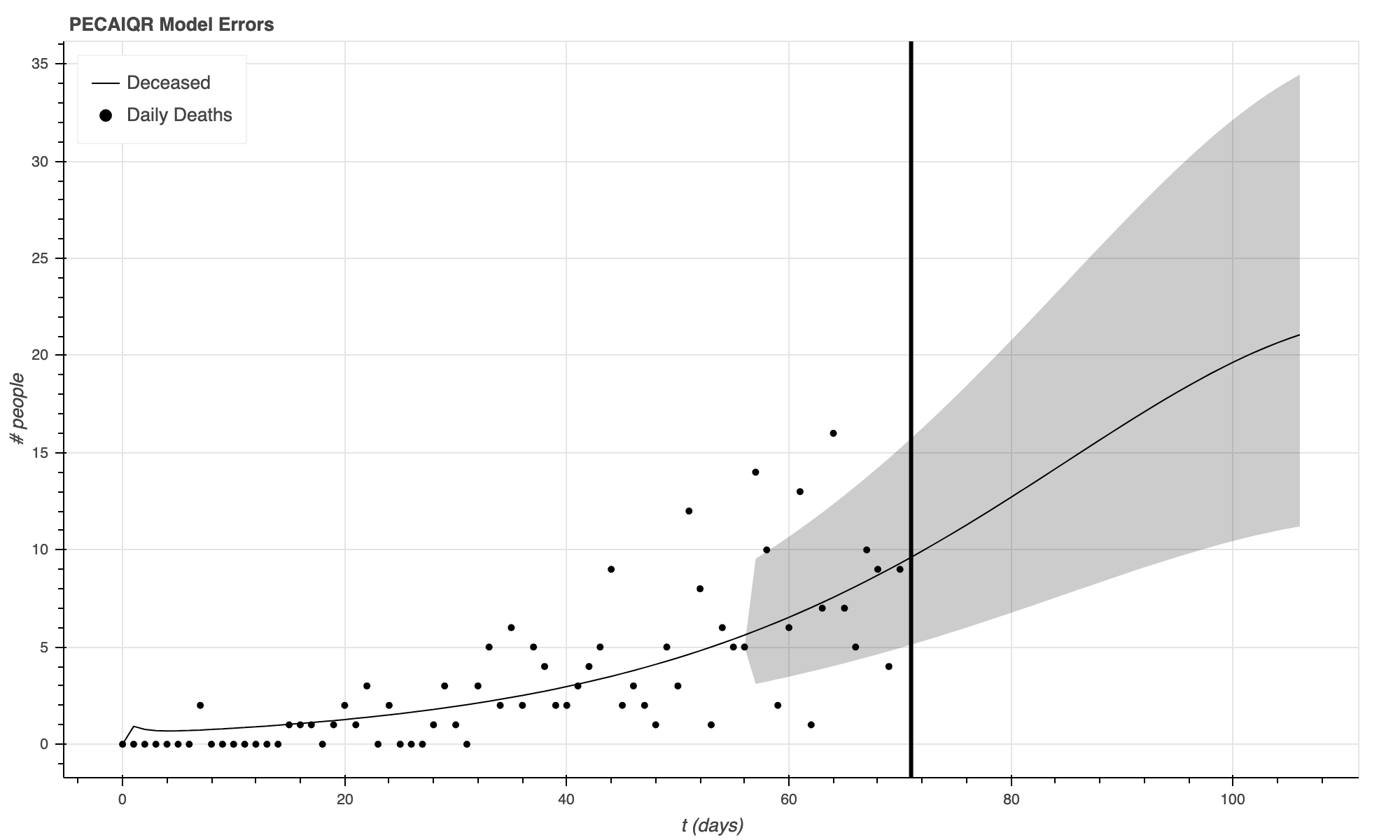}}}%
    \qquad
    \subfloat[]{{\includegraphics[width=7.5cm, height=4cm]{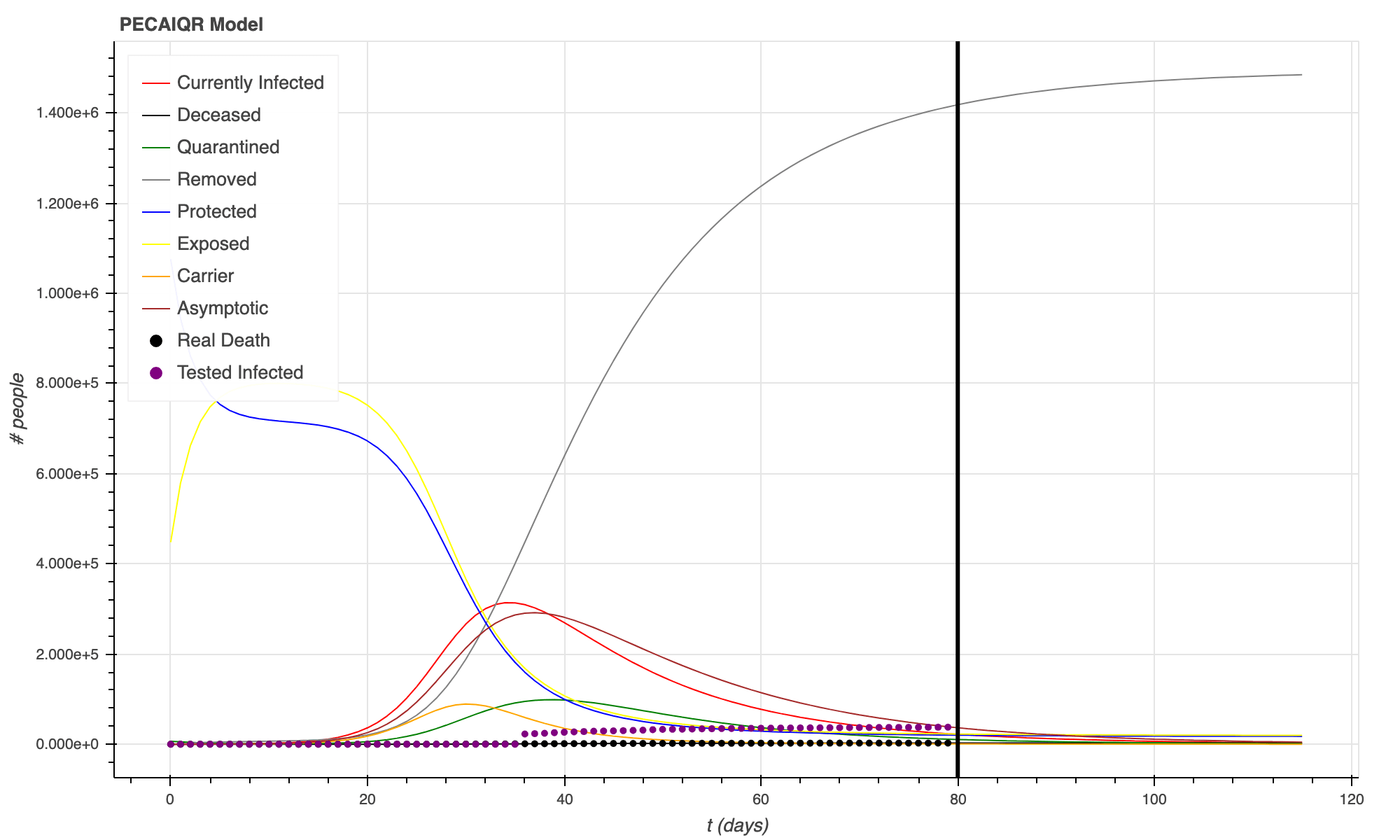}}}%
    \qquad
    \subfloat[]{{\includegraphics[width=7.5cm, height=4cm]{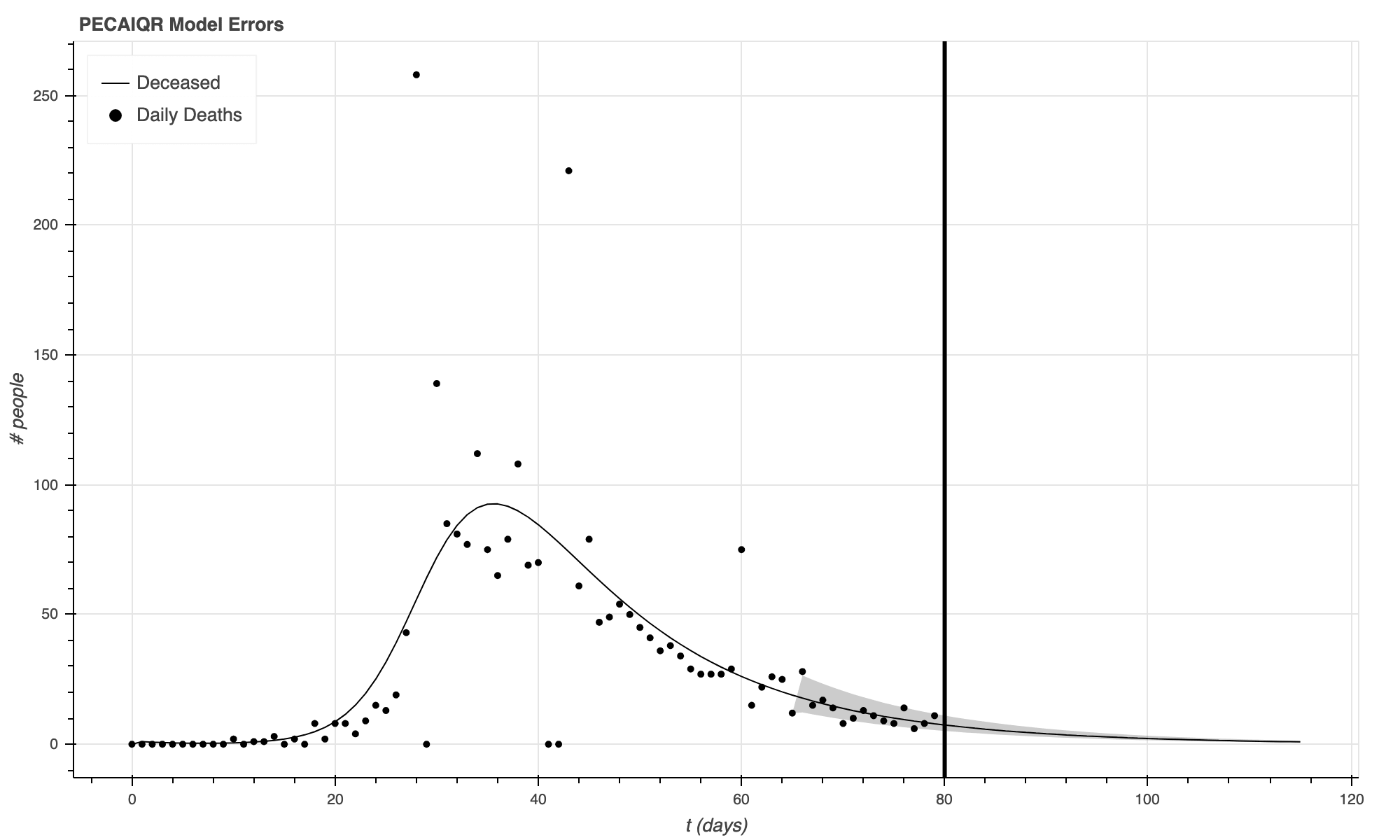}}}%
    \qquad
    \caption{Using the "fitting on tail" method. The first row of figures is county 36061. The second row is county 27053. The third row is county 39049. The fourth row is county 36059.}%
\end{figure}
\noindent The curves for the PECAIQR variables in the left column of the preceding figures make sense intuitively. We expect expect to see that conversion between the protected and exposed populations, and that the removed population to eventually dominate. We also expect the carrier population to peak before the infected population, which should have a similar shape to the asymptomatic population, and lastly we expect the infected and asymptomatic populations to peak before the deceased population peaks. Lastly, the second figure, which uses the "fit on tail" method, more closely captures the uncertainty of the tail end of the curve, as expected.\\    

\subsection{Parameter Visualization}
\noindent For the "nonconvergent counties," we attempted to skip over the parameter fitting step by guessing parameters based on the parameters of similar counties. In order to do this, we need to first establish proof of concept that there is some correlation between the non-covid features of a county and its PECAIQR parameters. So we visualized the parameters for all the "convergent" counties by using dimension reduction algorithms. We used Principal Component Analysis \cite{pca}, Singular Value Decomposition \cite{svd}, and t-Distributed Stochastic Neighbor Embedding\\
\begin{figure}[H]%
    \centering
    \subfloat[]{{\includegraphics[width=7.5cm, height=5cm]{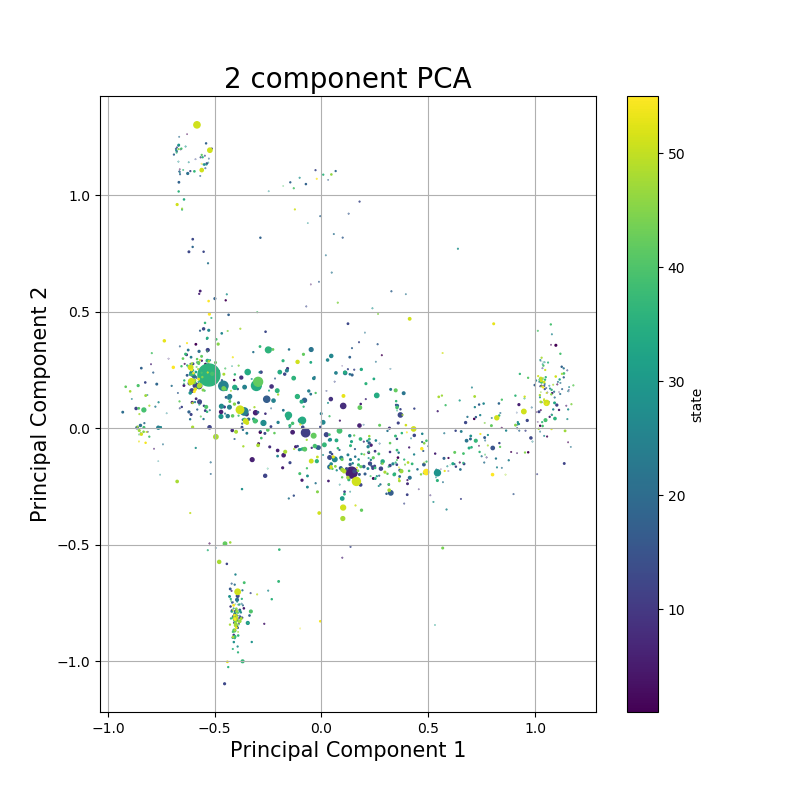}}}%
    \qquad
    \subfloat[]{{\includegraphics[width=7.5cm, height=5cm]{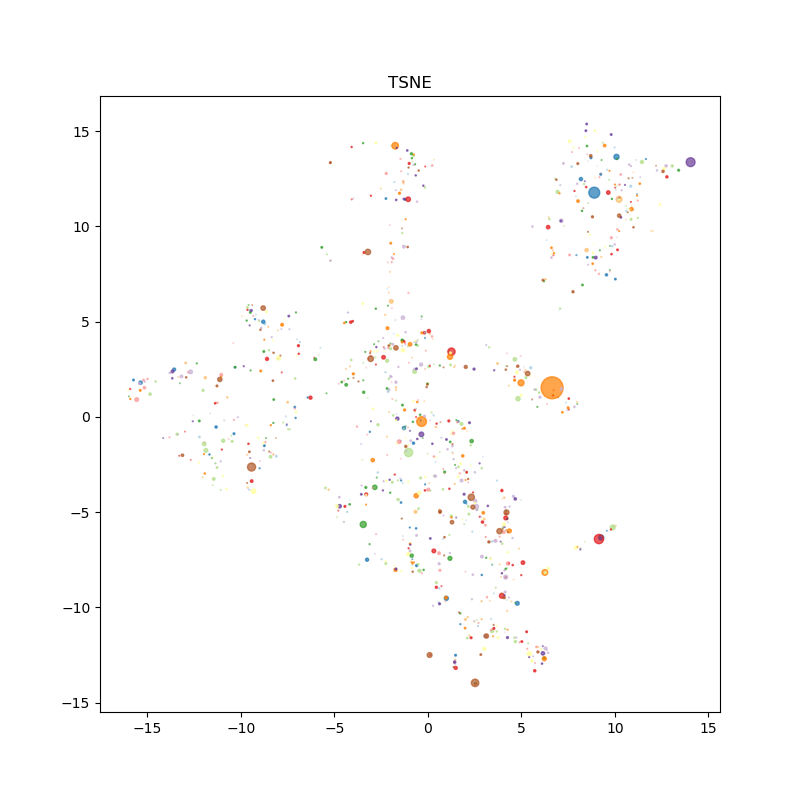}}}%
    \qquad
    \subfloat[]{{\includegraphics[width=7.5cm, height=5cm]{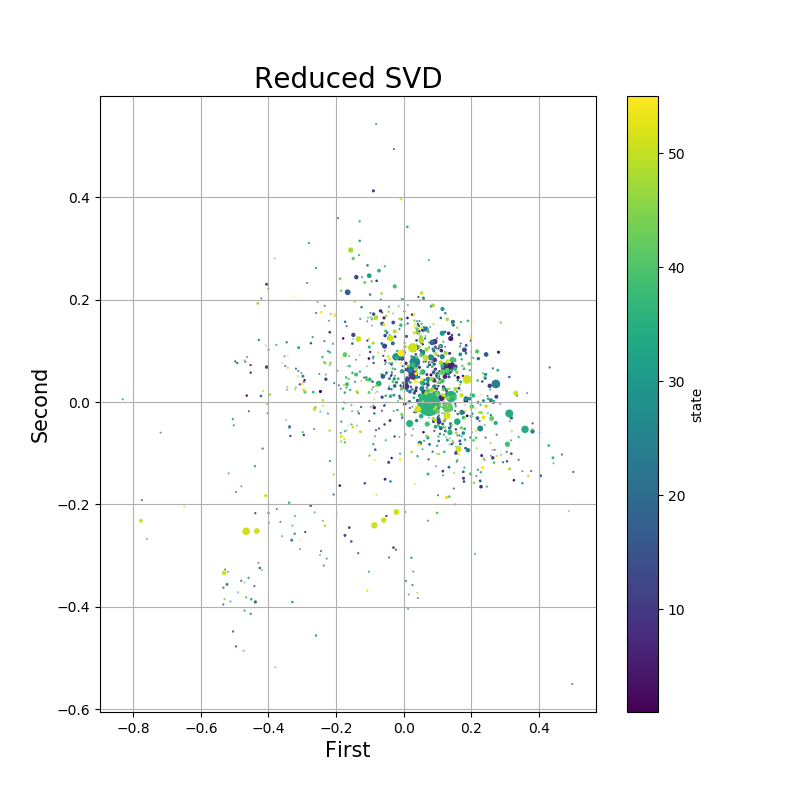} }}%
    \qquad
    \subfloat[]{{\includegraphics[width=7.5cm, height=5cm]{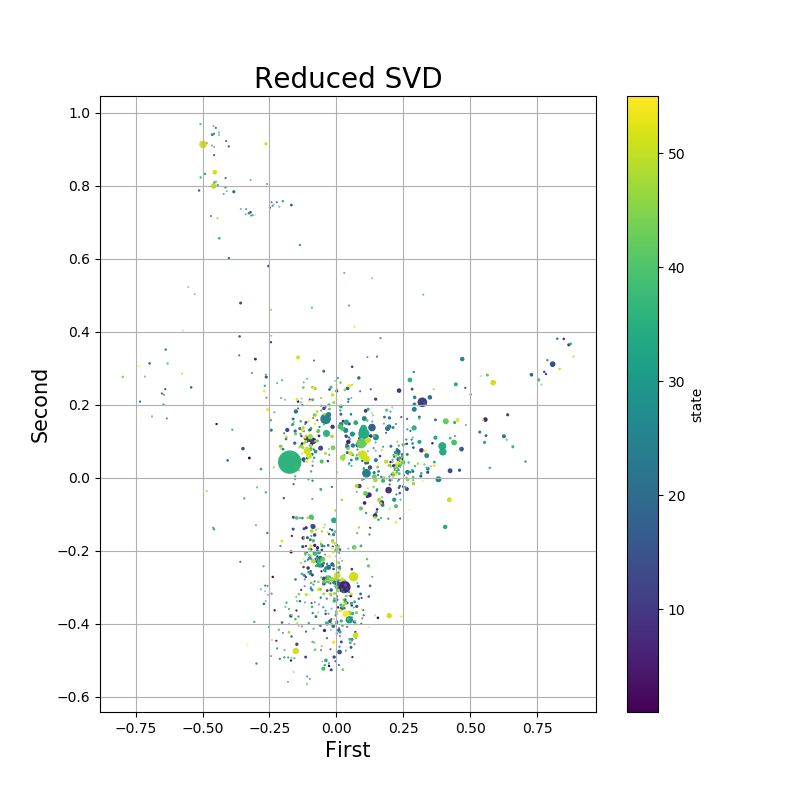}}}%
    \qquad
    \caption{Several different visualizations of the parameter space using dimension reduction algorithms. Each point represents a count, and its size correlates with the county population.}%
\end{figure}
\noindent For SVD, we plotted the first two columns of the left matrix against each other, which produces an visualization how similar the counties are by distance, based on their abstract preferences in the parameter space. Although we did not have time to further explore the possible correlation between parameter clusters and non-covid county-specific features, we were excited to see that there are some distinct intrinsic patterns in the parameter space. Further research would involve classifying these clusters with a clustering algorithm, and then attempting to predict the cluster labels or quantified distances of specific counties with a regression algorithm from the non-covid feature space.\\
\subsection{Degeneracy and Stiffness}
\noindent We also realized that, even for the larger counties, there was some uncertainty caused by the complexity of our model. Since our parameter space has so many dimensions, there is an issue with degeneracy. The least squares minimization settles on a parameter set that yields a minimum in the least squares error metric, but this is not necessarily the only minimum or even the best minimum. We were convinced of this when we discovered that for each county, we could get distinct parameter regimes with different initial guesses, and that these solutions are similarly valid. We will show this in the figures below, by repeating the procedure from the previous section, but with a different set of parameters as the initial guess for the least squares minimization.\\
\begin{figure}[H]%
    \centering
    \subfloat[]{{\includegraphics[width=7.5cm, height=4cm]{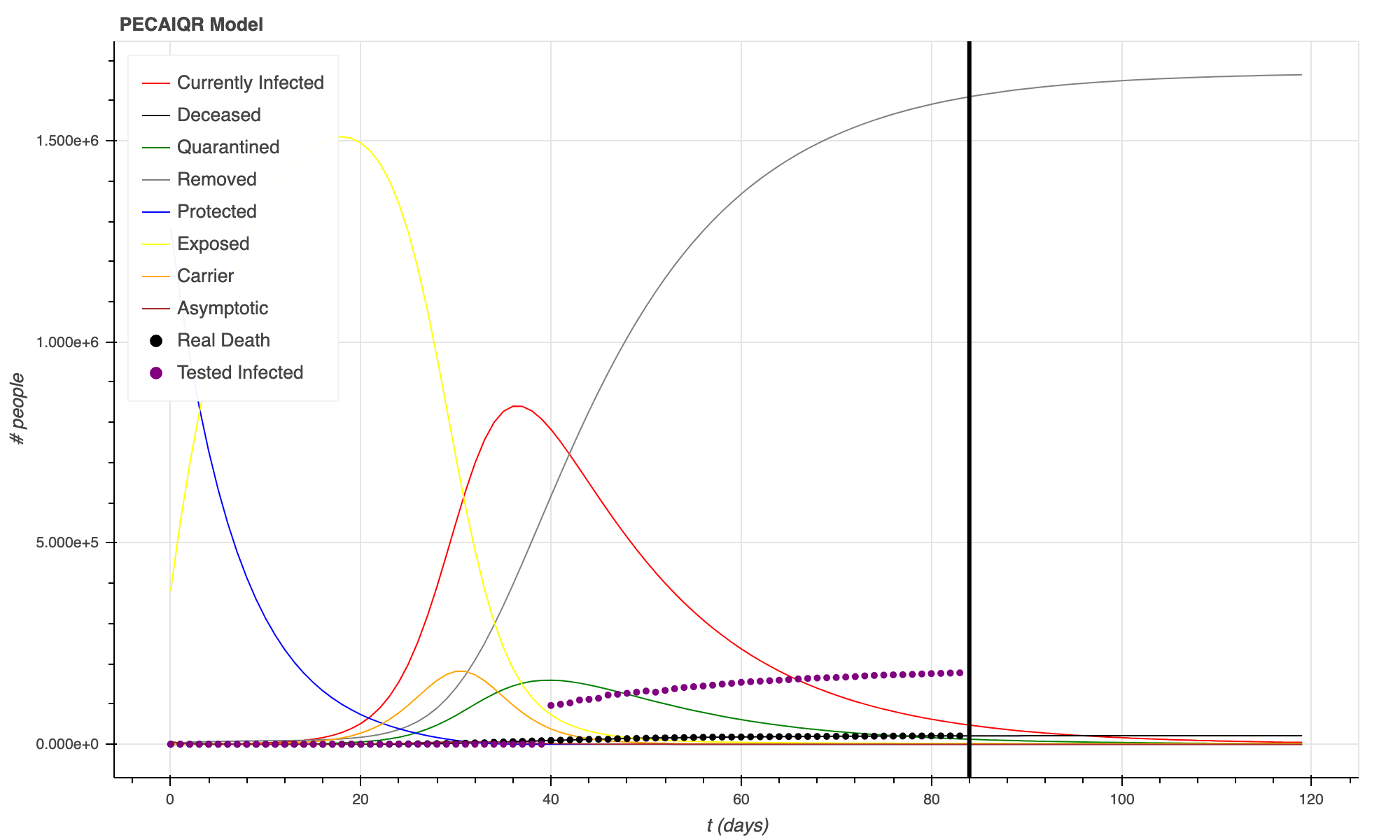}}}%
    \qquad
    \subfloat[]{{\includegraphics[width=7.5cm, height=4cm]{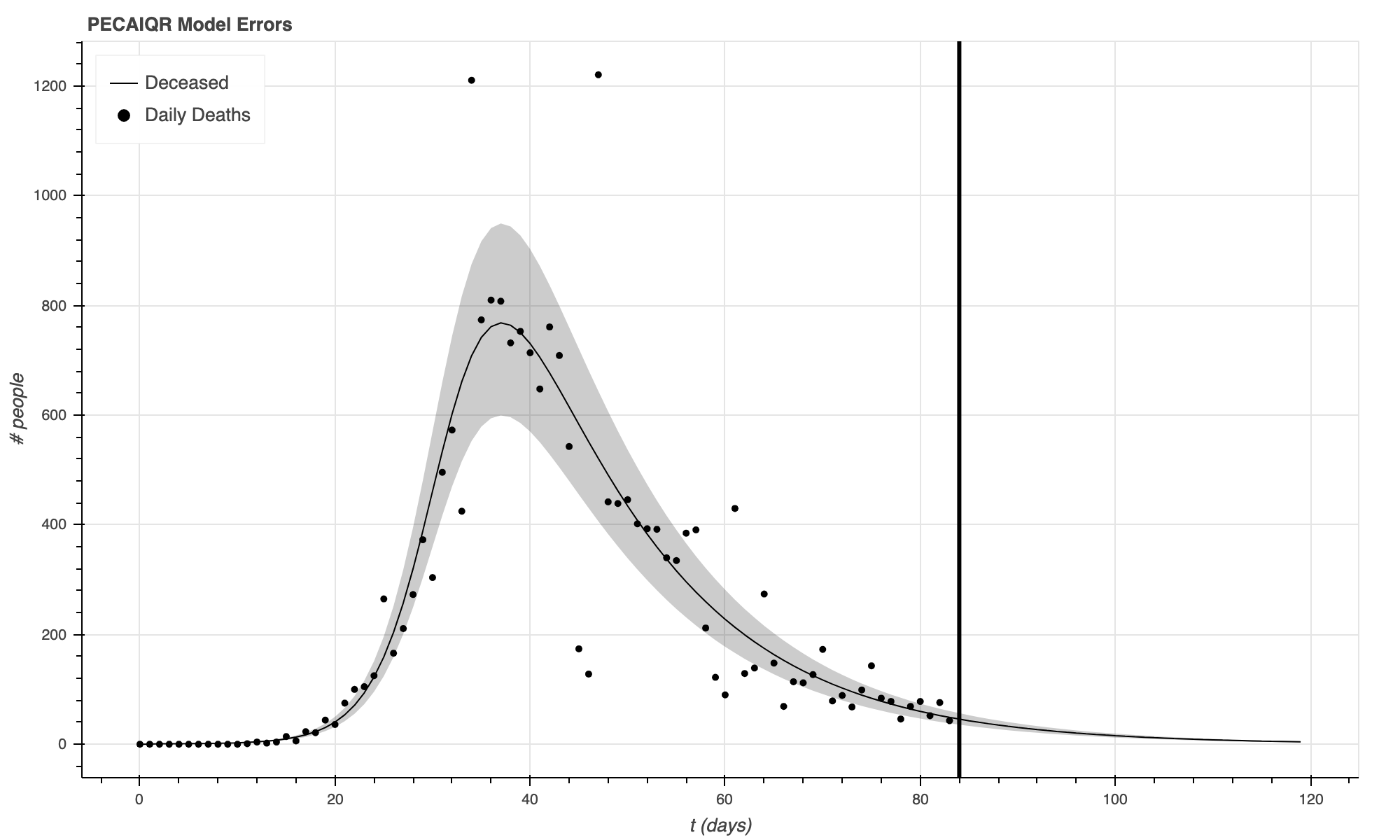}}}%
    \qquad
    \subfloat[]{{\includegraphics[width=7.5cm, height=4cm]{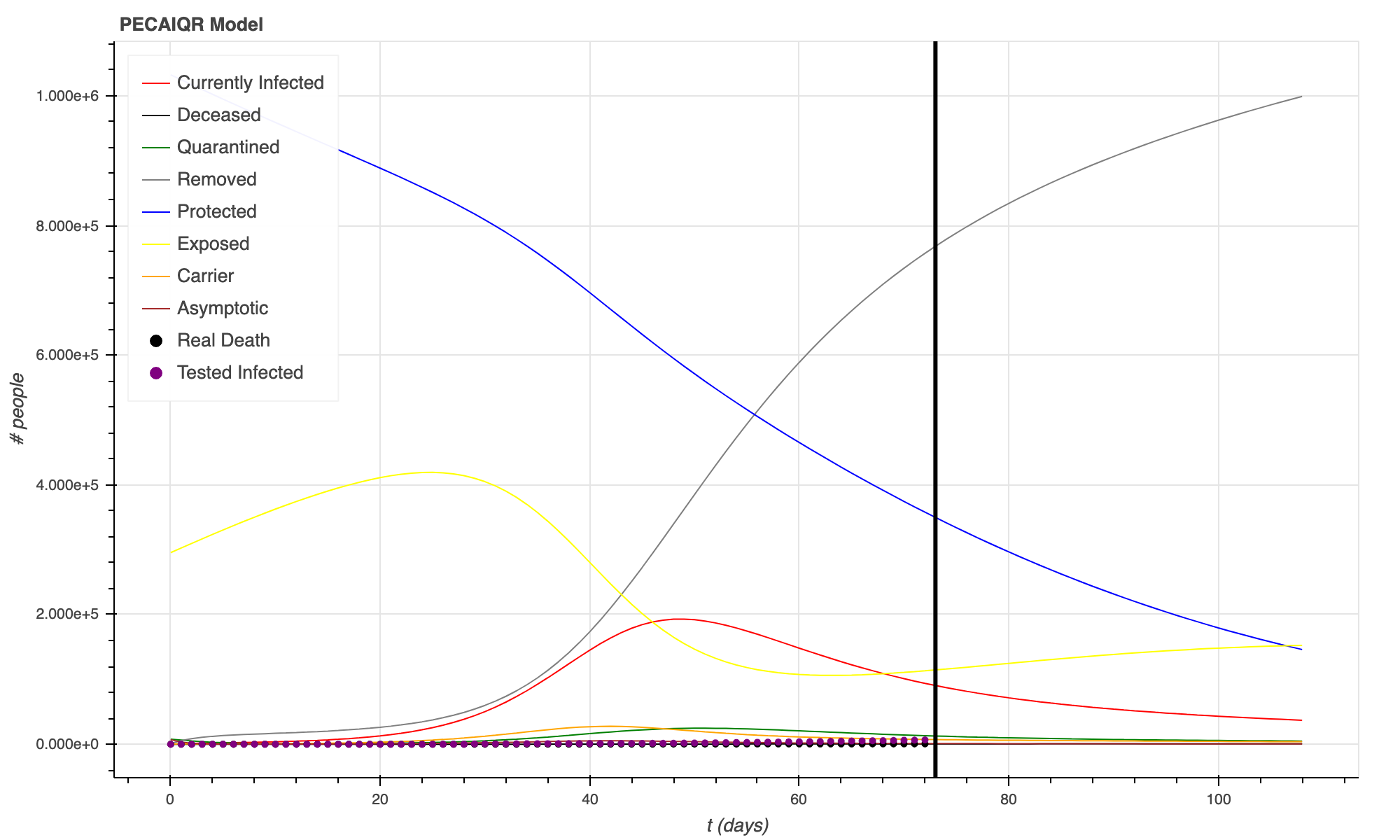} }}%
    \qquad
    \subfloat[]{{\includegraphics[width=7.5cm, height=4cm]{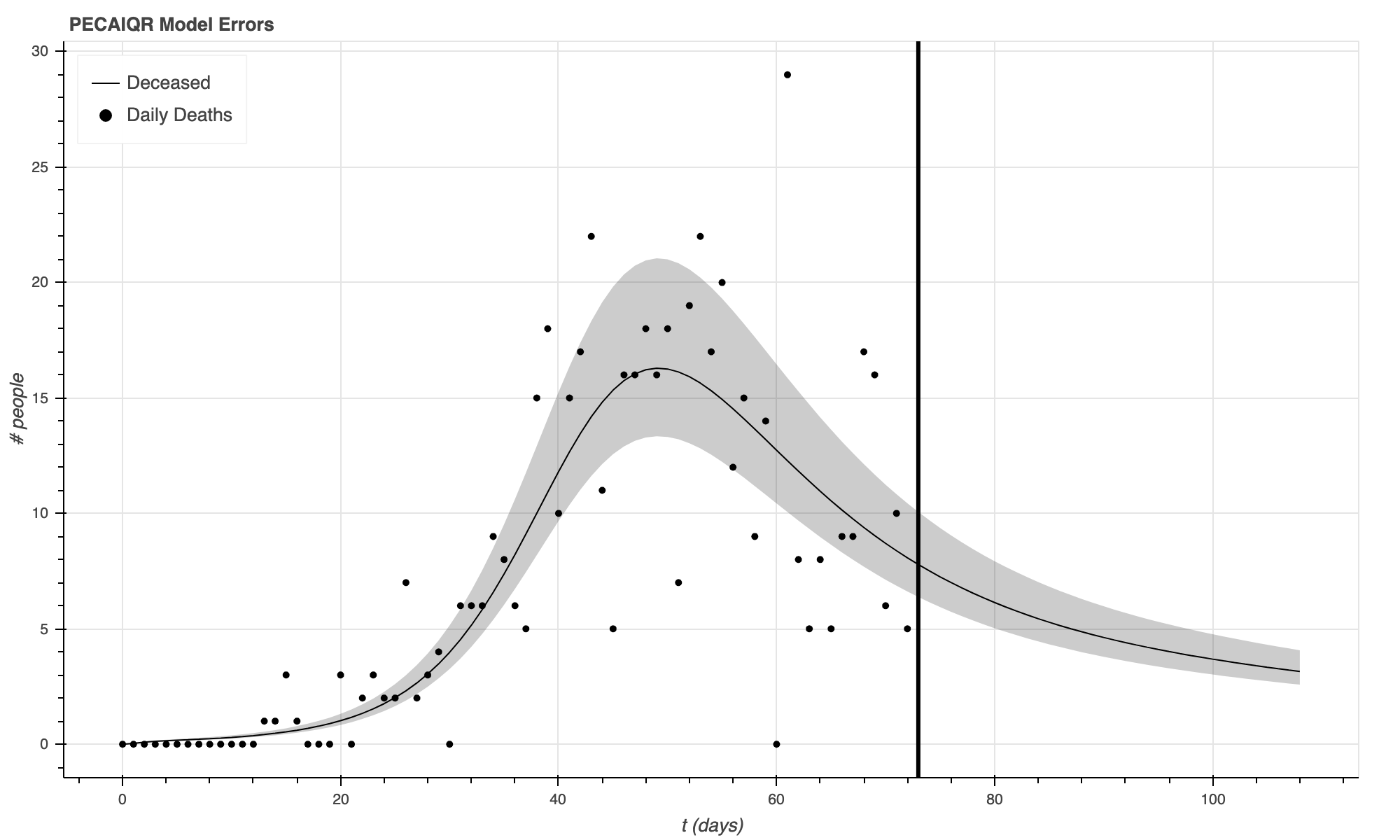}}}%
    \qquad
    \subfloat[]{{\includegraphics[width=7.5cm, height=4cm]{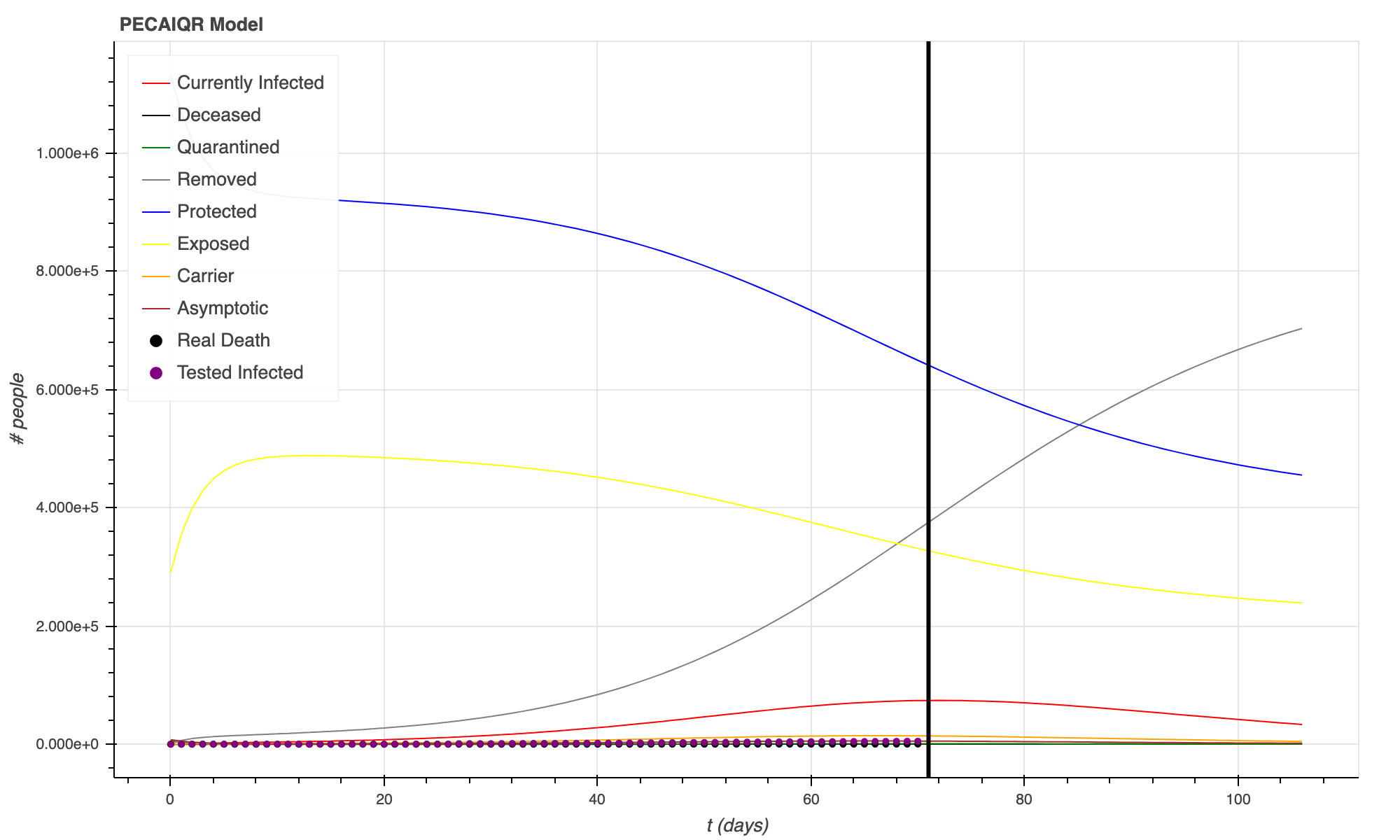}}}%
    \qquad
    \subfloat[]{{\includegraphics[width=7.5cm, height=4cm]{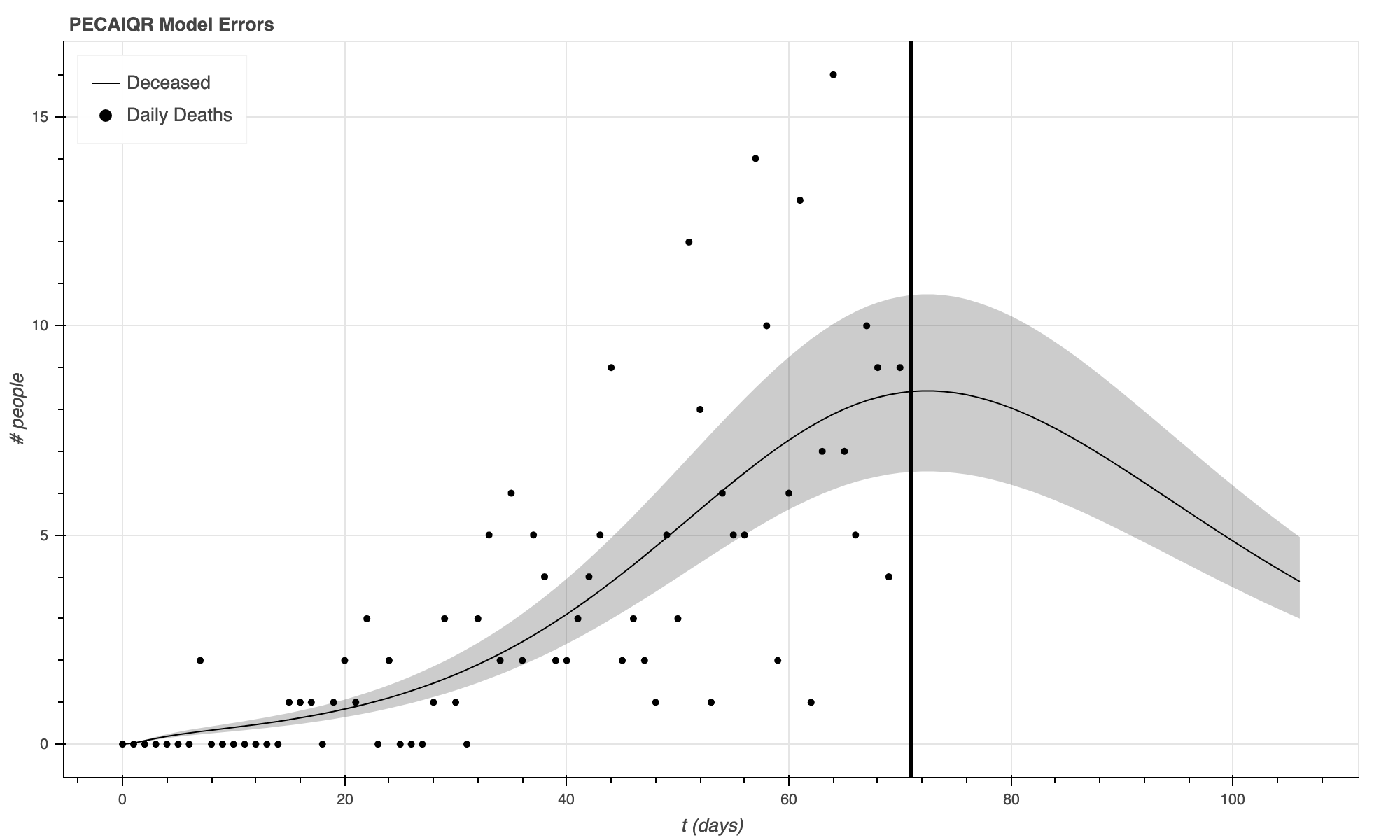}}}%
    \qquad
    \subfloat[]{{\includegraphics[width=7.5cm, height=4cm]{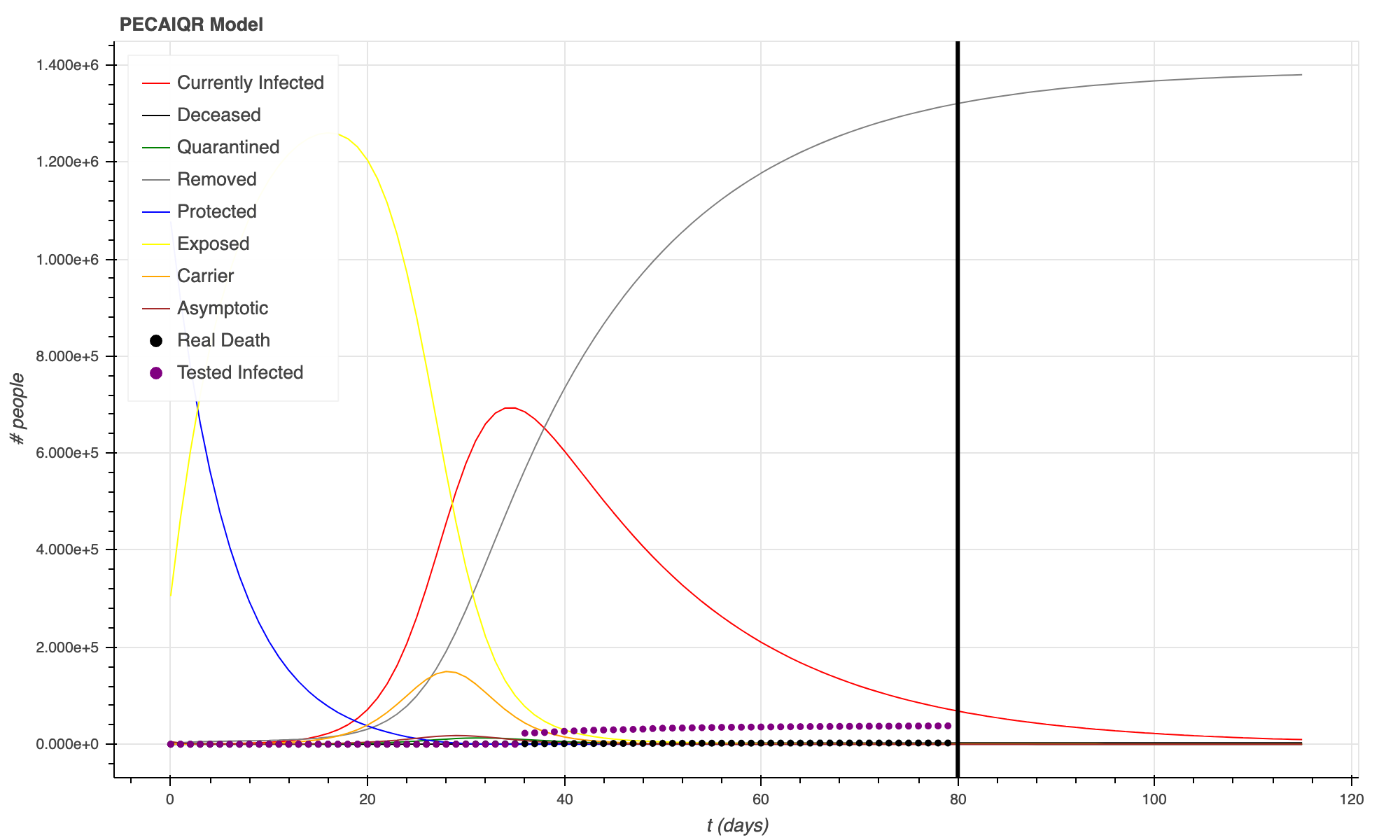}}}%
    \qquad
    \subfloat[]{{\includegraphics[width=7.5cm, height=4cm]{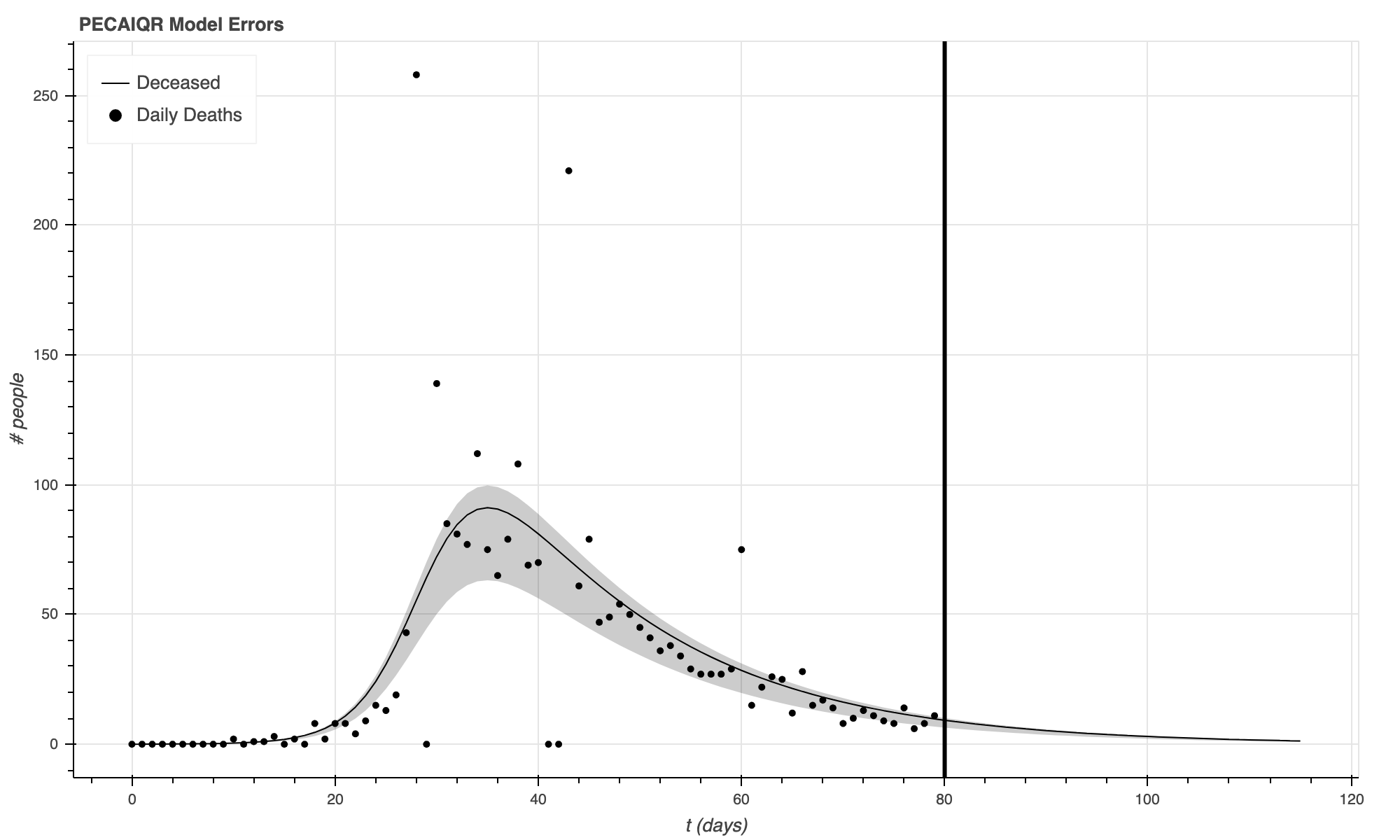}}}%
    \qquad
    \caption{The first row of figures is county 36061. The second row is county 27053. The third row is county 39049. The fourth row is county 36059.}%
\end{figure}
\begin{figure}[H]%
    \centering
    \subfloat[]{{\includegraphics[width=7.5cm, height=4cm]{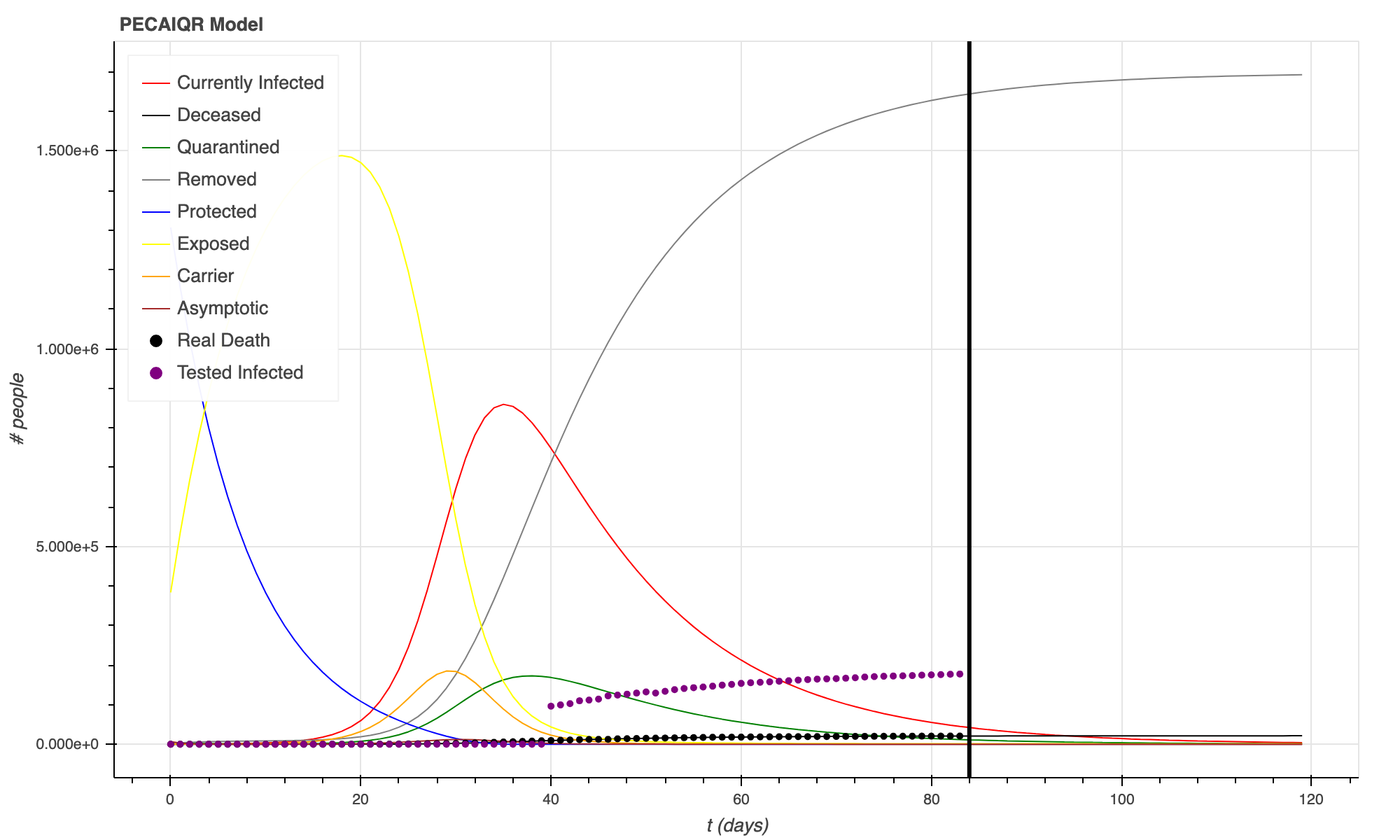}}}%
    \qquad
    \subfloat[]{{\includegraphics[width=7.5cm, height=4cm]{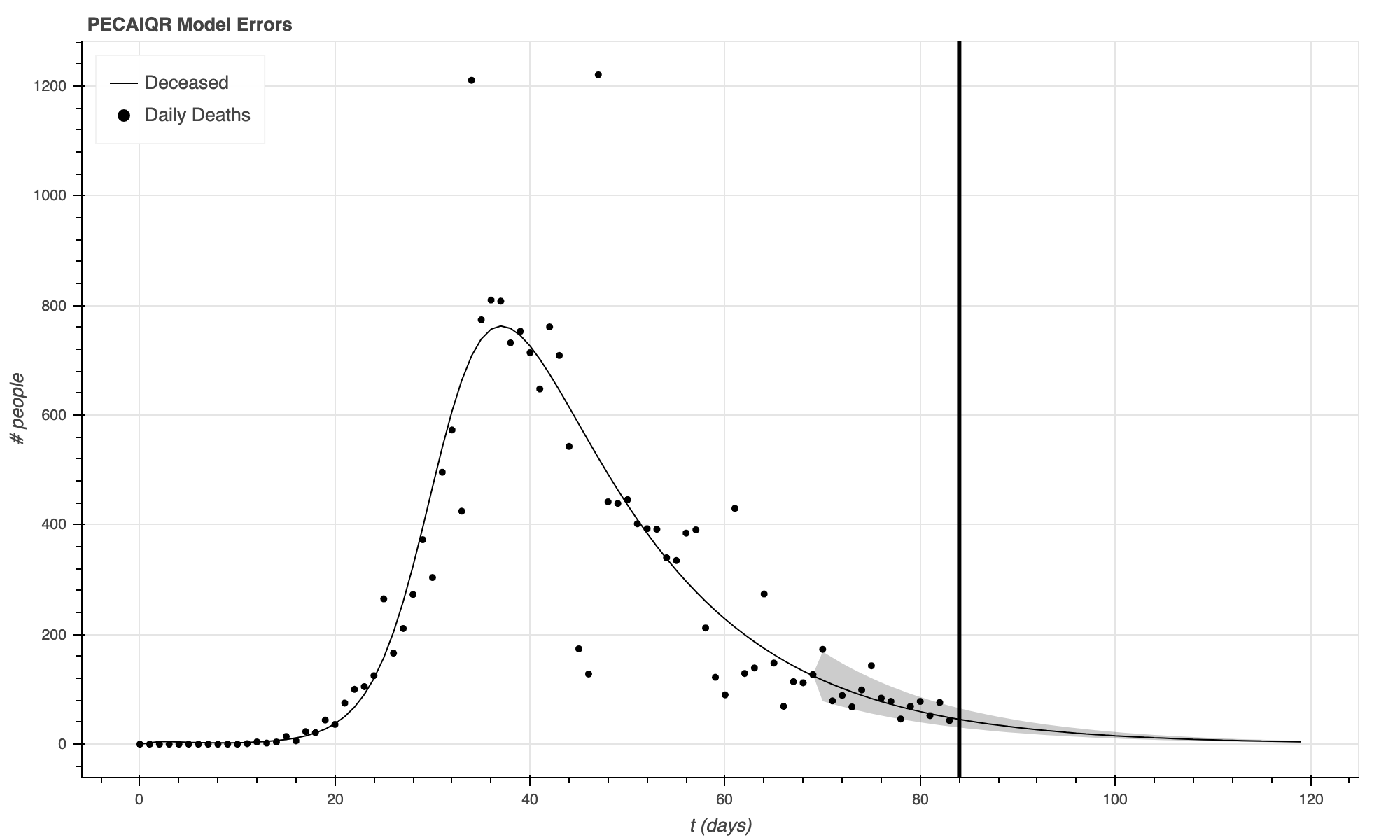}}}%
    \qquad
    \subfloat[]{{\includegraphics[width=7.5cm, height=4cm]{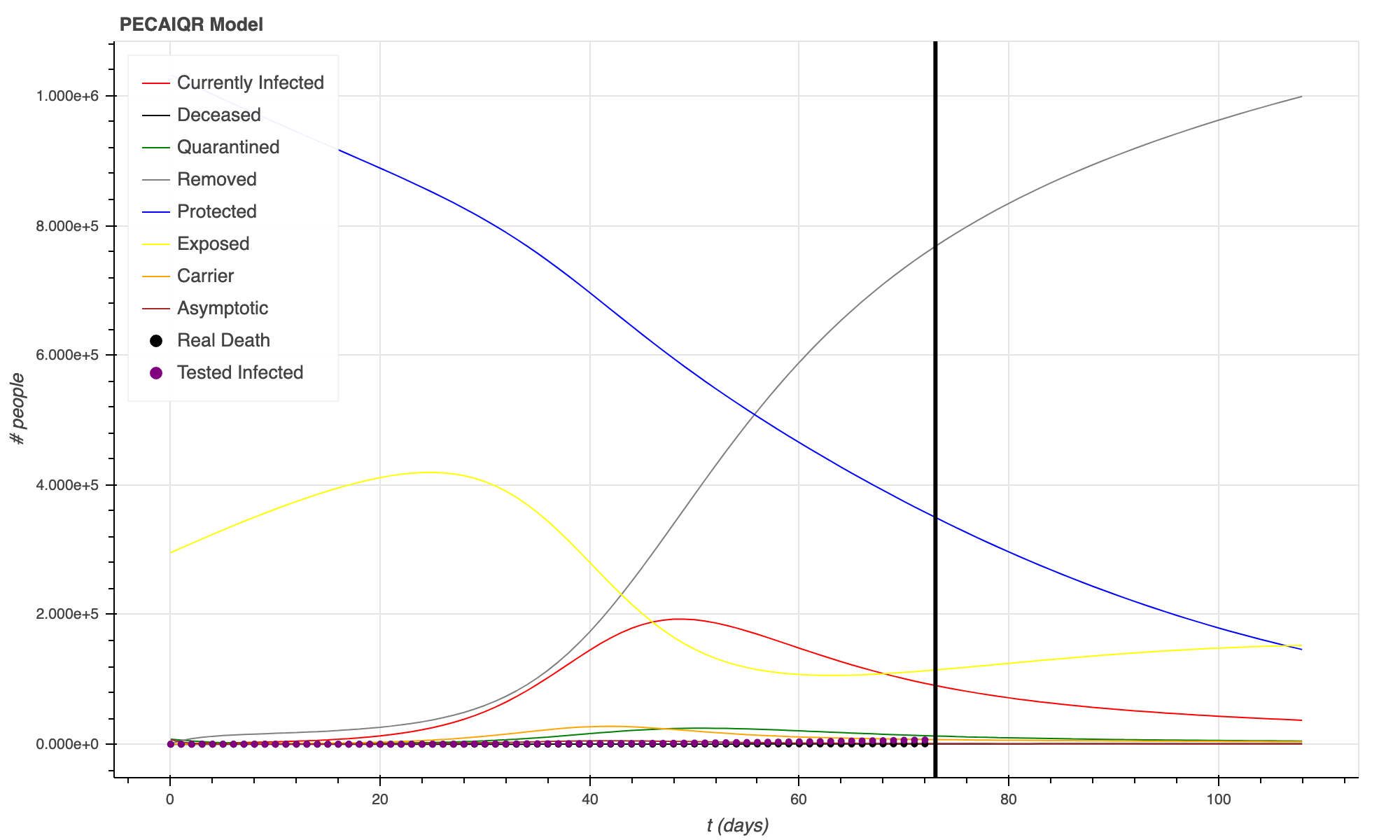} }}%
    \qquad
    \subfloat[]{{\includegraphics[width=7.5cm, height=4cm]{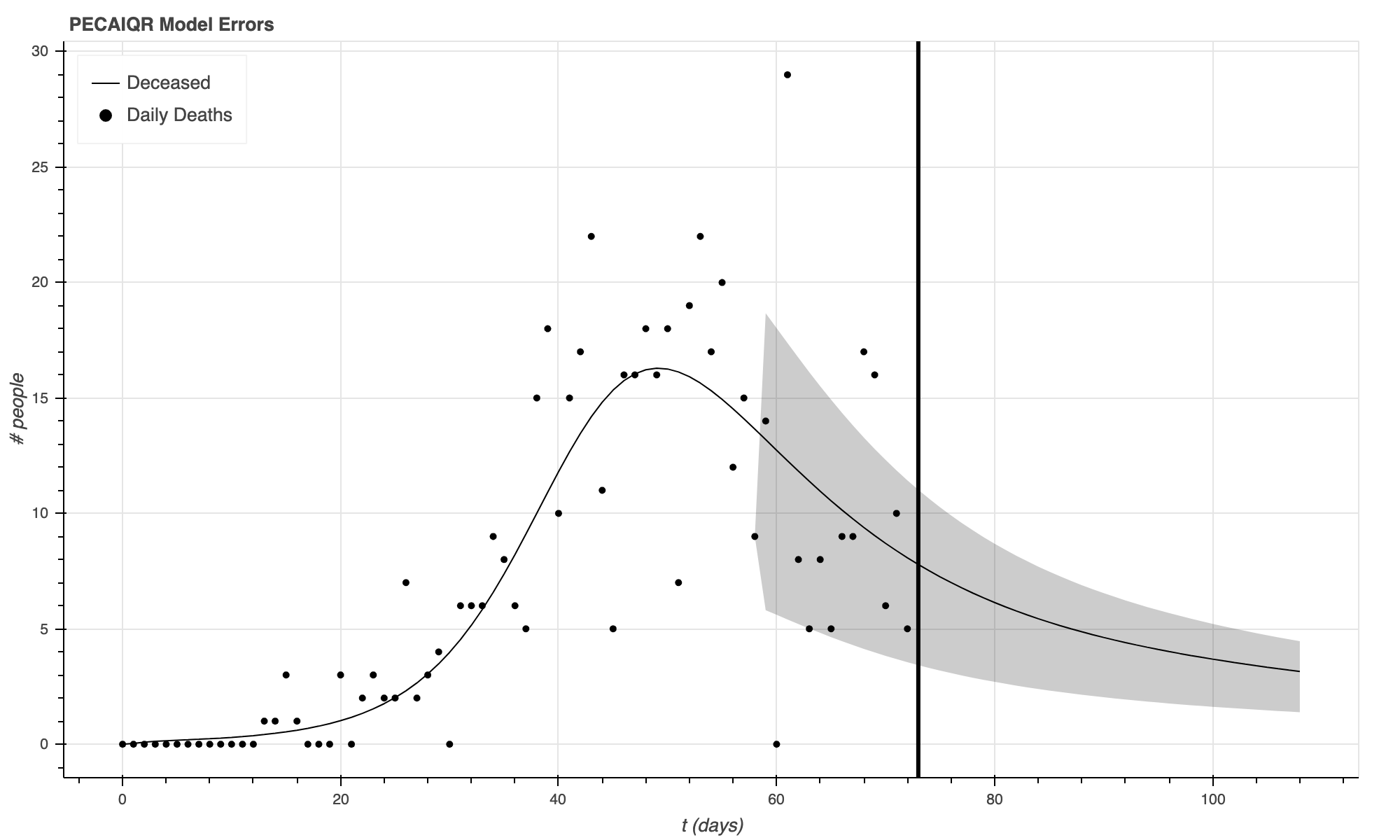}}}%
    \qquad
    \subfloat[]{{\includegraphics[width=7.5cm, height=4cm]{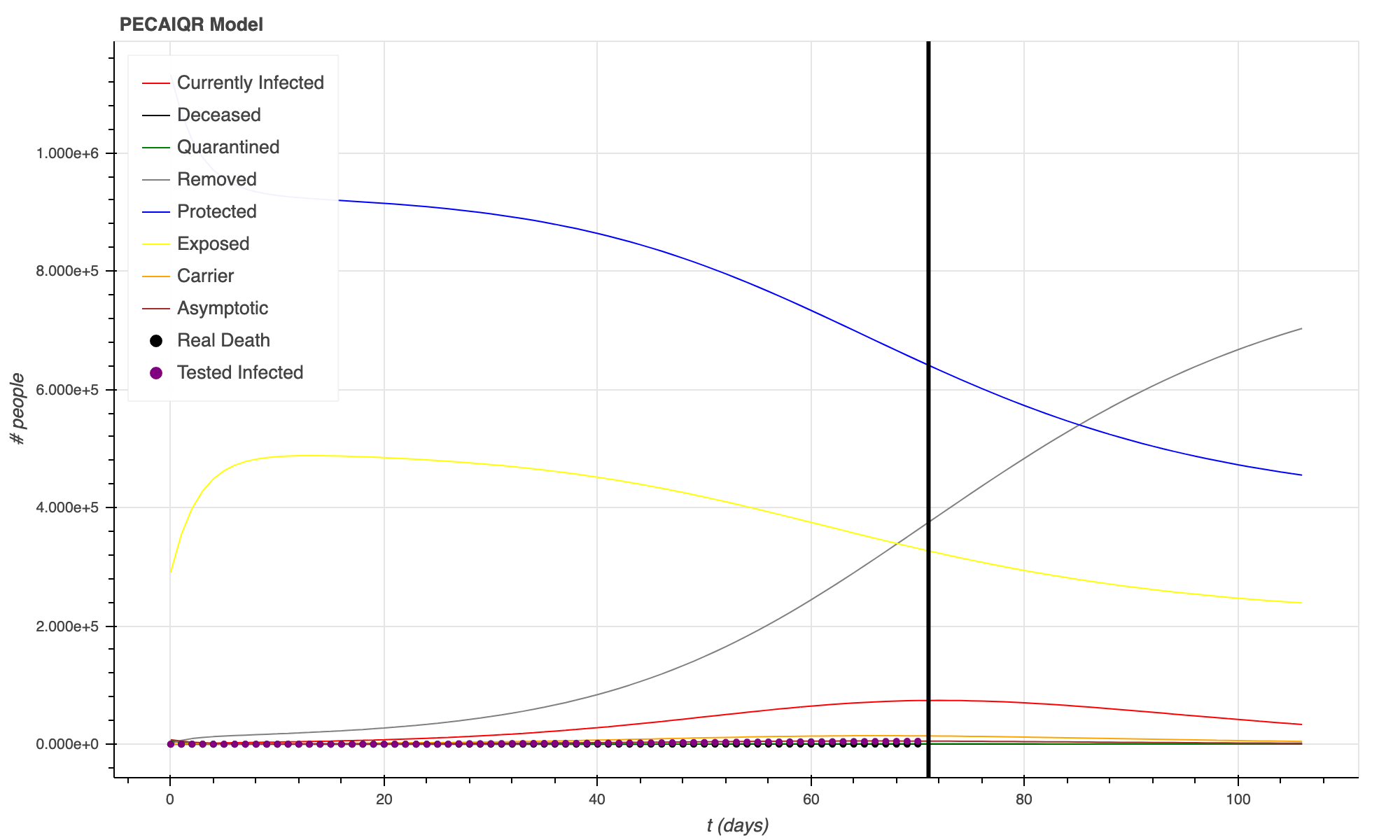}}}%
    \qquad
    \subfloat[]{{\includegraphics[width=7.5cm, height=4cm]{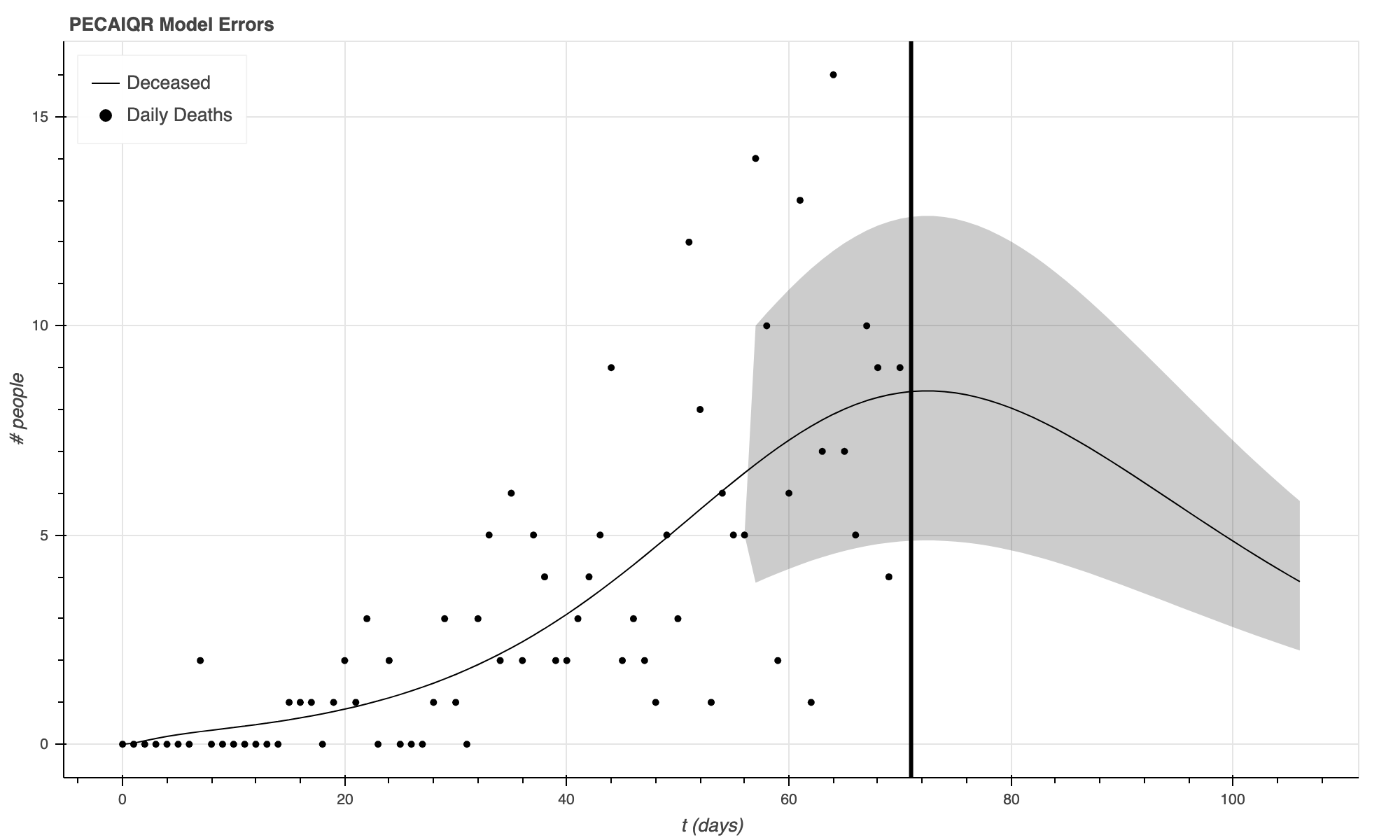}}}%
    \qquad
    \subfloat[]{{\includegraphics[width=7.5cm, height=4cm]{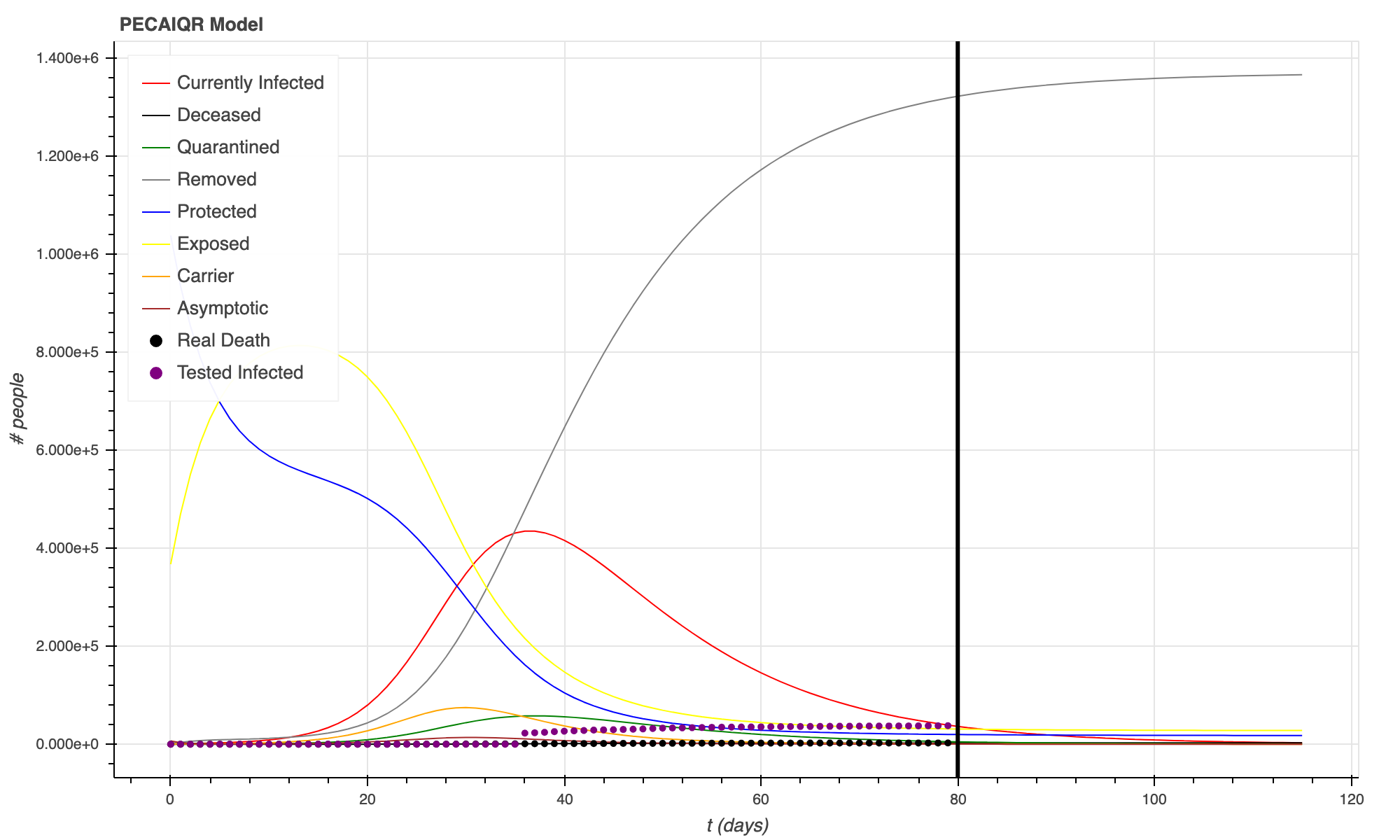}}}%
    \qquad
    \subfloat[]{{\includegraphics[width=7.5cm, height=4cm]{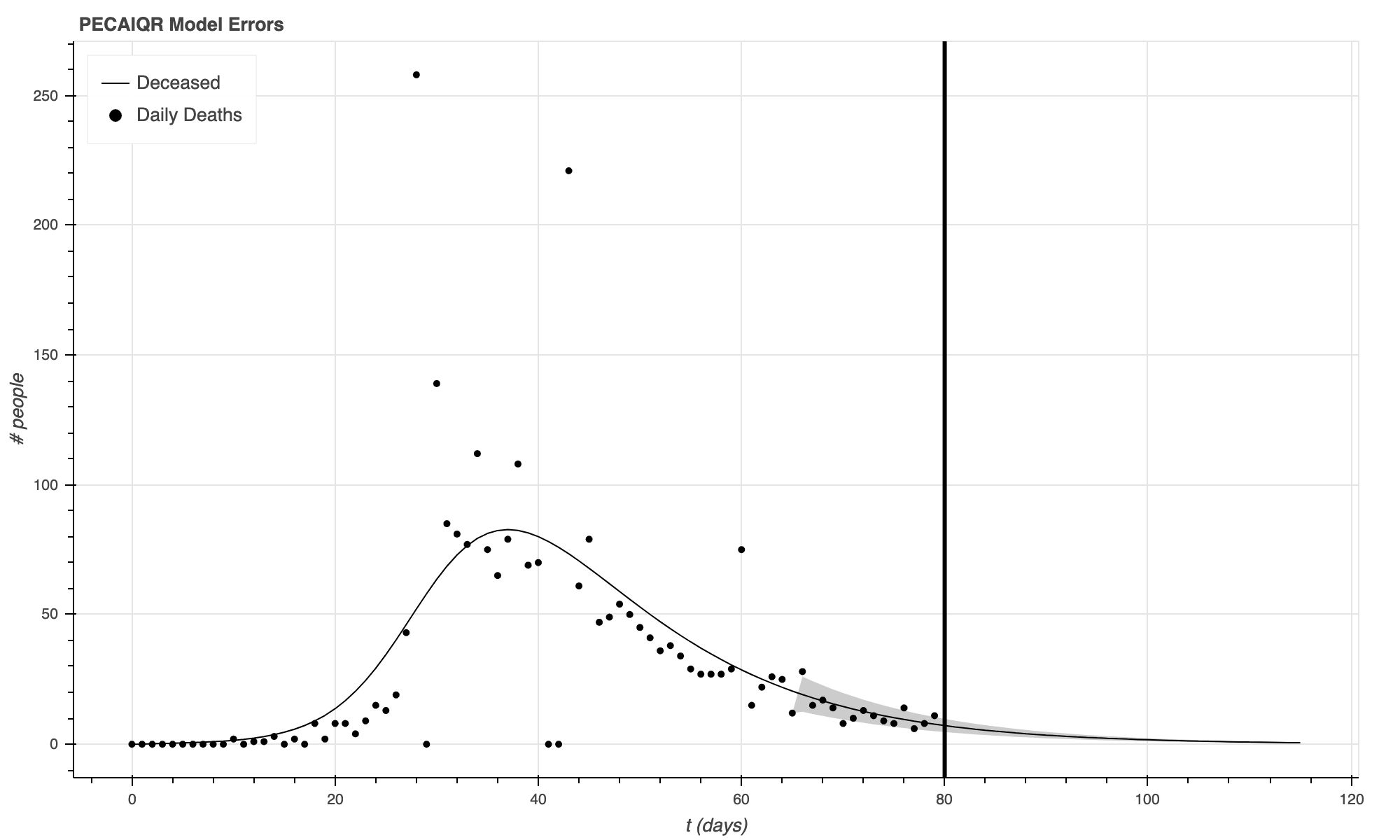}}}%
    \qquad
    \caption{Using the "fitting on tail" method. The first row of figures is county 36061. The second row is county 27053. The third row is county 39049. The fourth row is county 36059.}%
\end{figure}

\noindent Following these observations, we hypothesized that there could be better parameter guesses that we have not encountered. In order to get better parameter guesses, further research should focus on deriving reasonable values for the more intuitive parameters of the model, such as those concerning the rate of infection, from existing data. This also illuminates the issue of ODE stiffness. We believe the functionality of the scipy odeint package is quite limited. Given more time, we would explore other statistical packages like Stan, which has a state of the art implementation of 4th order Runge Kutta numerical integration for stiff ODEs \cite{stiff}. Another improvement would be to use a numerical integrator that solve handle delay differential equations, which would allow us to have more control over the time delay in expression between infected and death states.\\
\subsection{Advantages of the PECAQIR Model}
\noindent We believe that overall the PECAIQR model is very promising, and reveals the benefit of attempting more ambitious, complex epidemiological models. The epidemiological model's differential equations establishes intuitive rules by which it operates, so it has more long term predictive power than most other models. The shape of the solutions of the PECAIQR are also quite interesting, as they resemble heavy tail distributions similar to the Frechet/Weibull distribution. The heavy tail is especially important for pandemic forecasting, as we expect the daily deaths to fluctuate around a low value for some time, instead of decaying immediately to zero. In this respect, the tail end of the curve is almost stochastic in nature, once the infection curves lose enough momentum.\\  
\section{Failed Models}
\subsection{HMM Supervised Learning Prediction}
\noindent Due to the high variability found in the data for Reported Deaths for each county, with some counties reporting unrealistic jumps in the numbers of deaths, it seemed sensible to try using a stochastic model to make some predictions of death counts for counties. In order to best preprocess the county data for training, we explored possible clustering methods that would be able to cluster time series of varying lengths. Dynamic-Time Warping (DTW) was picked to be the distance measure of choice as it is able to compare time series of different lengths and provide a “warped distance” between each series of Daily Reported Deaths. \\

\noindent The Daily Reported Deaths were preprocessed by first removing the series with all 0’s, and then performing a z-normalization on each data series. The data series were then compared with Dynamic-Time Warping, and a hierarchical clustering was developed based on the DTW distance matrix. The number of clusters was determined using the elbow method. The series that had all 0’s are then added back into the clustering list, with cluster id ‘0’.\\ 

\noindent It was desired that further clustering be done, specifically clustering involving HMM-Based Clustering. The ideal setup would be a clustering based on an iterative DTW-HMM clustering algorithm, in order to fully extract similarities between county Reported Deaths. Unfortunately we did not have time to fully develop this idea.\\\\
\includegraphics[width=11cm, height=4.5cm]{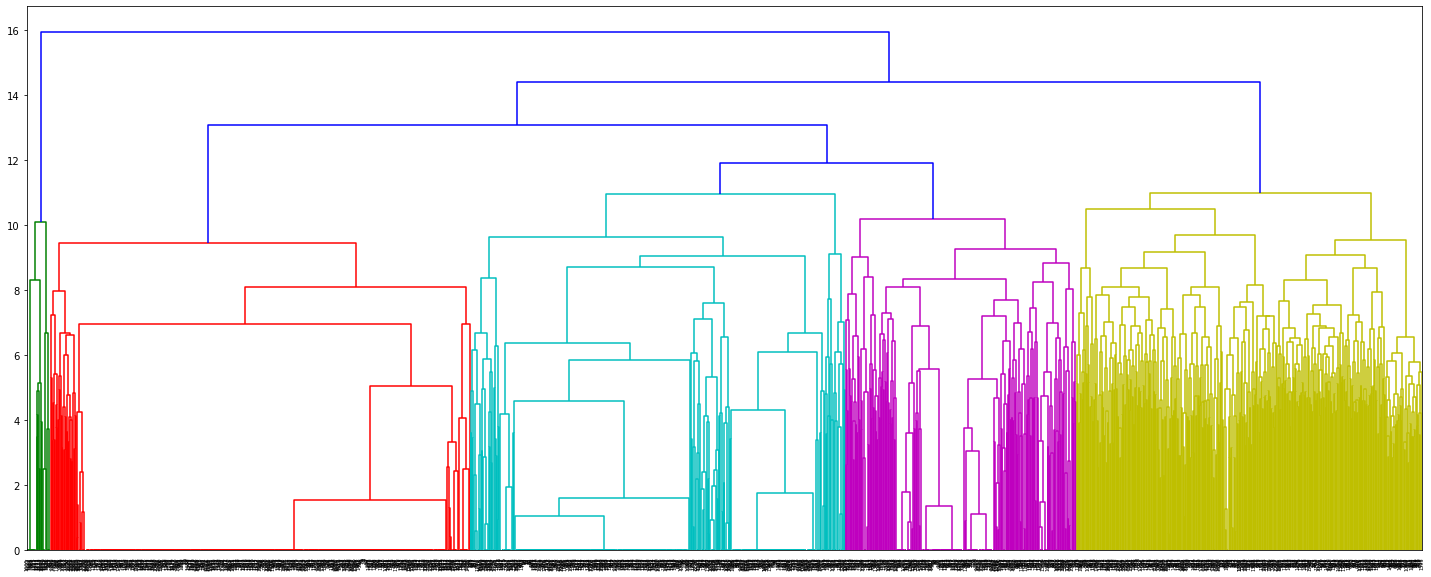}
\includegraphics[width=6cm, height=4.5cm]{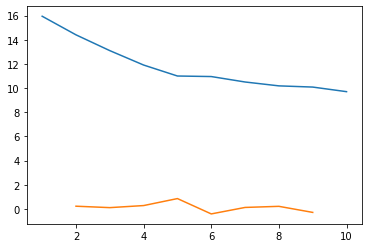}

Dynamic Time Warping Clustering of Daily Reported Deaths \quad \quad \quad \quad Plot of Loss vs Number of Clusters\\

\noindent Once the clusterings were developed, each cluster was used to train an HMM model. We used HMMLearn’s GaussianHMM model in order to have the Gaussian emissions needed for this kind of data (a multinomial model with discrete output would be stretched thin with $\sim$100 states). The number of states each GaussianHMM was given per cluster was chosen by doing a modified version of the elbow method, to ensure that the HMM is complex/simple enough to match the variance in Reported Deaths. Once these HMMs were trained, we developed a method to initiate an HMM close to a particular starting emission and allow it to generate 14 subsequent emissions, to make 14-day predictions for the counties in the cluster. These predictions could be made thousands of times for each HMM, allowing each cluster to effectively create a probability distribution of predictions.
Further work needed to be done to make this stochastic model effective. As the model only allowed for predictions based off of an entire cluster, there needed to be a secondary layer of scaling to allow a cluster prediction to be mapped to each individual county. An idea of attempting to do some time-series comparison and do some sort of series stretching/scaling was developed, but was never fully formed. There are definite rooms for improvement and optimization in this model. One fundamental issue is that perhaps the small number of states of the GaussianMM limits the complexity of each HMM per cluster. While this is true, this level of simplicity mixed with stochasticity could capture the random element seen in the Reported Deaths data. If this random element is able to be removed or smoothed away,however, perhaps this model would not be so useful any longer.\\
\subsection{Gaussian Process}
\noindent 
When attempting to use statistical regression to model nonlinear correlations in data, a common approach is to employ a Bayesian non-parametric strategy, such as the Gaussian process. Bayesian non-parametric strategies like the Gaussian process are essentially extensions of Bayesian Inference on an infinite-dimensional parameter space. Very loosely, this allows us to model the data as the combination of many different Gaussians (each quite accurate in its local region), stitched together to create a single model \cite{PyMC3Video}. This technique was attempted in the latter stages of the course to create predictions for counties where the predictions from the PECAIQR did not converge, an issue at the time for counties with poor data. \\

\noindent We used a custom implementation of Gaussian Processes, using a mean function of 0 and an exponential squared kernel. The data used was the rolling average of deaths over a three day window vs time. In this model, there are three parameters: $l$ which is the length parameter for the kernel, $\sigma_f$ which is vertical variation parameter for the kernel, and $\sigma_y$ which is the noise parameter. For each county, the optimal hyperparameters for that county were found by searching for the set of parameters within a given range that minimized the error. Here, the bounds on the parameters $5.0 \le l \le 15.0, 0.1 \le \sigma_f \le \frac{m}{500.0}, 0.1 \le \sigma \le \frac{m}{10.0}$ where $m$ was the maximum value of rolling three day average deaths over the past 14 days. The error function was the root mean squared error (RMSE) over the last 30 days. These bounds and the error function were determined with validation procedures.\\

\begin{figure}[H]
\centering
\includegraphics[width=17.5cm, height=13.5cm]{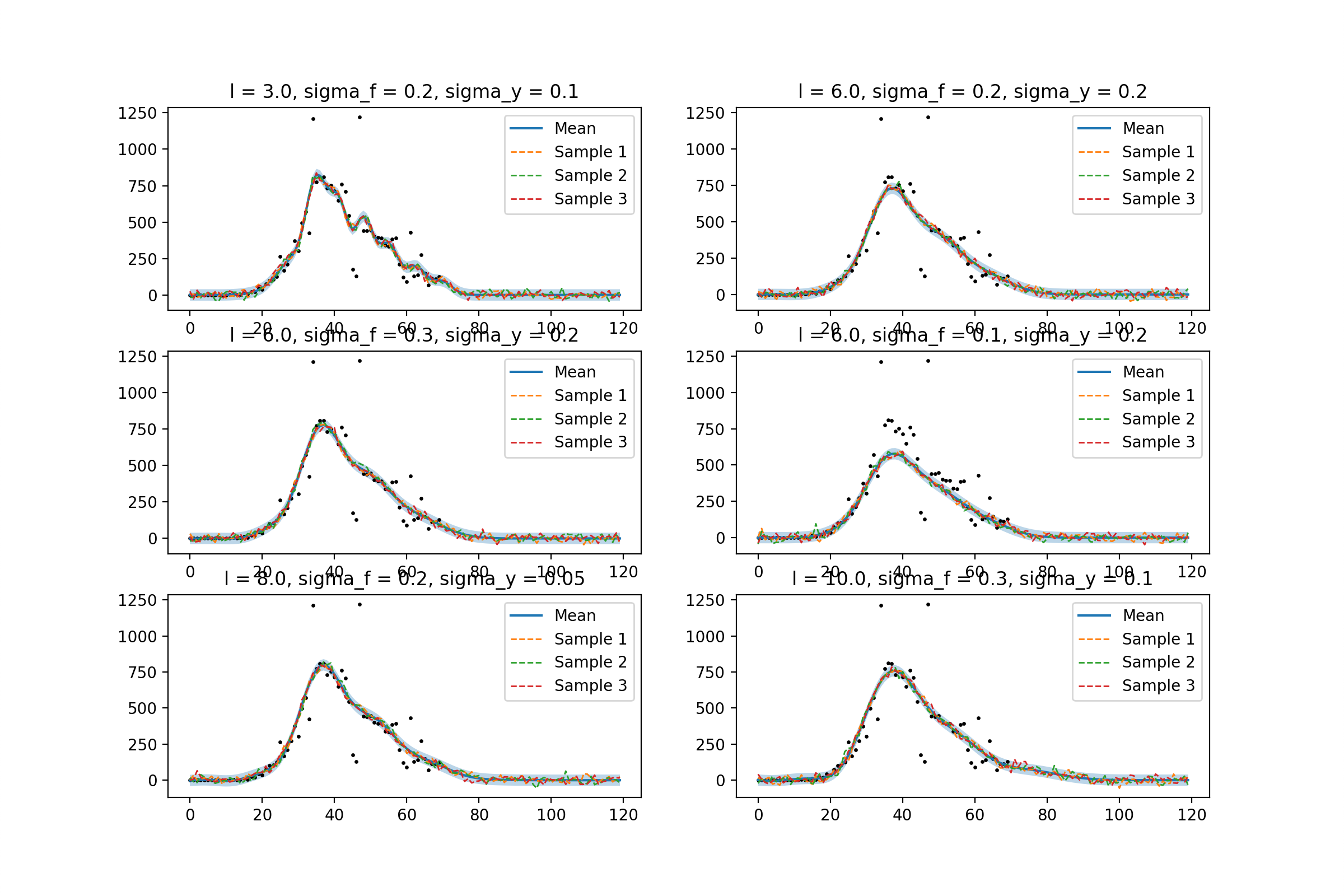}
\caption{Sampling of fits for county 36061 in the hyperparameter space}
\end{figure}

\begin{figure}[H]%
    \centering
    \subfloat[]{{\includegraphics[width=8cm, height=5cm]{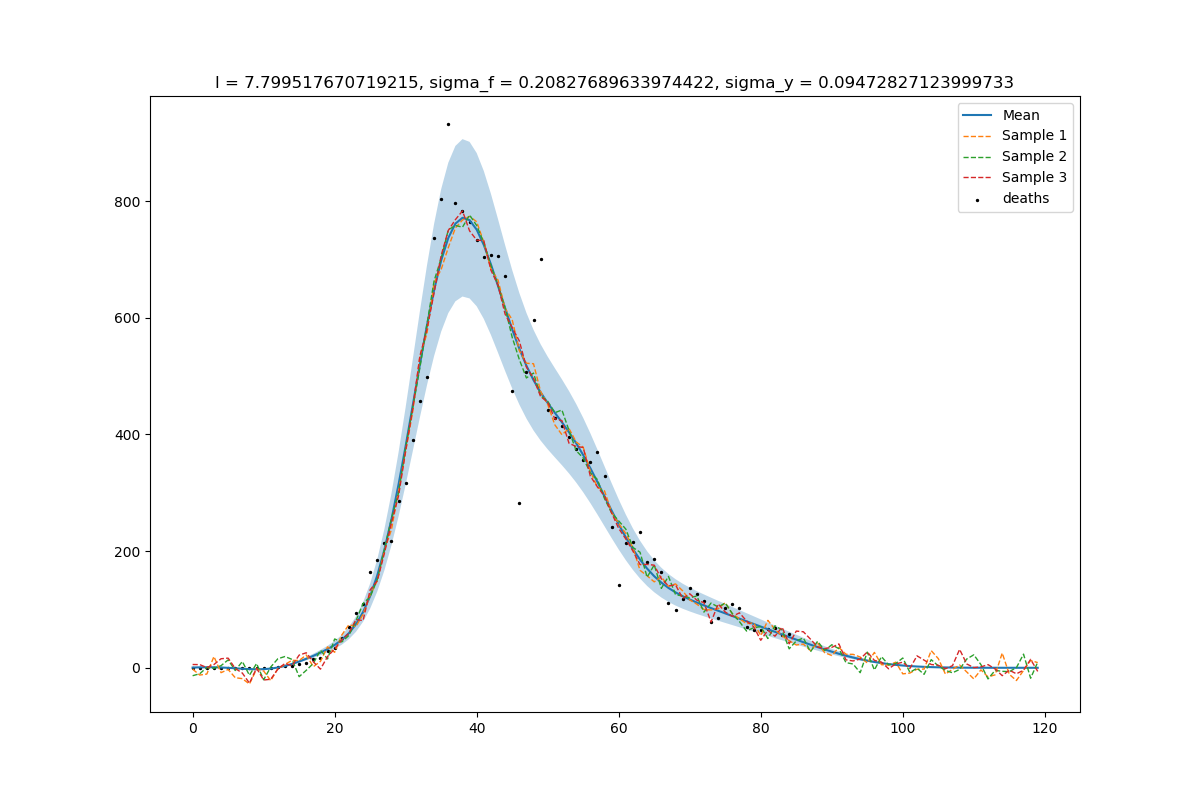}}}%
    \qquad
    \subfloat[]{{\includegraphics[width=8cm, height=5cm]{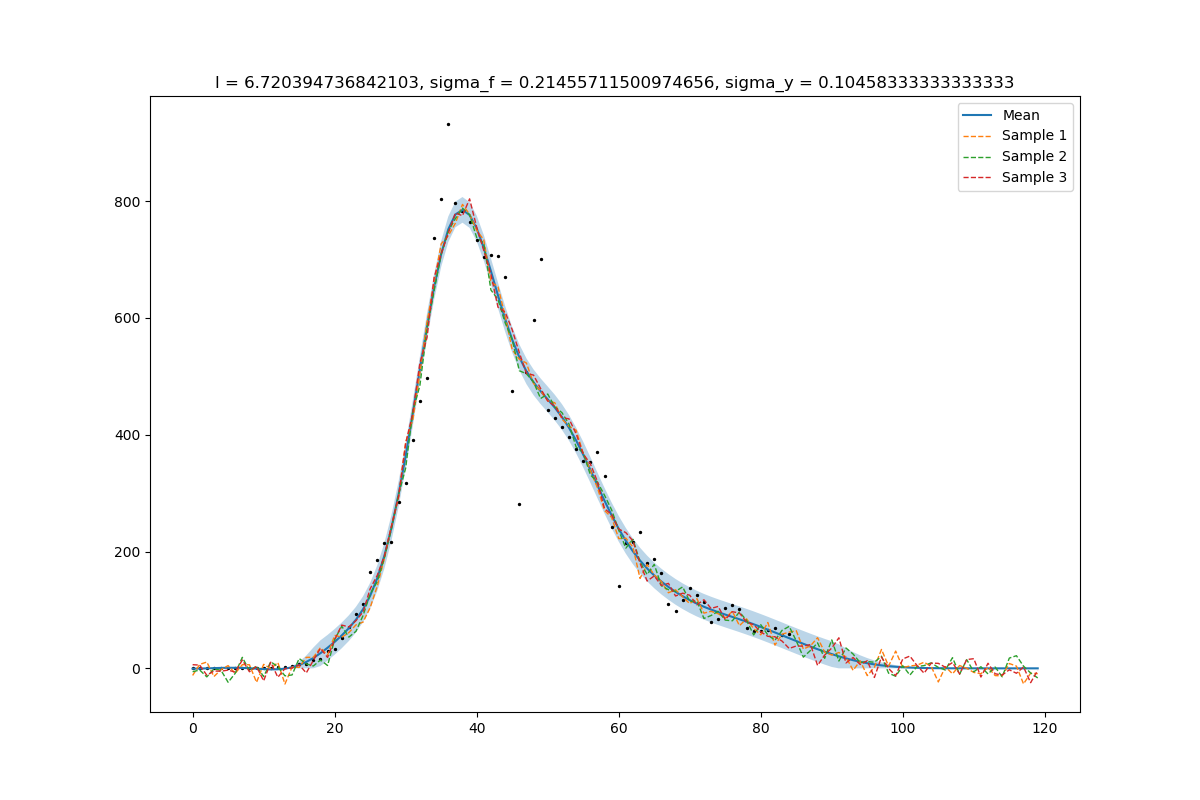} }}%
    \qquad
    \caption{Fit on county 36061 with optimized hyperparameters from Gridsearch. The shaded region shows confidence intervals from 10th to 90th percentiles. Sub-figure a) shows error calculated from the deviation of the residuals from the predicted fit, and Sub-figure b) shows error calculated from the co-variance matrix.}%
\end{figure}

\noindent The Gaussian process seems to fit quite accurately in the short term, but the predictions quickly drop to zero, as demonstrated in the plots above. Unlike the PECAIQR model, it does not retain a heavy tail. Therefore, Gaussian process does not have long term predictive power. We can possibly improve the Gaussian process and overcome the issue by setting a custom mean instead of using the default zero mean, which likely contributes to the rapid decay of the tail. We did attempt to so, inspired by the PyMC3 Tutorial posted in the CS156b Piazza \cite{PyMC3Tutorial}. It was attempted since a custom mean function could be used \cite{PyMC3Mean}; similarities were noted between the graph of the number of deaths vs time in New York County and the Weibull distribution, so the Weibull distribution was this custom mean. However, it was not pursued further since it could not successfully run -- when training the model on just one particular county, the program timed out and crashed Jupyter on the computer it was run on. Since the final deadline was already quite close, this attempt was abandoned. Given how this model can fit to a custom mean, if it was attempted earlier it might have proved to be useful.

\begin{figure}[H]
\includegraphics[width=16cm, height=13cm]{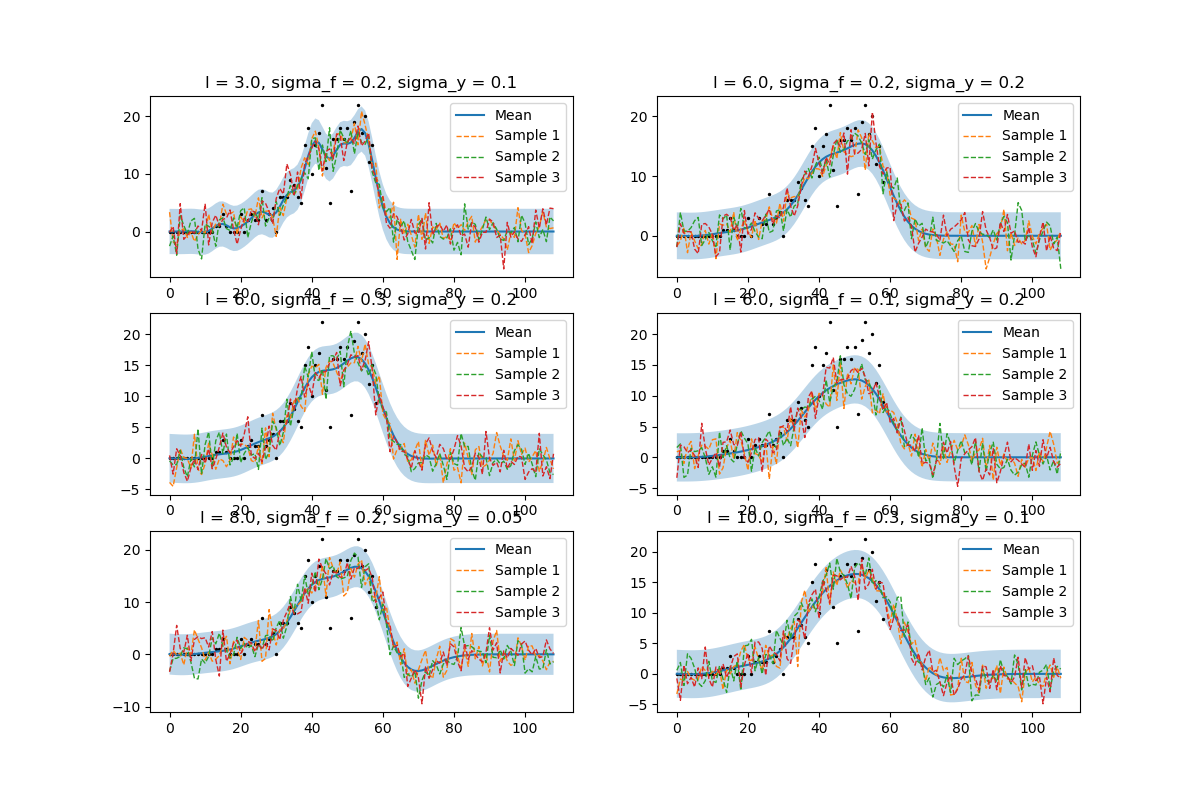}
\caption{Sampling of fits for county 27053 in the hyperparameter space}
\end{figure}
\noindent We verify that the issue of a rapidly decaying tail is not specific to county 36061 as shown above. We can conclude that the Gaussian process fits very well on the training data, but fails to present long term predictive power.\\

\noindent Again, these failures in the Gaussian process model may be overcome by setting a custom mean. But we believe that the epidemiological model, though less accurate, has intrinsic advantages that the Gaussian process cannot match. This is because the epidemiological model has some sort of intuition in the form of the rules established by its differential equations, while the Gaussian process is purely curve fitting. Perhaps, given more time, we could have combined these two models to utilize both the short term accuracy of the Gaussian process and the long term predictive power of the Epidemiological model.\\
\clearpage

\section{Longer Time Horizon Predictions}
\begin{figure}[H]%
    \centering
    \subfloat[]{{\includegraphics[width=7.5cm, height=4cm]{figures/36061_may25_regular_error.png}}}%
    \qquad
    \subfloat[]{{\includegraphics[width=7.5cm, height=4cm]{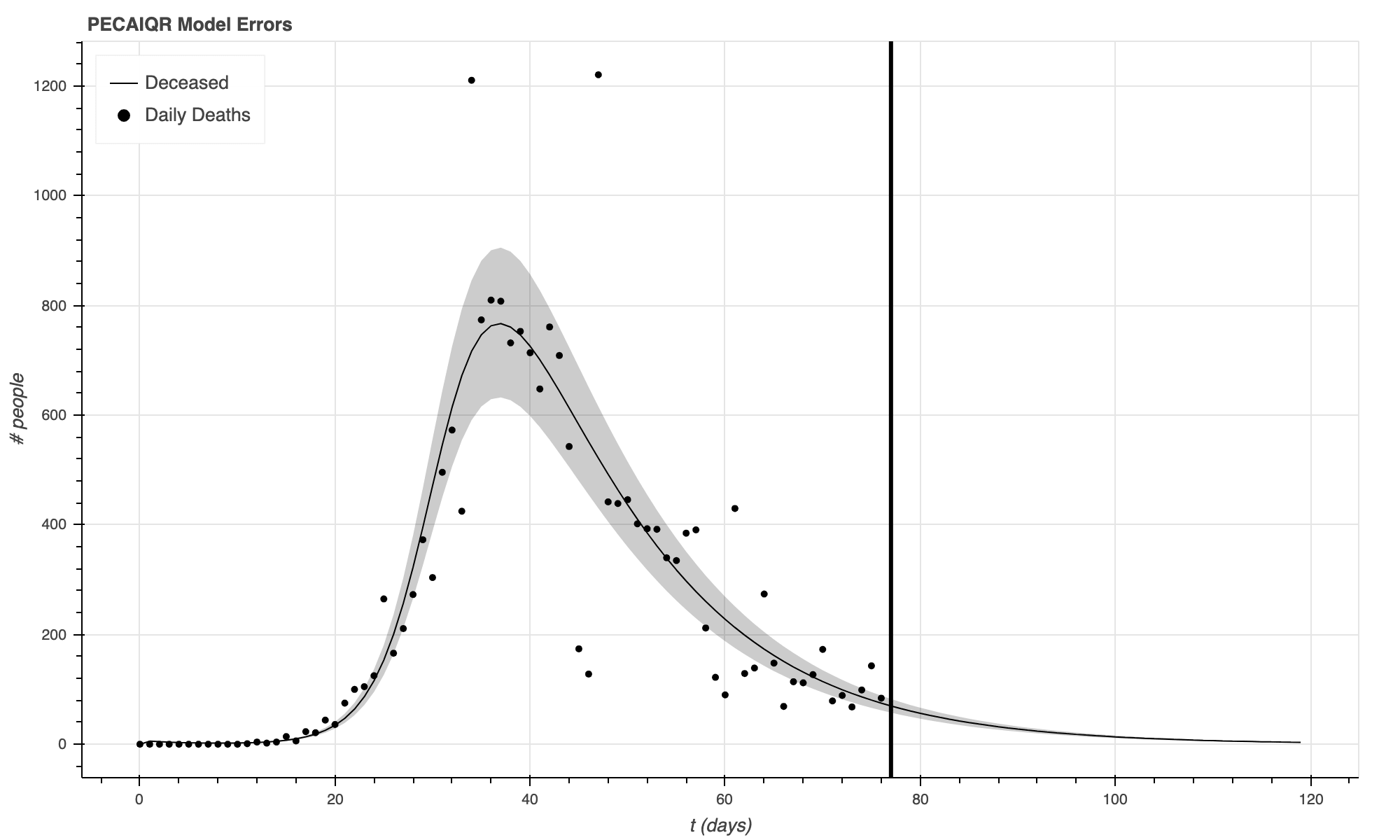} }}%
    \qquad
    \subfloat[]{{\includegraphics[width=7.5cm, height=4cm]{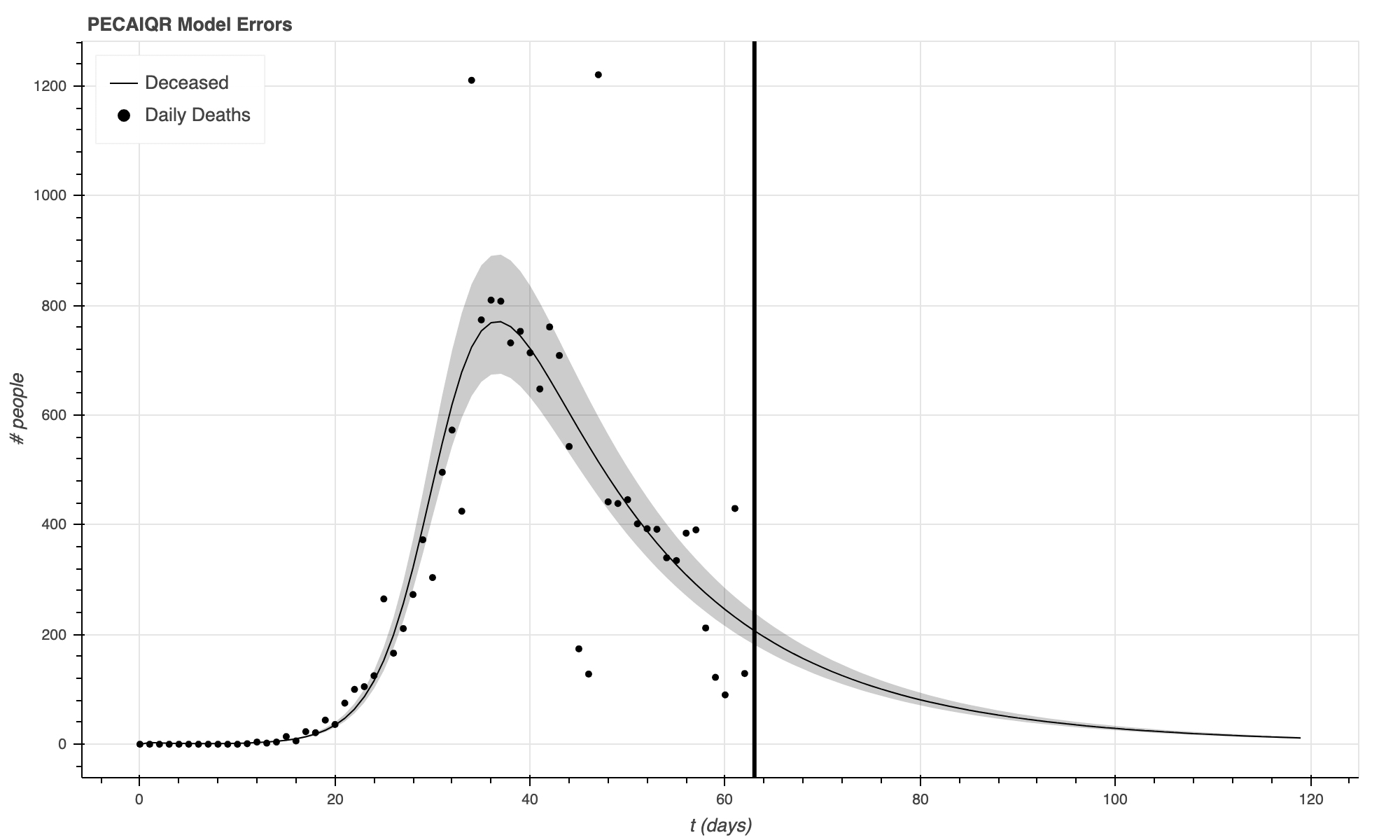}}}%
    \qquad
    \subfloat[]{{\includegraphics[width=7.5cm, height=4cm]{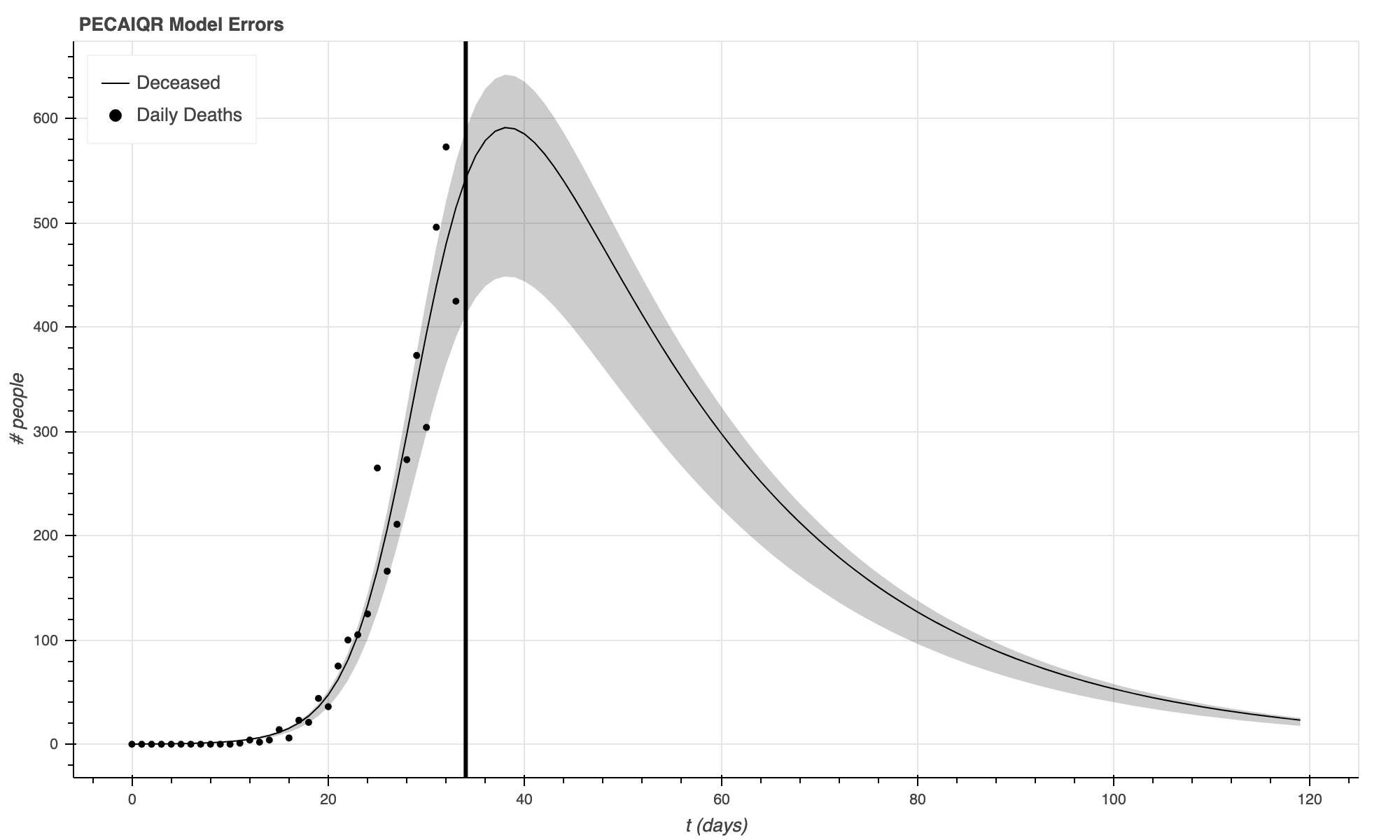}}}%
    \qquad
    \subfloat[]{{\includegraphics[width=7.5cm, height=4cm]{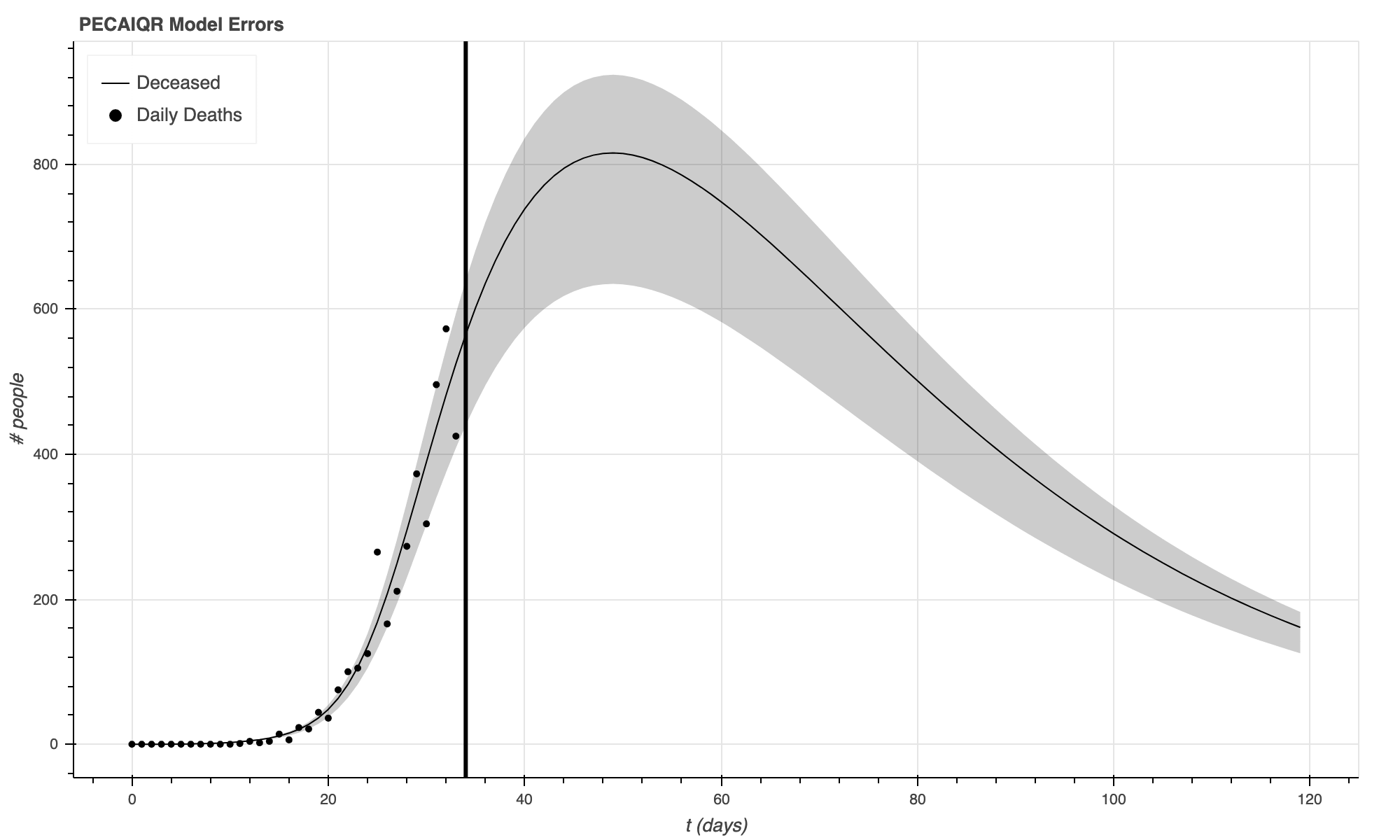}}}%
    \qquad
    \caption{Predictions on county 36061 with training date cutoffs at May 25 a), May 18 b), May 4 c), April 5 d) and April 5 e). Sub-figures d) and e) have the same cutoff, but different hyper-parameters}%
\end{figure}
\noindent Analysis of model predictions for different cutoff dates in the training data shows that the model is quite stable and consistent when predicting on the region past the peak. However, predicting before the peak is much harder, as we are no longer operating with the assumption that the infection curves are dying down. In sub-figure d) we see that the peak daily deaths value predicted by the model is significantly less than the actual peak that is revealed with more data. However, the location of the peak is correct. We realized that the training data at this early cutoff is largely dominated by data in the regime before the first stay at home order, but we were training using an initial parameter guess that accounted for the effects of a stay at home order, and so in sub-figure e), we tried a different set of parameters for the initial parameter guess in the fitting procedure, inspired from the parameters obtained from a fit on similar early curves in Italian regions. This modification yielded a more accurate prediction of the peak daily deaths value, but a less accurate placement of the peak. The prediction curve becomes extended when we use this alternative parameter set because the model no longer assumes that there will be a stay at home order, and therefore the curve will not flatten to the same degree. 
\clearpage
\begin{figure}%
    \centering
    \subfloat[]{{\includegraphics[width=7.5cm, height=4cm]{figures/27053_may25_regular_error.png} }}%
    \qquad
    \subfloat[]{{\includegraphics[width=7.5cm, height=4cm]{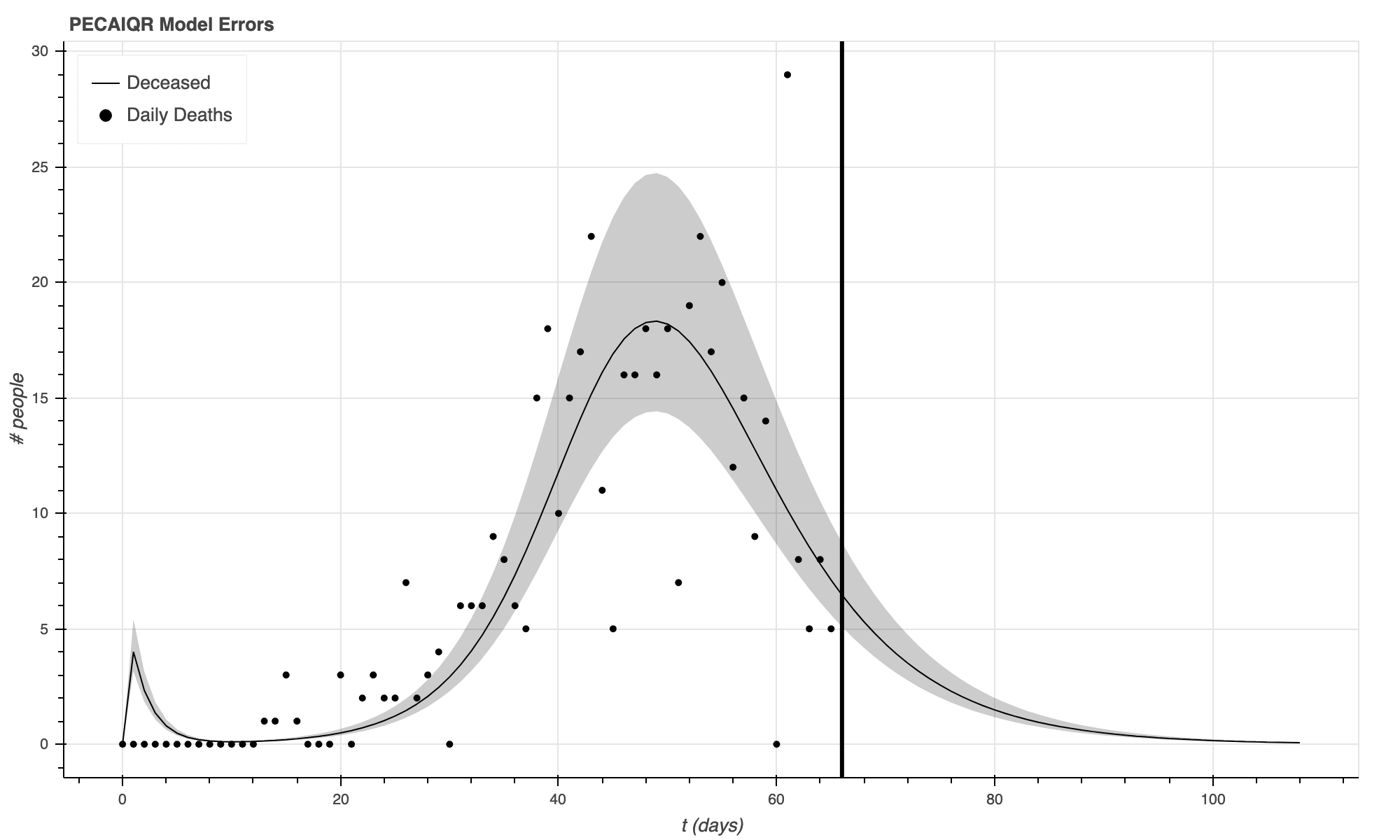}}}%
    \qquad
    \subfloat[]{{\includegraphics[width=7.5cm, height=4cm]{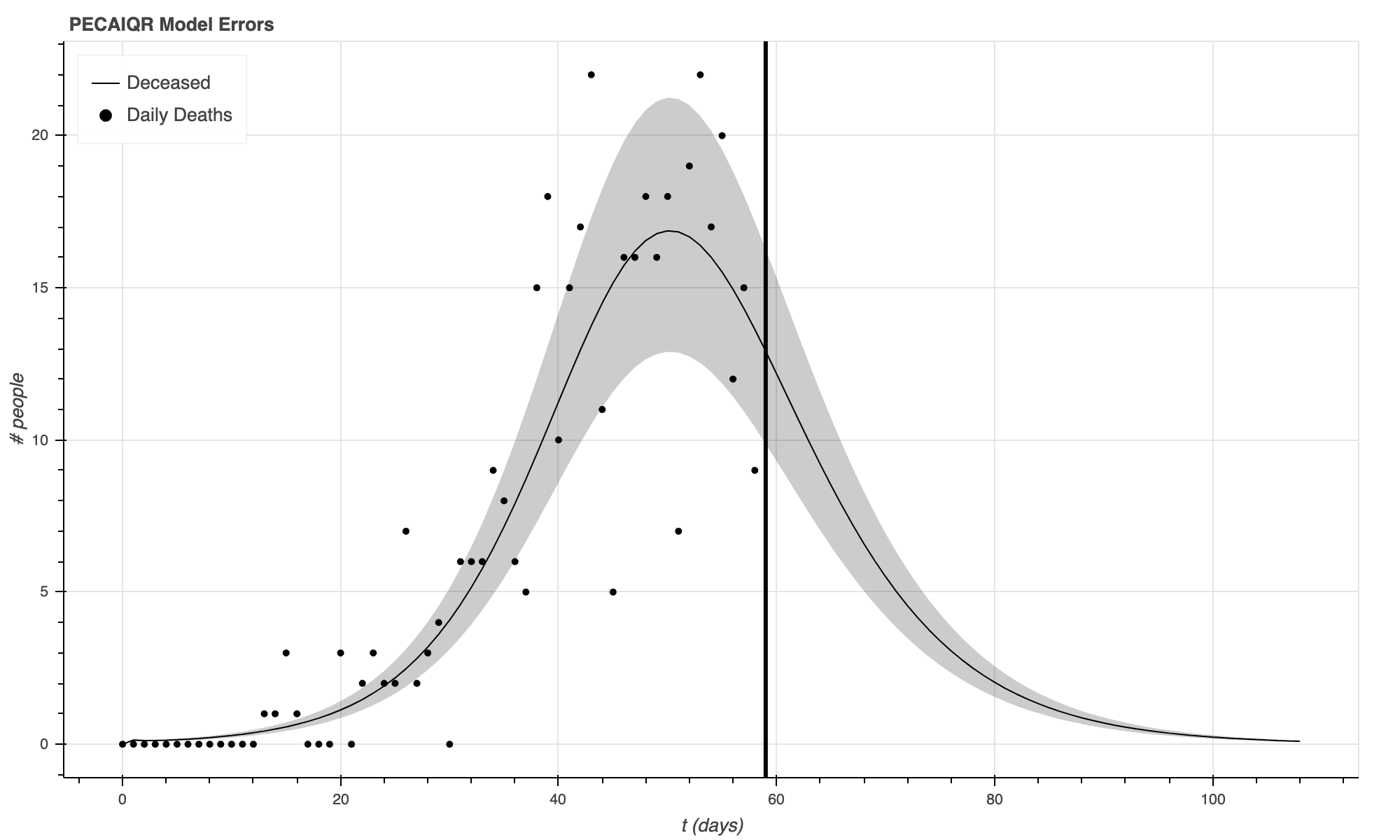}}}%
    \qquad
    \subfloat[]{{\includegraphics[width=7.5cm, height=4cm]{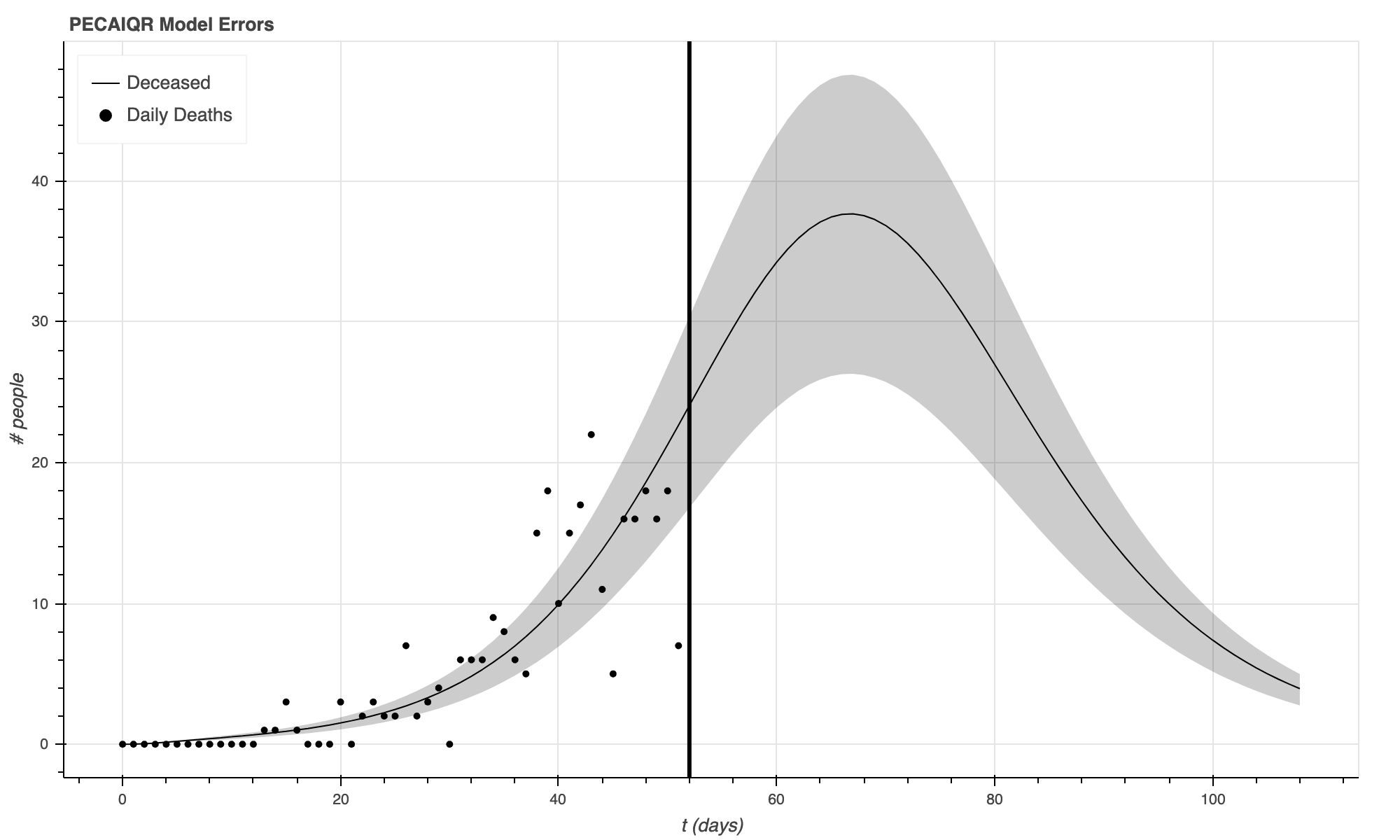}}}%
    \qquad
    \subfloat[]{{\includegraphics[width=7.5cm, height=4cm]{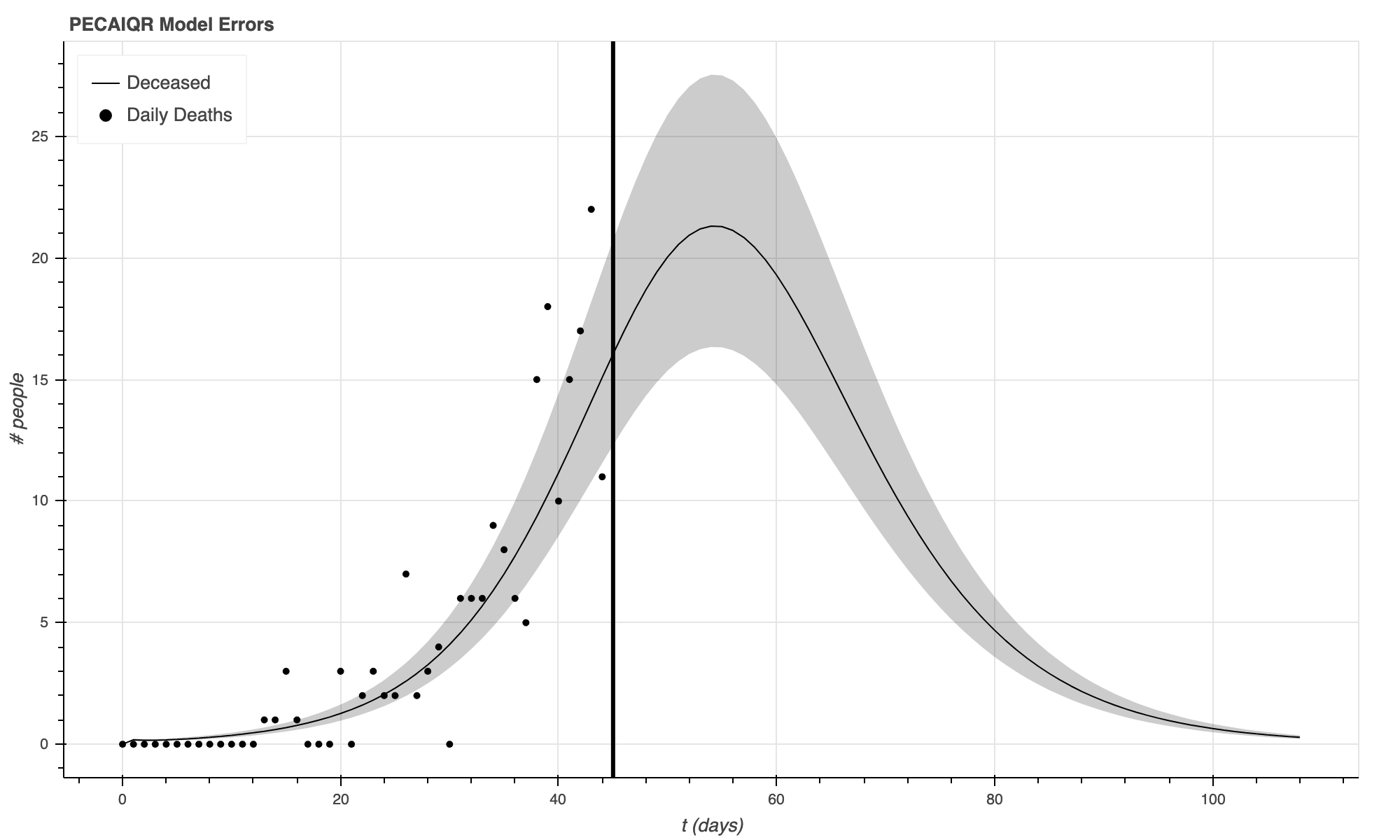}}}%
    \qquad
    \caption{Predictions on county 27053 with training date cutoffs at May 25 a), May 18 b), May 11 c), May 4 d) and May 4 e). Sub-figures d) and e) have the same cutoff, but different hyper-parameters.}%
\end{figure}
\noindent Analysis of model predictions for different cutoff dates in the training data shows that the model is quite stable and consistent when predicting on the region past the peak. One thing to note is that sub-figure a), trained on the most recent data, seems to have a much longer tail. This is a result of the high level of noise in the more recent data points, which are not included in the other cutoffs.\\

\noindent Predicting before the peak is much harder, as we are no longer operating with the assumption that the infection curves are dying down. In sub-figure d), we see that the fit is quite different from the fits with later cutoffs. The data before this cutoff is quite noisy, so the model cannot accurately predict when the infection curves will begin to die off. To fix this, in sub-figure e) we fit the model on active cases as well. The model is able to use the active case statistic to determine that the infection curve dies down earlier than it otherwise would predict. 
\clearpage
\section{Policy Effect Predictions}
\begin{figure}[H]%
    \centering
    \subfloat[]{{\includegraphics[width=7.5cm, height=4.5cm]{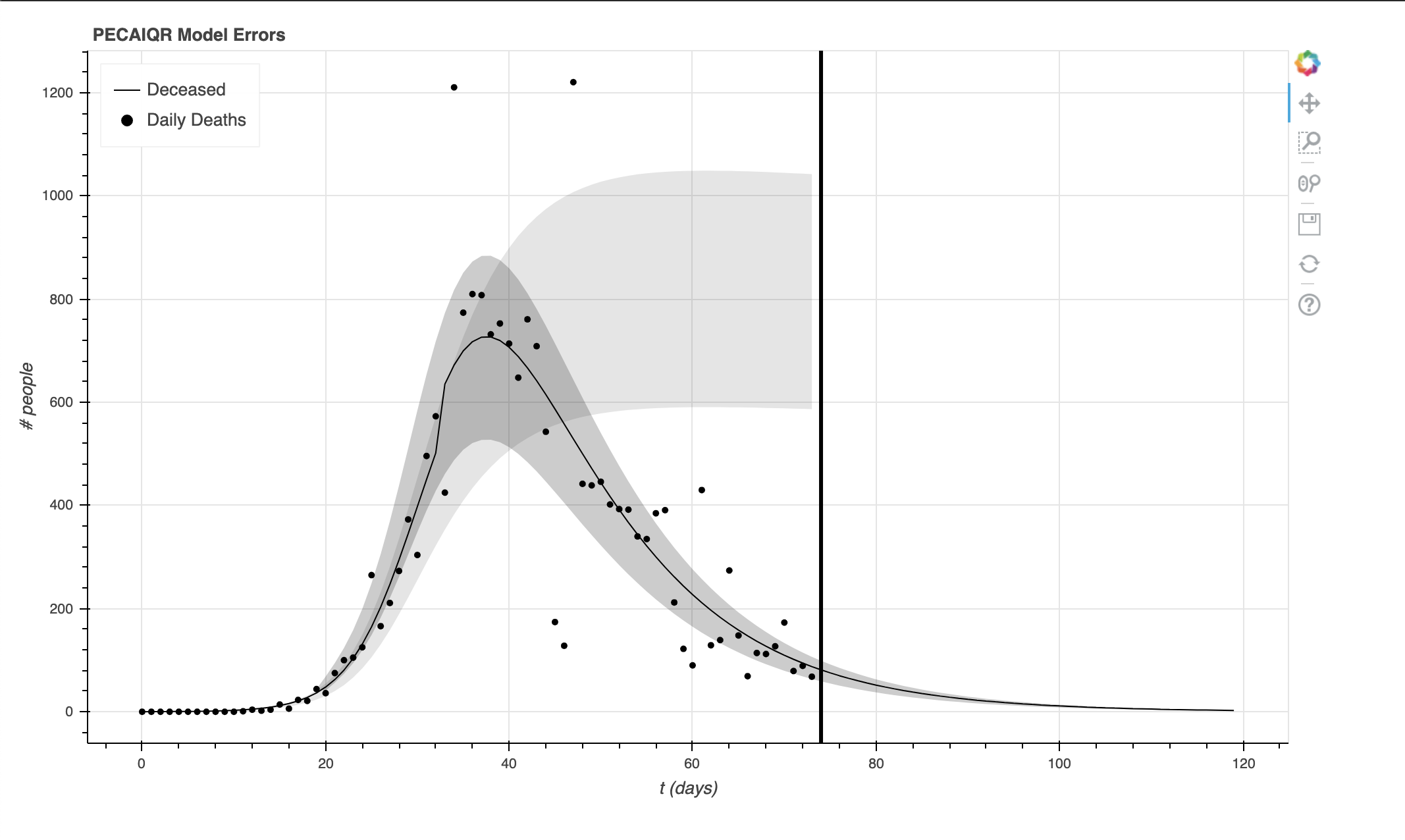}}}%
    \qquad
    \subfloat[]{{\includegraphics[width=7.5cm, height=4.5cm]{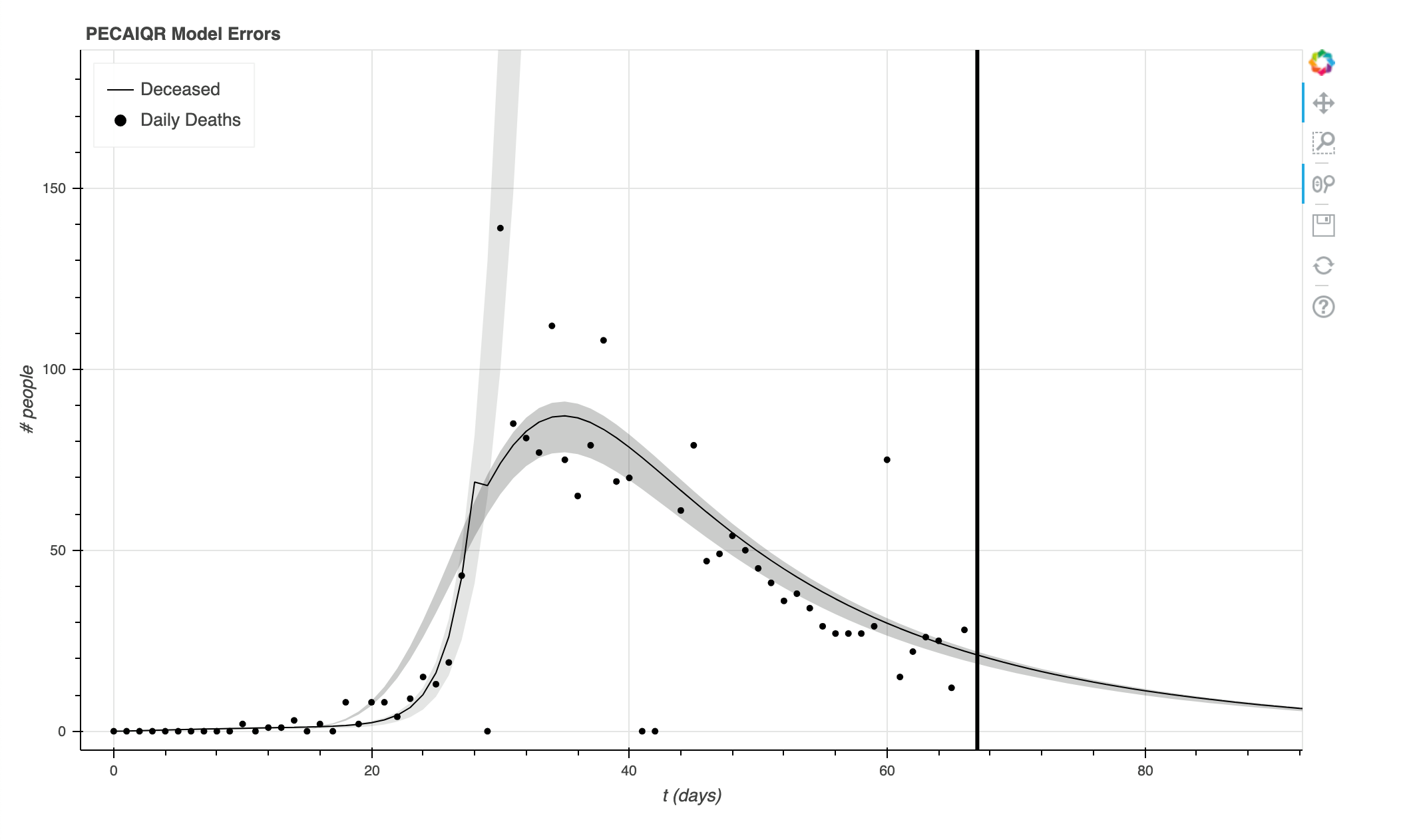}}}%
    \qquad
    \subfloat[]{{\includegraphics[width=7.5cm, height=4.5cm]{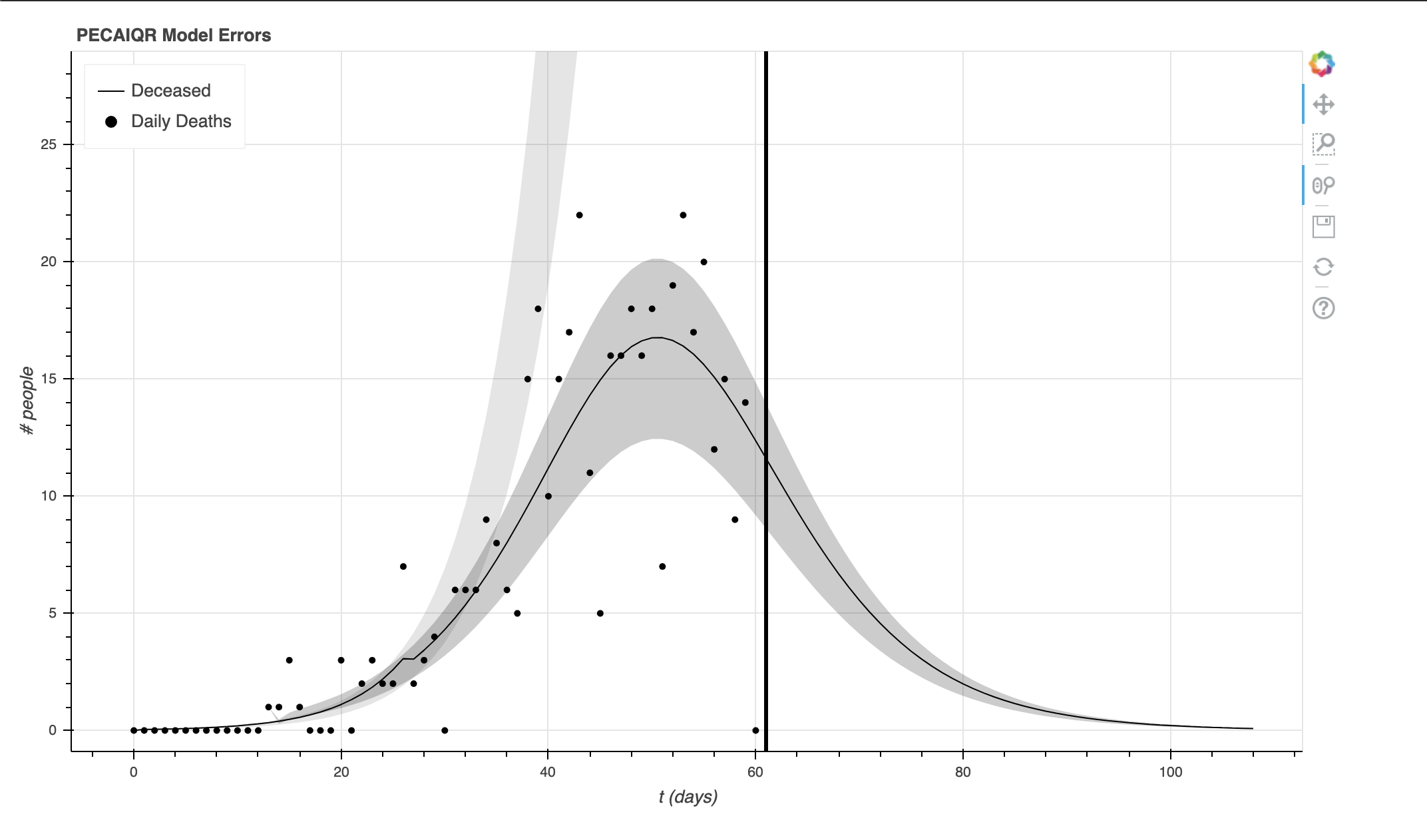} }}%
    \qquad
    \subfloat[]{{\includegraphics[width=7.5cm, height=4.5cm]{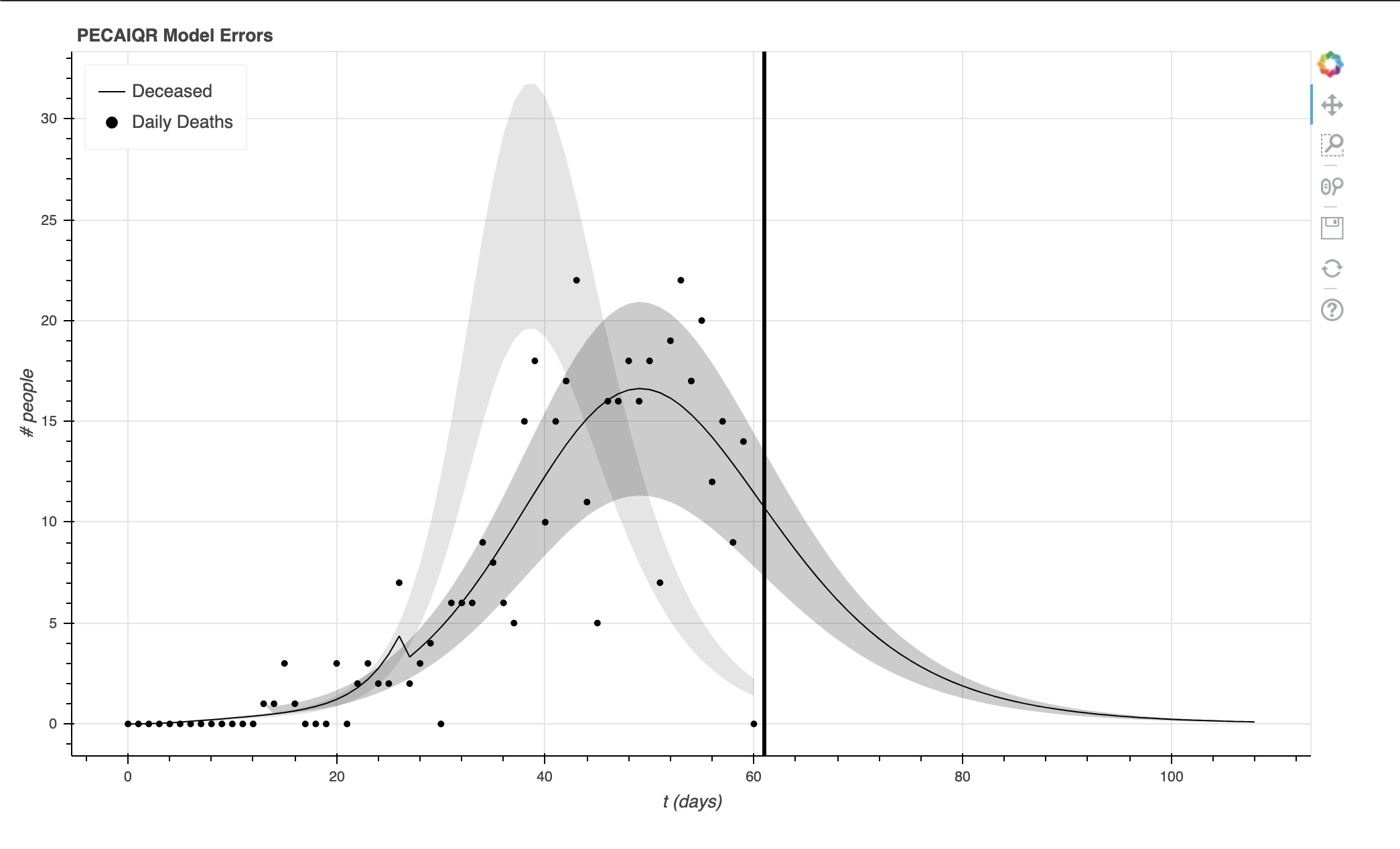} }}%
    \qquad
    \caption{Policy regime predictions on county 36061 a), county 36059 b), county 27053 c), and county 27053 with a moving average c).}%
\end{figure}
\noindent Here, we fit each of the three counties using the policy regime method described in section 3.1 \textbf{Fitting}. The faint gray curve shows the predicted infection curve on the data regime before the date of the stay at home order, plus the time of death, and the dark gray curve shows the predicted infection curve on the data regime after the stay at home order. In counties 36059 and 27053, the stay at home order seems to have flattened the curve, but in county 36061 the effects are more ambiguous. The is likely relates to factors specific to New York that worsened the outbreak, such as the high population density of New York City, and the shortage of hospital resources later on.\\
\section{International Comparisons}
\begin{figure}[H]%
    \centering
    \subfloat[]{{\includegraphics[width=7.5cm, height=5cm]{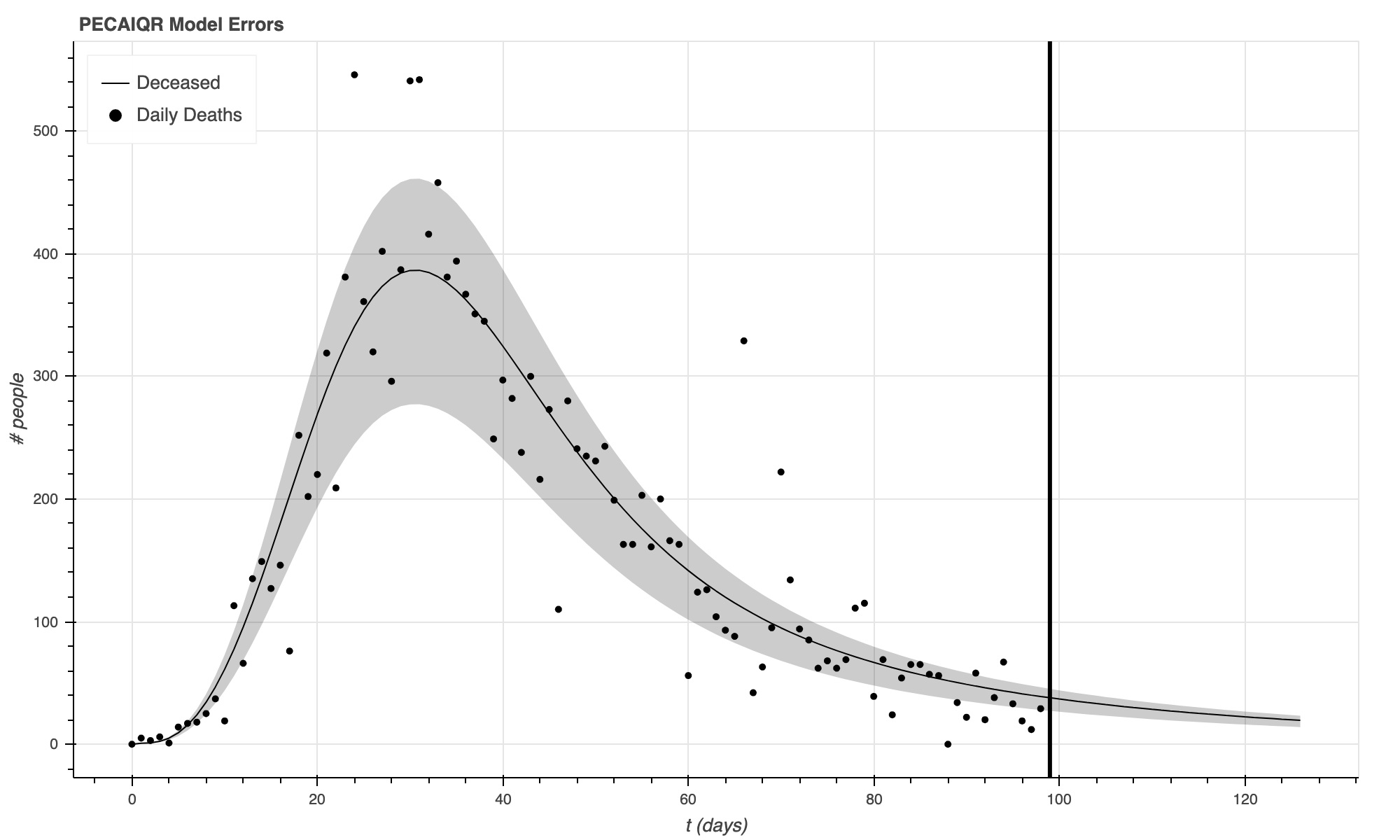} }}%
    \qquad
    \subfloat[]{{\includegraphics[width=7.5cm, height=5cm]{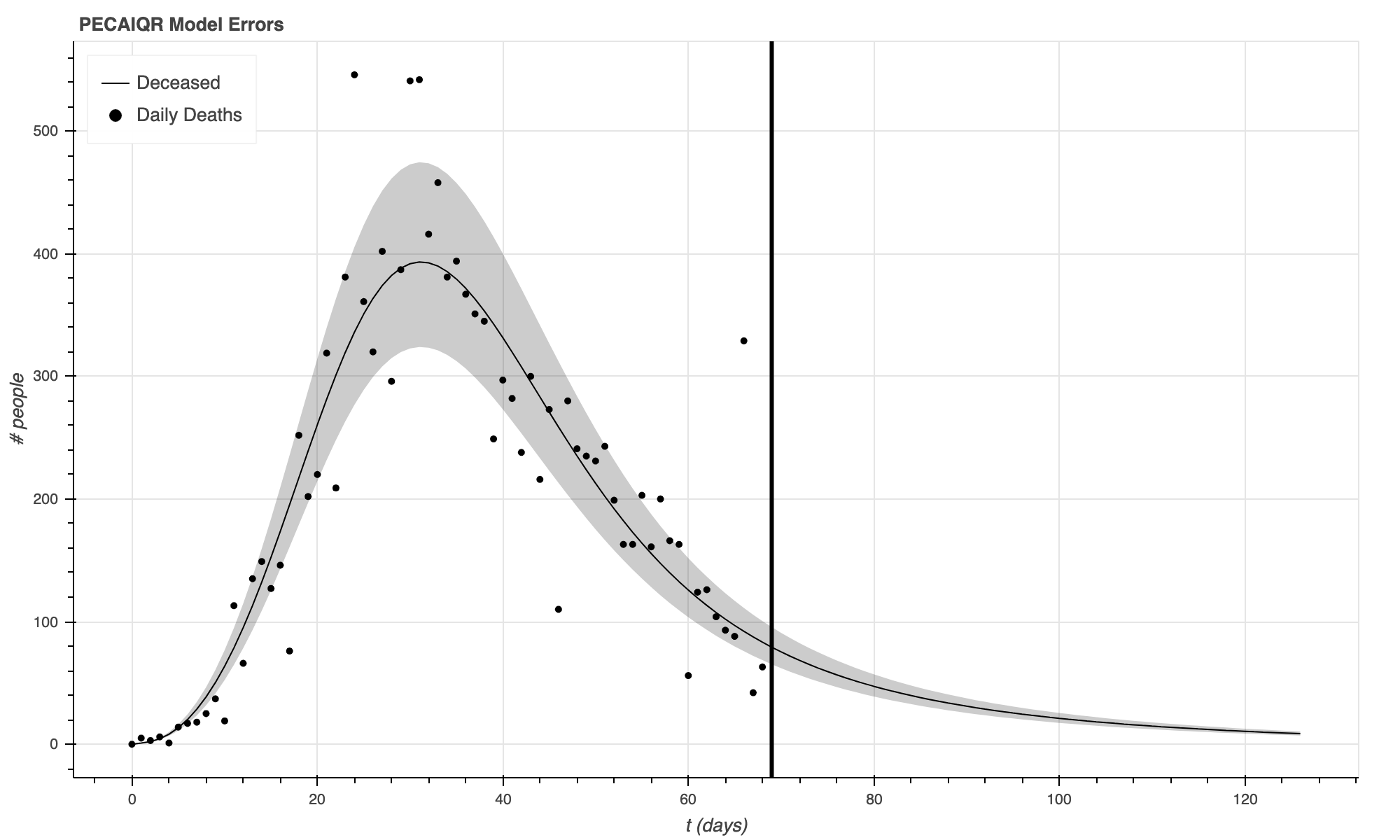}}}%
    \qquad
    \caption{Predictions on Lombardia with training date cutoffs at June 3 a) and May 4 b).}%
\end{figure}
\noindent The model can be easily adapted to train on International data. At the bare minimum, the model only requires death reporting and population. The predictions made from the May 4th cutoff, roughly a month before the June 3rd cutoff, shows that the prediction fit is quite stable and is consistent.\\
\section{Resource Use Predictions}
\begin{figure}[H]%
    \centering
    \subfloat[]{{\includegraphics[width=7.5cm, height=5cm]{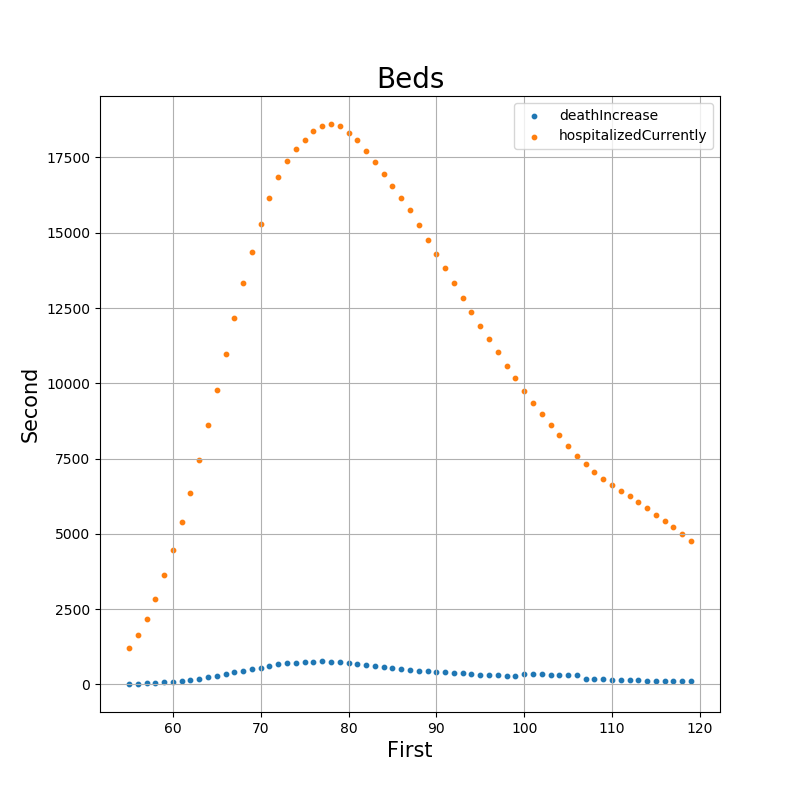} }}%
    \qquad
    \subfloat[]{{\includegraphics[width=7.5cm, height=5cm]{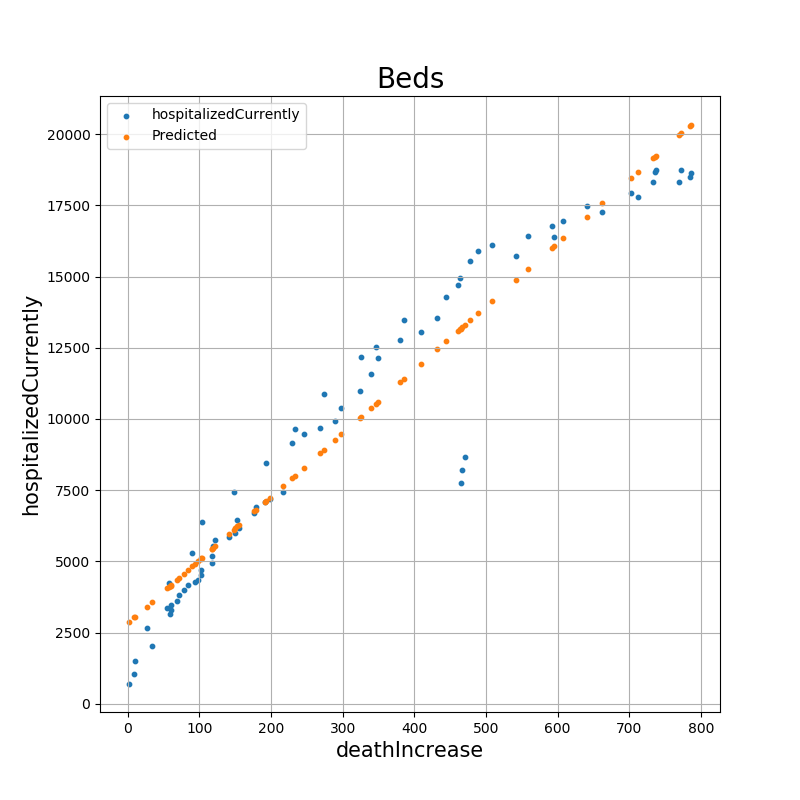} }}%
    \qquad
    \subfloat[]{{\includegraphics[width=7.5cm, height=5cm]{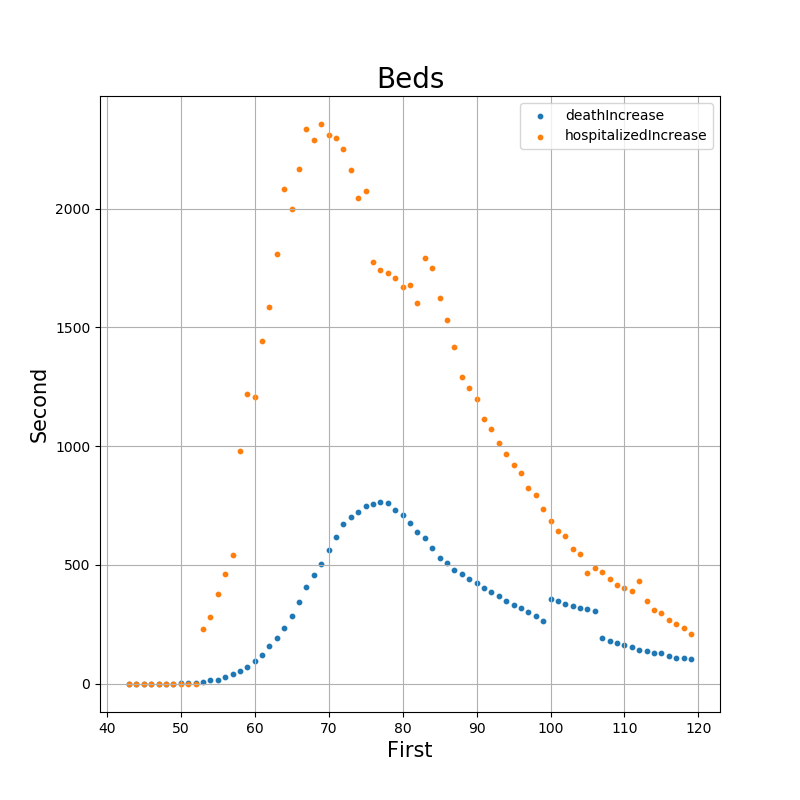}}}%
    \qquad
    \subfloat[]{{\includegraphics[width=7.5cm, height=5cm]{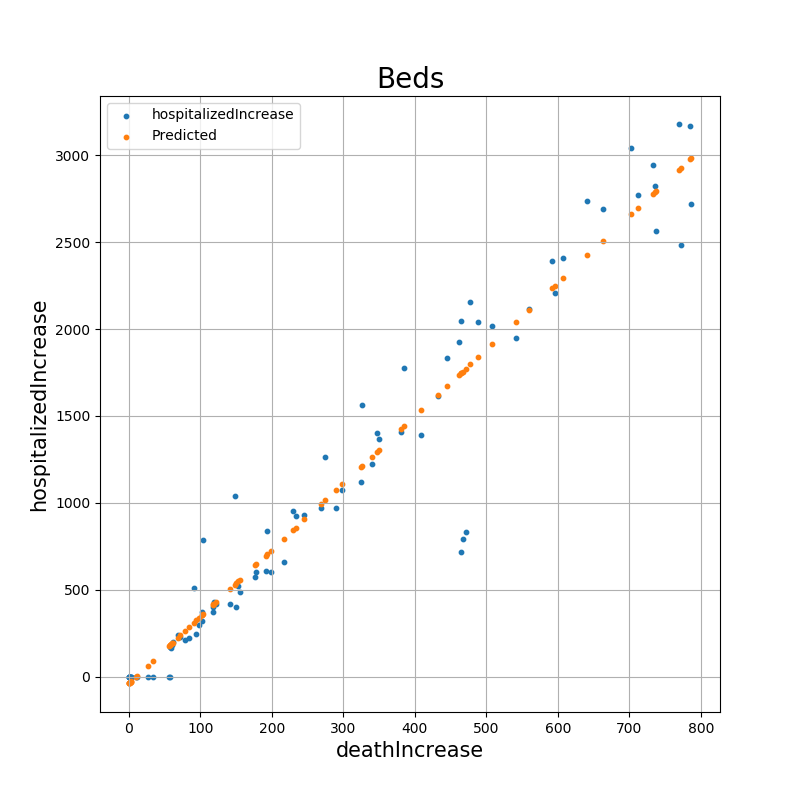}}}%
    \qquad
    \subfloat[]{{\includegraphics[width=7.5cm, height=5cm]{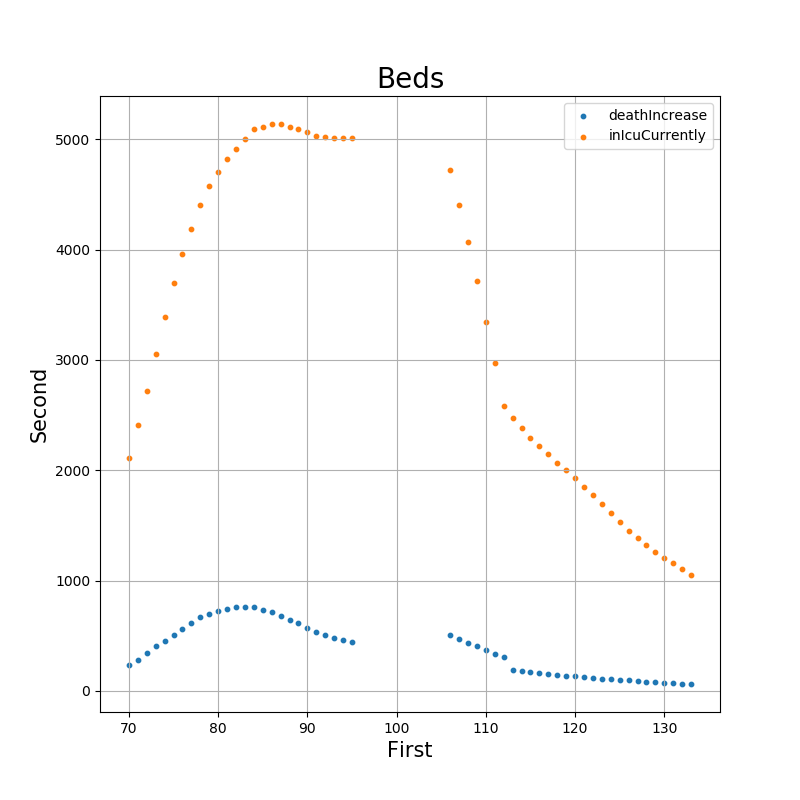}}}%
    \qquad
    \subfloat[]{{\includegraphics[width=7.5cm, height=5cm]{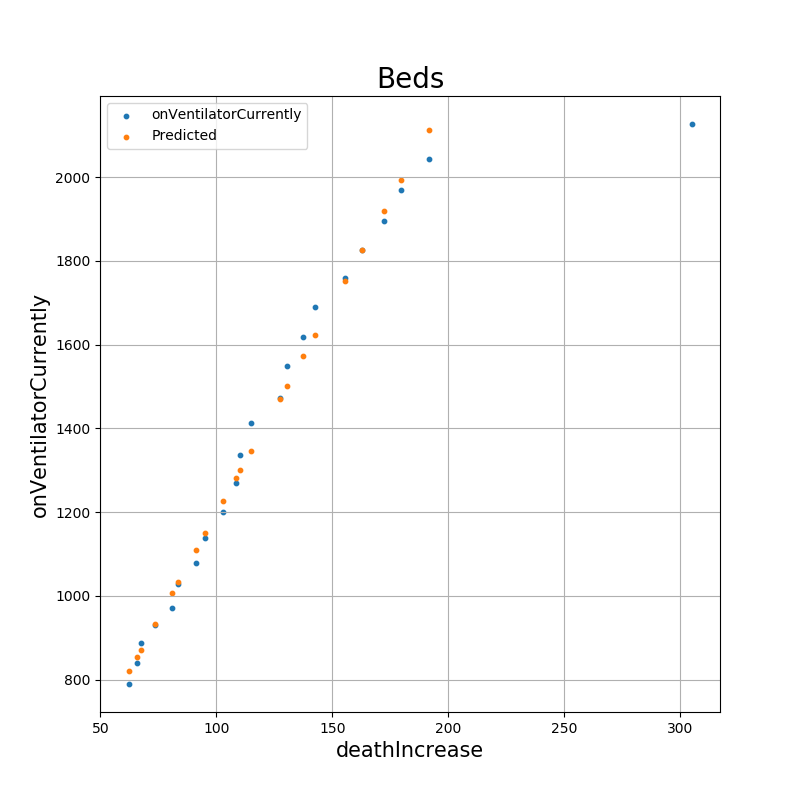}}}%
    \qquad
    \caption{The first column of figures shows daily deaths and currently hospitalized, daily hospitalized, and currently in ICU plotted, respectively, plotted against time. The second column shows the linear regression for daily deaths vs currently hospitalized, daily deaths vs daily hospitalized, and daily deaths vs currently on ventilator, respectively.}%
\end{figure}
\noindent To smooth out the data, we used a moving average with a window of 7 data points. We then aligned the peaks for each hospital statistic to match the peak of the daily deaths, in order to account for the offset term that corresponds with the average time between hospital admittance and death. The first column shows the hospital statistics before they were aligned with the death statistic, and there is clearly an shift, although only by a few days. This makes sense, as patients who are admitted to the hospital are likely patients who have already developed a severe condition.\\

\noindent Note that the hospital statistics and the death statistics are both Gaussian. This suggests that we can find some scaling factor once we align their peaks. For this, we perform a linear regression against the death statistic, revealing a definitive linear correlation for all the statistics. Note that there is a slight nonlinearity in the regression between currently hospitalized and daily deaths, as the daily deaths increases beyond a certain point. This may indicate that we are nearing the hospital capacity, and so the change in number of currently hospitalized patients begins to lag behind the change in daily deaths. Also note that there is an extreme outlier in the regression between currently on ventilator and daily deaths at the peak daily deaths value. Similarly, this is also likely caused by some limit on the number of ventilators available. Indeed, there seems to be a ceiling to the number of people on ventilators past a certain number of daily deaths.\\

\noindent Using the linear regression fit, we can now directly convert any of our model predictions, as well as their confidence intervals, to predictions for hospital resources.\\
\section{Data Challenges}
\noindent We noticed that the vast majority of counties have inconsistent reporting of deaths. Some counties even had cumulative death statistics that decreased on certain dates. Clearly this is not possible, as death is permanent. Other counties report constant values for cumulative deaths (zero values for daily deaths) for an extended period of time, followed by a quick spike. It is doubtful that all deaths suddenly occur in a single day, so this observation suggests that the deaths reporting might not be distributed correctly, perhaps due to administrative lag. For counties with low numbers of deaths, this creates large amounts of variance in the data, which makes it hard to fit. It seems like the deaths reporting also seems to be correlated with the day of the week. For many counties there is an interesting trend that the daily deaths increase over consecutive days before the weekend. Again, this might be due to administrative lag in the hospitals that report deaths.\\

\noindent The active cases statistics also seem to be poorly recorded, and this makes sense given how ambiguous the classification of an active case can be, especially with the limitations on testing. Many counties completely lack useful active case statistics, and in other cases, it is only available very late into the infection curve. In general, statistics that attempt to quantify the number of cases, whether active, cumulative, or daily, are intrinsically unreliable, as they depend on the availability of tests, which can vary greatly over time. Perhaps a more reliable statistic is the ratio of positive tests to administered test on any given day.\\

\noindent Another issue was with the non time-series data. Some files, such as the Age Race file from \cite{Age_Race} and the Aggregate JHU file from \cite{killeenCountylevelDatasetInforming2020}, had valuable information but also numerous holes. While this data was not used in creating the predictions, it was used heavily in Section 11 \textbf{Creative Data Visualizations} which details an attempt on clustering based on categories in these datasets. Such holes proved difficult to fill in. The filling in was done by using the data from the nearest county if a given entry was missing for a certain county. However this could be ineffective if this certain county and its nearest neighbor are very different in nature, as then the data for this certain county would not be particularly representative. While the data holes were not particularly impactful for the models described above, they certainly could be for models that took many non time-series features from those datasets into consideration. \\  
\clearpage
\bibliographystyle{bibft}\it
\bibliography{bibfile}

\end{document}